# Beyond relationalism in quantum theory:
## A new indeterminacy-based interpretation of quantum theory

Francisco Pipa[1]

Department of Philosophy, University of Kansas

Abstract

The received view in foundations and philosophy of physics holds that if we reject supplementing quantum theory (QT) with certain kinds of hidden variables and consider that unitary QT is correct and universal, we must adopt a relationalist interpretation of QT. Relationalist interpretations relativize measurement outcomes to, for example, worlds, systems, agents, or reference frames. They include the Many-Worlds Interpretation, Relational Quantum Mechanics, and QBism. These interpretations have potential costs connected with their relationalism that make them unattractive. Thus, if there exists a non-relational non-hidden variable universal interpretation of QT, it should be taken seriously. I will present an interpretation of this kind called Environmental Determinacy-based or EnD Quantum Theory (EnDQT), which goes beyond relationalism, potentially circumventing its issues and showing that the received view is misleading. EnDQT circumvents relationalism by constructing an account of indeterminate and determinate values and underlying quantum properties that isn't relational while maintaining unitary non-hidden variable universal QT. In situations where relationalists assume that measurement outcomes are relativized, such as in the extended Wigner's friend scenarios, there aren't determinate outcomes but systems with non-relational indeterminate values. In this interpretation, certain systems acquire determinate values at some point in time, and the capacity to give rise to determinate values through interactions propagates to other systems in time and space via local interactions, which can be represented via certain networks in space evolving over time. When there is isolation from the rest of the systems that belong to these networks, such as inside the friend's lab in the extended Wigner's friend scenarios, indeterminate values non-relationally arise inside. I will discuss other independent good reasons for adopting EnDQT, including providing a local causal explanation for Bell correlations and novel empirical posits represented by these networks.

---

[1] franciscopipa@ku.edu

# 1. Introduction

The extended Wigner's friend theorems (e.g., Bong et al., 2020; Brukner, 2018; Frauchiger & Renner, 2018) present a dilemma involving different interpretations of quantum theory (QT).[2] Let's analyze a simplified account of the scenarios underlying these theorems (Gao, 2018, Dieks, 2019).[3]

Consider the following EPR-Bell's-like scenario where we have Alice in an isolated laboratory so that there is "no leakage of information" arising from the interaction between the lab and the open environment.[4] Then, the contents of the lab can be coherently manipulated by performing arbitrary quantum operations on them, treating these contents as pure states. Space-like separated from Alice, there is Bob, who shares with Alice a pair of systems in the following singlet state,

$$|\Psi(t)>_S = \frac{1}{\sqrt{2}}(|\uparrow_z>_A|\downarrow_z>_B + |\downarrow_z>_A|\uparrow_z>_B).$$

---

[2] By interpretations of QT, I mean (sloppily) what we typically call interpretations of a theory, but also new theories.
[3] For the sake of brevity and to aid intuition, I will consider this simplified account throughout this article. However, a more precise statement of the dilemma presented here could be given, for example, by looking at some of the extended Wigner's friend theorems. Roughly, in the simplest theorems (e.g., Bong et al., 2020; Brukner, 2018), typically there are two space-like separated friends sharing an entangled pair and two Wigners performing operations on the friends plus the friend's measured target system. Then, based on the assumption that these four outcomes were (absolutely) obtained together plus certain assumptions regarding the statistical independence and locality, a joint probability distribution is derived, and an inequality violated by quantum correlations.
[4] See, e.g., Zurek (2003).

Suppose that Alice and Bob perform randomly selected spin measurements on their systems in different directions. Next to the laboratory of Alice, there is Wigner, also space-like separated from Bob. After Alice measures her system, Wigner, who treats the interactions between Alice and her system as evolving unitarily, performs certain operations on Alice plus Alice's system that reverses their quantum states to the previous ones before her measurement.

Let's consider two situations where Alice measures her system in a spin-z direction in two relativistic inertial reference frames. Let's consider frame 1 (lab frame). The quantum framework predicts that Alice will obtain spin-z up with probability ½ and spin-z down with probability ½. Then, Alice's result is reversed by Wigner, and Bob makes a measurement. QT leads to the prediction that he will obtain spin-z up with probability ½ and spin-z down with probability ½, conditioning on the state of Alice. Let's call Alice and Bob "friends."

Because the friends are space-like separated, according to relativity, we can choose another inertial reference frame, frame 2. In this frame, Alice measures the spin-z of her system, then (contrary to the frame 1 case) Bob measures his system, and then Alice's measurement is undone by Wigner. In this situation, if Alice obtains, for example, spin-z up, the quantum framework predicts that Bob, conditioning on the outcome of Alice, will obtain spin-z down with 100% probability if he measures it on the same basis as Alice. However, in frame 1, according to the predictions of standard QT, Bob's result doesn't need to be spin-z down.

So, QT can yield two contradictory predictions. This leads to the following dilemma, assuming that we hold A-E) where

A) situations like the one above are physically possible or at least instructive (more on this below),

B) unitary QT (no modification of the dynamical equations of QT). A modification of the dynamical equations would lead to ammmm threshold at which the state of Alice inside her isolated lab would collapse,

C) we don't want to add certain kinds of "hidden" variables, such as the ones involved in future boundary/teleological conditions,[5] retrocausal or superdeterministic interpretations, that could perhaps solve the contradiction;[6]

---

[5] See, e.g., Kent (2015).

D) QT is universal, i.e., it applies in principle to any physical system,

E) We respect relativity by not choosing a preferred frame (unlike in the case of Bohmian mechanics, which violates this condition by adopting non-local hidden variables),[7]

Then,

F) we should adopt what I will call a relationalist interpretation of QT, explained below.

Let's call this dilemma the Wigner's Friend dilemma (which may be considered a version of the measurement problem).[8] The received view in foundations and philosophy of physics typically holds something like this dilemma, as one can see by inspecting the Extended Wigner's Friend Theorems and prominent literature about those theorems.[9] For instance, Brukner (2022) writes,

"The startling conclusion [of the extended Wigner's friend scenarios] is that the existence of 'objective facts' shared by Wigner and his friend is incompatible with the predictions of quantum theory as long as assumptions of 'locality' and 'freedom of choice' are respected. Experiments were conducted that confirmed the predictions of the no-go theorems, albeit with photons as a primitive model of Wigner's friend. We can conclude that although the facts of the friend and those of Wigner cannot be considered 'jointly objective', they preserve the objectivity 'relative to the observation or observer'."

Holding the lesson above, there are multiple views that provide a *relationalist interpretation of* QT that allow them to deal consistently with the above dilemma without falling into A)-E), and this conservatism of relationalist interpretations is often regarded as advantageous. This is because the options A)-E) come accompanied by certain mostly well-known and often undesirable consequences, which I will not explore here. Relationalist interpretations include Everett's relative-state formulation of QT,[10] the Many-Worlds Interpretation (MWI),[11] relational quantum mechanics (RQM),[12] QBism,[13] Diek's perspectival modal interpretation,[14] and Healey's pragmatism.[15] They involve different strategies, all of which have potential issues that may make them undesirable. I will present some of these strategies and some of their drawbacks.

---

[6] In this article, when I refer to hidden variables, I will be referring to hidden variables of this kind and of the non-local kind.

[7] Another interpretation called consistent histories (see, e.g., Griffiths, 2019) appeals to certain rules that are violated when we consider situations where systems (such as Alice) can be both in a superposition and having a determinate outcome. The projector onto the state in which she is in a superposition with the system doesn't commute with the projector onto the state that represents the situation where she is observing a determinate outcome. However, this interpretation per se is insufficient to address this dilemma because the lab could be as big (or macroscopic) as we want so that from the point of view of inside the lab, we would be pressed to represent the state of affairs inside of it by a state corresponding to Alice having observed a determinate outcome. What is the correct state to attribute to Alice?

[8] One could argue that the Wigner's friend dilemma is an expression of the measurement problem whose more traditional formulation is given in Maudlin (1995) and Myrvold (2018). However, it differs from this formulation by, among other things, not appealing to talk concerning the completeness of the description provided by quantum states/wavefunctions.

[9] See e.g., Bong et al. (2020), Brukner (2018, 2020, 2022), Leegwater (2022), and Frauchiger & Renner (2018).

One strategy, which I will call the *privatistic strategy*, is to argue that outcomes are relative to certain entities such as systems or agents. When entities, such as systems, are isolated from each other and perform certain measurements, for a system S, the other system S' doesn't obtain any (determinate) outcomes, and vice-versa. So, entities may disagree about the outcomes obtained. This strategy is followed by, for example, at least a more traditional version of RQM (Di Biagio & Rovelli, 2021), as well as QBism. No facts or determinate outcomes occur inside Alice's sealed lab for Bob; hence, we cannot predict Bob's outcome by conditioning on Alice's outcome. Facts/determinate outcomes are relative in these contexts, and therefore for Bob, there is no fact/no determinate outcome about Alice; although relative to Alice, there is a fact about her outcome/determinate outcome. To be consistent, these views consider that facts/determinate outcomes are fundamentally relative and private. There are many reasons to be unsatisfied with this strategy. One of them is that it becomes harder to justify something we take for granted in doing science, which is the objectivity of measurement outcomes or even scientific objectivity in general. The other is that it represents the world in a complicated and fragmented way, full of multiple private perspectives, becoming difficult to see how precisely they agree and how to justify that agreement.[16]

---

[10] See, e.g., Barrett (2018) and references therein.
[11] See, e.g., Wallace (2012).
[12] See, e.g., Di Biagio & Rovelli (2021) and Adlam & Rovelli (2022). It might seem that in this later version of RQM, facts are absolute in certain contexts where decoherence occurs. However, as also recognized Healey (2022), when facing scenarios like the ones above, this later version is pressed to assume relative outcomes to deal with the Wigner's friend dilemma.
[13] See, e.g., Fuchs & Stacey (2019).
[14] See Dieks (2019) and references therein.
[15] See, e.g., Healey (2017) and Healey (2022).
[16] There are at least two versions of this strategy. In one, determinate outcomes are relative, but the systems aren't. In another one, even the existence of systems is relative.

Another strategy that I will call the *single-world multiplicity strategy*, which in a sense may be combined with the previous strategy, holds that systems may have obtained outcomes relative to each other, but outcomes vary according to entity E. So, within a single world there are a multiplicity of outcomes varying according to E. For Bob, Alice inside the lab may have obtained a determinate outcome and vice-versa, at least in certain circumstances. However, determinate outcomes in a world vary according to E. For example, as we have seen above, some single-world relationalists (Dieks, 2019) could claim that the different outcomes of Bob can vary relative to different simultaneity hyperplanes.[17]

Unfortunately, even if this relationalist strategy works for some entity E, situations like those above can quickly become complex.[18] For instance, they can involve an arbitrary number of N friends measuring systems and an arbitrary number N' of Wigners that perform measurements on these systems plus the friend on a different basis, where the Hilbert space of the quantum states of the entangled systems could have an arbitrary number of dimensions! By considering the possible different orders between events that arise, we can see that these relationalist interpretations require a complicated story about how the myriad of compatible and incompatible outcomes form different sets of "relative facts" depending on the hyperplanes, or more generally depending on the different instances of entity E, and which sets there are.

Yet, another strategy is the *many-worlds multiplicity strategy*. In situations like this one and even others, multiple outcomes occur, each relative to different worlds. This is famously the strategy adopted by the MWI of the splitting variety. They postulate a branching into different worlds in Alice's open or closed lab (if it's big enough to give rise to decoherence). Then it's impossible to condition Bob's outcome based on Alice's (single) outcome since Alice has multiple outcomes relative to different worlds. The main issue of this view is the well-known issue of justifying probabilities in QT.[19]

Notice that in order for Wigner to reverse Alice's and her system's state, he has to be able to correctly treat them and their interaction as evolving unitarily. This contrasts with how Bob treats Alice as obtaining an outcome inside her lab. Thus, in this scenario, there is a certain perspectivalism relative to the physical state of Alice. Such perspectivalism is accepted by relationalist views that may consider determinate outcomes as arising from a certain perspective, such as for a certain system S that could be within a world, but from another perspective, S plus the system that S measures are still correctly described by a superposition of states.

---

[17] Healey (2022) presents a related view based on contexts of assessments.
[18] See, e.g., Bong et al. (2020) and Leegwater (2022) for slight complexifications of the above scenario.
[19] See, e.g., Albert (2010); Kent (2010) and Price (2010).

One option when facing these scenarios is to be skeptical of their physical realizability and even physical possibility, adopting option A). However, besides being an ad-hoc way of setting aside these at least logical possibilities, this option neglects another important foreseeable possibility given future technological developments: one day, we might be able to implement extended Wigner's friend scenarios in quantum computers involving quantum agents as friends that make operations on quantum systems. Conceivably, these computers could be as large or complex as we want.[20] Also, what to say in the case the agent is a quantum computer, talks to us, and gives us determinate answers to our questions, even if its quantum circuits process qubits in a superposition?[21] So, even granting that these scenarios are impossible, we still have to have a clear answer about how to interpret what it means for a system to be in a superposition and when and how they give rise to determinate outcomes. Even if we invoke decoherence of the contents of the lab by its environment E or of the quantum computer by its environment E as a criterion for determinacy, we could imagine another powerful quantum computer E' that uses the system-environment E entangled states to make quantum computations, even if the system is being decohered by E. In a sense, we would have decoherence, but still, quantum superpositions that have a clear function of performing computations (and again, we have the Wigner's friend dilemma). So, a view that doesn't face the above more idealized scenarios might have a certain explanatory deficit in accounting for certain more realistic scenarios or hard-to-judge cases.

Thus, taking seriously extended Wigner's friend scenarios, if a non-relationalist interpretation is found that preserves standard QT, not falling into A)-E), relationalist accounts may be regarded as undesirable because non-relationalist interpretation may not fall into the problems of relationalist interpretations. I will argue that EnDQT is that alternative interpretation. Therefore, given the potential issues of current relationalist interpretations, EnDQT should be taken seriously as a viable new interpretation of QT. Moreover, EnDQT possesses other potential benefits that can make it even more attractive, as I will discuss in the last section (section 4). Let's see below how to understand EnDQT strategy.

---

[20] See, e.g., Wiseman et al. (2022) for an extended Wigner's friend theorem, where the friend is a quantum computer and Wigner is a user of this quantum computer. It is also presented experimental proposal concerning these scenarios.
[21] See Appendix 2 for EnDQT account on these scenarios.

Instead of opting for relationalism to deal with the above scenario, another possibility is to adopt an *indeterminacist strategy*. This strategy imposes certain constraints that allow us to deny that Alice, in her sealed lab or any system in those circumstances, obtains determinate outcomes, and contrary to relationalist interpretations, not having determinate outcomes is objective and not relative to Bob or anyone (such as systems, agents, worlds, etc.). Hence it is denied that we can condition the outcomes of Bob on the outcomes of Alice in any isolated-lab circumstance. So, the assumption that we can collect statistics or assign probabilities (which are based on determinate outcomes) or quantum states (representing systems with determinate values) or determinate values or elaborate frequencies based on the state of affairs of Alice in her sealed lab is denied, and no relative outcomes exist because there are no determinate outcomes at all. This is generalizable to any extended Wigner's friend scenario since all of them make these assumptions.

Collapse theories are an example of an interpretation that adopts a particular indeterminacist strategy in certain circumstances. In this case, a system absolutely lacks a determinate value of some observable up until a certain collapse of the quantum state, which is given by a certain modification of the Schrödinger equation. However, this particular strategy falls into B).[22]

A tempting naïve indeterminacist strategy is to use decoherence to establish the conditions under which the system loses the ability to give rise to interference, yielding classical behavior and, arguably, having determinate values of certain observables during the interactions that give rise to decoherence. Then, one can claim that there are no outcomes or determinate outcomes inside the lab because there is no decoherence. Decoherence appeals only to standard QT, and that might be regarded as a virtue.

However, this attempt fails for at least two reasons. I will go over these reasons because they will motivate the strategy adopted by EnDQT.

The first reason is that decoherence is usually expressed in the Schrödinger picture. In this picture, systems interact locally, becoming entangled, and continue to be

---

[22] Another indeterminacist strategy is the teleological one of Kent (2015). Basically, there aren't determinate outcomes in the present inside the lab because they are erased by Wigner and this prevents the existence of a (determinate) record of these outcomes in a certain special moment in the future. See Leegwater (2022) for a more detailed explanation of how this view deals with these scenarios via this strategy in this teleological way. Note that there is something conspiratorial about this view. The outcomes of space-like separated friends being indeterminate or not in a certain present depends on the future choices of the Wigners to erase these outcomes or not, which can be assumed to be random. However, in order to be the case that in the present there are determinate (indeterminate) outcomes, the future measurement choices of Alice and Bob, and what determines those choices, have to conspire to render them determinate (indeterminate). Another one is the bare theory, which is the standard quantum theory with no collapse, and assuming the eigenstate-eigenvalue link (see section 2.2). This interpretation has been shown to be problematic on several grounds (see, e.g., Albert, 1992; Barrett, 2001, 2019)

that way even if space-like separated. This strategy would press us to fall into a MWI relationalist strategy, or introduce hidden variables or collapse dynamics to explain how measurement outcomes are obtained via these non-locally entangled systems, and we want to circumvent that. A well-known strategy to adopt without abandoning the Schrödinger picture is to assume that quantum states in a superposition don't literally represent physical facts one by one but work more as indirect predictors and inferential tools inserted in the Born rule to yield predictions and make inferences about the features of systems. Then one interpret decoherence

"as a condition for when we can impose an explicit collapse rule without empirically contradicting quantum theory. But that 'condition' is precisely the condition in which we can get away with treating the quantum state probabilistically, and from that perspective, 'collapse' is just probabilistic conditionalization." (Wallace, 2019, pp. 22)

So, by adopting this strategy, quantum states shouldn't be regarded as things in the world. Moreover, the measurement of Alice at one wing of a Bell scenario shouldn't be considered to influence or collapse the state of Bob at the other wing, or vice-versa. Alice just learns about her system and updates her information about it, updating her state. She may also learn about Bob's system at the other wing. EnDQT will adopt this strategy to some extent.

One of the strategies that could complement the above one is to keep track of local interactions (i.e., between non-space-like separated systems) through, for example, a network representing the different local systems qua relata of these interactions that give rise locally to decoherence at different times and over space. So, in an EPR-Bell scenario, Alice should only take into account the entanglement of the above quantum state $|\Psi(t)>_S$ with her (local) environment if she wishes to also use decoherence to account for the measurement account. The same for Bob.[23] Let's call this strategy and the one mentioned in the previous paragraph, the informed indirect-predictor strategy.

We might think that by adopting this strategy, quantum states cannot assume other representational roles. However, we can have our cake and eat it too. Given an accepted ontology and adequate representational tools, the interpreters of a theory can stipulate the best representational role of its elements in the process of attaining knowledge about quantum entities and their properties. Accordingly, I will view quantum states as still representing some ontology via what they tell us about the so-called quantum properties instantiated by systems (which are further represented via observables) and how systems interact with each other, which is further represented through that network. It's often defended in philosophy that we should be realists about quantum states due to their predictive and explanatory role. However, this role of quantum states in standard QT typically comes together with their observables and other considerations. It's unclear that in this relation between quantum states and observables, we need to reify quantum states rather than relate them to represent something else. In our case, they will represent certain so-called quantum properties of quantum systems and their values.

---
[23] More on this in section 4.

This strategy may lead to progress, but it's still insufficient because of the second reason the naïve attempt above fails. Decoherence is insufficient to address the above problem. One way to see this is to note that the friend's lab can be as big as we want so that we would say that there is decoherence arising from the interaction between the system and its environment inside the lab, and so we would be pressed to regard the friend as obtaining determinate outcomes. Adopting an indeterminacist strategy, we seem to need to supplement standard QT with some criteria that allow us to distinguish between the decoherence occurring inside the isolated lab that cannot be used to account for certain observables of the system having determinate values during its interactions with the measurement device, i.e., the so-called "virtual-decoherence" (which is "a kind of decoherence" that can always be reversed, such as by Wigner); and the "real" decoherence inside non-isolated labs that may be used to account for the determinate values of certain observables of the system during its interactions (and which cannot be reversed at least effectively such as by Wigner in this example). So, it seems that we should include this criterion in the informed indirect-predictor strategy by, for example, using the resources that the ontology represented by the above-mentioned networks provides.[24]

So, a more successful strategy could use the global features of the local *environment* inside and outside the lab to constrain and influence a system having determinate values of some observables inside the lab during interactions, but without modifying standard quantum theory, adding hidden variables (of the kind mentioned above), or making quantum theory not universal. Like certain relationalist interpretations, the systems "would have determinate values of certain observables" only in interactions with other systems. However, contrary to relationalist interpretations, both having determinate values and not having determinate values is something absolute and objective due to how certain systems interact (and not relative to anything); and having these determinate values depends on certain global features of the environment whose origin dates back to the past (but not determined by the future and so, no hidden variables). I will call this particular indeterminacist strategy an *environmental indeterminacist-based strategy*.

---

[24] See, e.g., Zeh (2003) for the distinction between virtual and real decoherence.

This is the strategy adopted by the view that I will call Environmental Determinacy-based Quantum Theory, which I will shorten as EnD Quantum theory or EnDQT. The point of the name is that this view takes very seriously the idea that the environment of a system can affect the determinacy of values of quantum systems. As we will see, by adopting this strategy, EnDQT won't fall into A)-F).

EnDQT proposes a local ontology with the above features that we gain knowledge about through QT, in agreement with the indirect predictor strategy. This ontology involves systems that have two kinds of properties, quantum properties and value properties (henceforward, values), where the values are understood via the more fundamental quantum properties. These later properties are properties that fundamentally quantum systems have.[25] Quantum properties may differ from each other via their so-called degree of differentiation. For instance, one system might have a certain quantum property momentum with a certain degree of differentiation D*', and another system can have a quantum property momentum with a higher degree of differentiation D*'', i.e., D*'' > D*'. The amount of real or virtual decoherence roughly quantifies the degree of differentiation of certain quantum properties designated as D*-P that a system instantiates, which is quantified via the amount of distinguishability of the quantum states of the environment that gets entangled with the quantum states of the system, where the latter are eigenstates of the self-adjoint operator concerning P. P could be energy, momentum, a spin direction, etc. So, certain environments that give rise to real or virtual decoherence affect the degree of differentiation of quantum properties.

Similarly, values come in terms of different degrees of determinacy. Only when a system has a value (e.g., the value momentum) with a maximal degree of determinacy is its value (property) represented by certain values (e.g., momentum equal to 10 kg m/s) or value intervals. Values are functionally defined in terms of the system having certain quantum properties under interactions with certain specific systems. More precisely, for a system to have a value with a certain degree of determinacy D is to have a certain quantum property with a certain degree of differentiation D*' where D=D*' under interactions with these specific systems.[26]

---
[25] Or at least more fundamentally than values.

As I will argue, the first systems with determinate values arose in the past[27] through certain special systems with certain quantum properties (which I call initiators). Moreover, these systems started certain chains of local interactions over time and space and maintain the existence of these chains. These chains allow determinate values to persist over time and space, and to propagate, giving rise to other systems having determinate values. They can be represented by certain evolving networks where the networks will represent how systems relate in terms of frequent interactions in space and over time.

More concretely, they involve certain systems S that influence other systems S' and, as I will say, *value-determining* their quantum properties in space and over time, i.e., making S' have a quantum property with the highest degree of differentiation and allowing determinate values to arise during interactions. Moreover, importantly, for EnDQT, value-determination also allows S' to give rise to determinate values in interactions with other systems. Without value-determination, S' cannot give rise to determinate values in other interactions. So, in turn, S' interacts with systems S'' value-determining their quantum properties in space and over time, and also allowing determinate values to arise and S'' to give rise to determinate values in interactions with other systems, and so on. These chains will be called stable differentiation chains (SDCs) and the process of value-determination will be typically represented with the help of decoherence (see Figure 1), being presupposed by this technique when assuming EnDQT as I will argue.

So, as I will explain in more detail, like the origin of matter or spacetime in the early universe, there was also "the origin of determinacy." Also, like the expansion of the universe, "there is also the expansion of determinacy." Systems instantiating quantum properties that don't belong to certain SDCs with respect to some quantum properties belong to the so-called unstable differentiation chains (UDCs) with respect to these quantum properties, where both SDCs and UDCs belong to the so-called differentiation makeup (DM) of the world.[28] Members of UDCs during a time interval can be, for example, isolated systems or any system that doesn't "connect back" directly or indirectly (via interactions) to the initiators of an SDC during that time interval.

---

[26] Although one may be skeptical of its existence, I will take the perspective that *there is* such a thing as indeterminate values or quantum indeterminacy, and that it comes in terms of degrees. In my view, it seems more explanatory to assume that that there is indeterminate value and degree-based determinate values than determinate values arising from nowhere. However, there is a view in the vicinity that would deny the existence indeterminacy and indeterminate values and/or that quantum indeterminacy coming in degrees (or seeing it as just an epiphenomenon), and, for example, just assume the existence of something like determinate values in certain circumstances when the systems are not isolated. When the system doesn't have determinate values, it just has unstably differentiated or less stably differentiated quantum properties. EnDQT as an interpretation of QT is above and independent from these ontological debates and could be easily modified to adopt this anti-realism about value indeterminacy.

[27] I will argue in section 2.2 that they should have arisen in the early universe.

When the lab of Wigner's friend is isolated, the systems outside the lab cannot interact with the systems inside the lab, not allowing frequent interactions to proceed in that region of space and into the future. These interactions maintain the contents of the lab as part of an SDC most of the time in a time interval, such as the friend and his measurement device. Since they don't occur anymore, the systems inside the lab are not anymore part of any SDC. So, the friend cannot interact with the target system anymore in such a way that it gives rise to her system having determinate values (of spin in a certain direction in the case of the example above) during the interactions. Thus, no determinate values (concerning *any* physical property) can arise inside the lab, including the values of the spin-z of the system, but rather indeterminate values arise. Bob cannot condition his outcomes on the (determinate) outcomes of Alice, and Alice inside her isolated lab doesn't obtain determinate outcomes either. This fact is objective and *absolute* and, contrary to relationalist views, it doesn't vary according to systems, agents, etc. Bob can only condition his outcomes on the outcomes of Alice when the lab of Alice is open because in this case Alice and her system (via Alice's measurement device) are interacting with the elements of an SDC. The outcomes of Alice and Bob in these open-lab situations are also absolute. Therefore, as I will argue, relationalism is circumvented by adopting EnDQT.

---

[28] EnDQT is not necessarily committed to an ontology of systems with fundamentally quantum properties. Another possible alternative ontology is to consider that the DM is primary or the SDCs. Systems are secondary and are carved up from the DM or just the SDCs. I will opt for the systems-based ontology here because it's more intuitive and widely adopted. The DM-first or SDC-first deserves future development.

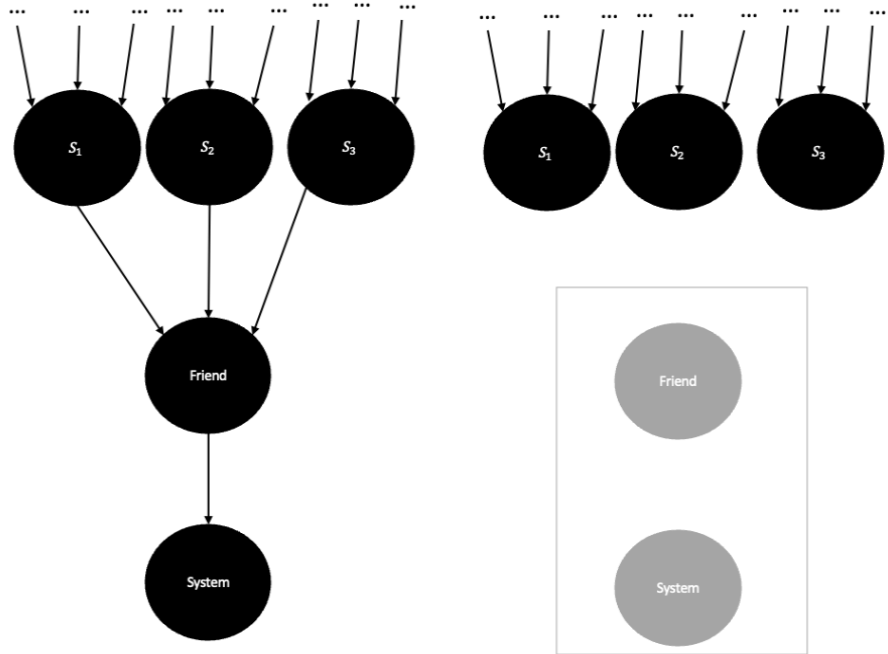

Figure 1: Local interactions in a moment in time over a two-dimensional spatial region in a Wigner's friend scenario. Dark arrows designate the value-determining relations (that can be represented via decoherence) between system X and Y. $X \rightarrow Y$ means that X interacts with Y, which allows Y to have certain determinate values. In turn, this constant interaction with X allows Y to give rise to certain determinate values under interactions with a system Z, where $Y \rightarrow Z$, and so on. So $X \rightarrow Y \rightarrow Z \rightarrow \cdots$ represents system X interacting with system Y giving rise to determinate values, which allows Y to give rise to determinate values under interactions with Z and so on in a region of space. This is a (toy example) sample of a network, involving non-space-like separated systems, that started in the past, persists over time, and extends over space, which I have called a stable differentiation chain (SDC). Dark nodes are systems that belong to this SDC with respect to certain quantum properties. Grey nodes are systems that don't belong to this or any (in the case of total isolation like in the Wigner's friend case) SDC, belonging to what I will call the unstable differentiation chains (UDCs). On the left, we have the situation where the Friend is connected to this network via interactions that allow her (or her measurement device, more precisely) to give rise to determinate values under interactions with her system. On the right, we have the situation where the lab is isolated, which isolates the friend from this network of interactions and doesn't allow her to give rise to her target system having determinate values under interactions with it. The friend rather gives rise to the system having indeterminate values in these interactions, and this fact is non-relative and objective. Thus, everyone will assign to the friend and her system the same state and agree on the state of affairs occurring inside the lab.

So, EnDQT considers that the criterion for a system having a property "with a determinate value" involves the system to be interacting in a certain way with elements of an SDC typically represented by decoherence. Decoherence can only be used to establish the criteria for determinacy when the interactions between systems that give rise to decoherence fit into these chains. It cannot be used as a criterion when the friend's lab is isolated. Moreover, according to EnDQT, no perspectivalism arises. Within this single world, Alice (or any other system) either gives rise to determinate outcomes, or it doesn't. Wigner, Bob, and anyone should treat Alice and Alice's system using the same quantum states.

There are two versions of EnDQT that vary according to how they regard the dynamics of the system or more precisely, what the time evolution in the equations of quantum theory represents and what occurs under interactions. I will go briefly over them here.

One view, which I will call the *deterministic EnDQT view*, considers that the evolution of systems is deterministic, where their evolution is represented via unitary evolution. Systems that belong to SDCs evolve deterministically, having determinate values that arise through interactions. Systems that don't belong to the SDCs also evolve deterministically, although they have indeterminate values and are represented in the standard way via quantum states that are in a superposition. Certain global quantum states will represent the possible evolutions of SDCs and provide, via the Born rule, probabilities for the possible determinate values of the system under interactions with systems that belong to an SDC. As we will see, probabilities will allow for the prediction of the possible developments of SDCs over time. The initial conditions that give rise to the possible SDCs can be represented by hidden variables. The deterministic EnDQT can appeal to these hidden variables and the selection of which one occurred in our actual universe via a chancy process, where these chances are Born probabilities, to justify the Born rule. The use of Born rule probabilities will be regarded at least in part as a consequence of the ignorance regarding the initial conditions, which were selected via the above chancy process. However, there are other more sophisticated and perhaps more satisfactory EnDQT deterministic views, with better resources to explain the use of Born probabilities. I will not give a complete account of the deterministic view here but sketch it in section 2.2, and discuss further in section 4 why it is different from other deterministic views, such as superdeterministic views and the MWI.[29]

Another view, which I will call the *probabilistic EnDQT view*, is similar to the above one but considers that the evolution of systems is always deterministic until certain interactions occur, giving rise during these interactions to determinate values probabilistically, where these are ultimately objective probabilities or chances. I say in certain interactions because when the system is measured on the same basis in which it is prepared, we still have a deterministic process.

Note that, contrary to collapse theories, both for the deterministic and probabilistic view there is no literal collapse of quantum states in a superposition during interactions because we are adopting the informed indirect predictor strategy, and thus there aren't non-local influences that are transmitted through quantum states of systems in a superposition. There is rather a state update of the original state of the localized target system that can be implemented upon decoherence. This stance is sometimes called subjective/epistemic collapse, although for EnDQT it will represent something objective. In the case of probabilistic EnDQT, it represents the result of a chancy process; and in the case of deterministic EnDQT it will represent the result of a deterministic process, i.e., the deterministic evolution of an SDC. Privatistic and single-world multiplicity strategies mentioned above adopt a similar account of the quantum state and dynamics, as well as arguably most physicists, although they adopt a more non-realist stance. Also, contrary to collapse theories, in the probabilistic EnDQT view, arbitrary systems can be placed in a superposition as long as they *don't* interact with elements of an SDC. Moreover, there is no need to modify the Schrodinger equation to represent and predict these chancy processes, a particular interpretation of decoherence and quantum properties is enough. I will say more about these views in section 2.2.[30]

---

[29] There are some hybrid EnDQT-Everett deterministic views. One example of a possible view would be splitting world-EnDQT view could consider that the branching into worlds just occurs over the evolution of the SDCs, where they integrate more systems into the SDCs, branching into different possibilities. For obvious reasons, I won't discuss these possibilities because they don't get us out of the dilemma.

[30] We can also adopt a non-fundamentalist stance towards differentiation. In this case, the degree of differentiation of properties of quantum systems can be further explicated by more fundamental features. If non-fundamentalists find a way to explain the above features, it shouldn't be in terms of the typical local (e.g., typically adopted by retrocausal/superdeterministic theories) or non-local ones (adopted by Bohmian mechanics) hidden-variables. A non-fundamentalist account may be attractive when assume deterministic EnDQT. Maybe differentiation can also be explained away by some collapse theory that achieves (at least most of) the virtues of EnDQT or an account where quantum theory is not universally applicable to every feature (e.g., to gravity). However, if the goals of EnDQT in terms of surpassing the Wigner's friend dilemma are achieved, and no other goals are added, it's unclear whether we need such hidden variables or collapse or non-universality. EnDQT also allows for the following non-committal stances: what there is in the world is a phenomenon called differentiation. How we decide to further understand it is a matter of future scientific developments or perhaps metaphysical speculation or preference, or even representational convenience.

What I will argue for in this paper will be mostly valid for both views, and I will often not distinguish one from the other except in certain circumstances.

It is important to be clear that EnDQT is not a MWI or an Everett's relative-state view. Besides not requiring the posing branching into worlds in measurement-like circumstances, EnDQT differs from MWI-like interpretations by *never* viewing a process representable by an effectively irreversible process of decoherence as a sufficient criterion for determinacy. In the Wigner's friend scenario, we could have a large lab where decoherence inside of it leads to decohered systems that persist over an arbitrary amount of time, which tempts an MWI proponent to consider that there are determinate properties that arise within each branch represented via the quantum states of such decohered systems. EnDQT rejects such a claim. *Importantly*, contrary to the MWI, decoherence will be *"just a tool"* to represent the differentiation phenomenon, not a branching of the wavefunction into multiple worlds' phenomenon. Rather than evidence for this phenomenon, for EnDQT, the Wigner's friend scenarios, and the related correlations will be empirical evidence for the *breaking* of SDCs in a certain region of spacetime.

EnDQT isn't also a "relative-states-like formulation." In these formulations, we are following only one of the possible branches of the wavefunction without the existence of splitting. First of all, it's unclear how views of this kind such as the so-called divergent view (see, e.g., Wilson, 2020) and the many-threads view (Barrett, 2019) deal with this scenario, since they will be pressed to assume that both the friend Alice and the Bob had absolutely determinate outcomes inside the actualized branch. However, this strategy can't deal with the Wigner's friend dilemma as we have seen. EnDQT circumvents that assumption.[31] Second and relatedly, contrary to these formulations, the relative states concerning what we typically call observers, such as the friend, *aren't* a sufficient criterion for determinacy.

To emphasize, note that some interactions, typically represented via decoherence, don't give rise to determinate values in any instance, such as interactions between the friend and her systems inside the isolated lab. The systems in these cases that give rise to decoherence will be said to have unstable differentiator quantum properties of quantum properties of other systems and will be called unstable differentiators.

---

[31] It seems that in order to deal with these theorems and scenarios, they will have to invoke a kind of non-locality that puts them in direct conflict with relativity.

I will start by explaining the basics of EnDQT (section 2). Section 2.1 provides an account of quantum properties and quantum indeterminacy based on decoherence. Readers familiar with decoherence and that don't want to delve into the details of how EnDQT represents quantum properties and more fine-grained details of this ontology may wish to skip most of section 2.1., reading only roughly the first four pages of this section. Sections 2.2 and 3 are the most innovative parts.

Section 2.2 explains SDCs and how they give rise to determinacy. Then, I will go into more detail about how EnDQT surpasses the Wigner's friend dilemma via certain constraints on the SDCs in our universe (section 3). This shows that the received view is misleading, that we can circumvent relationalist, and that we can potentially circumvent the issues tied to relationalism without assuming options A)-E).

In section 4, I will discuss other possible independent good reasons to accept EnDQT, which are works in progress, and suggest future developments. The potential payoffs are large and thus shouldn't be neglected. They include providing a local causal explanation of Bell's correlations; providing a different deterministic view of QT; providing a new solution to the problem of explaining the diverse temporal asymmetries; not having a problem of probabilities like the MWI; providing novel and distinct empirical posits from other interpretations of QT that perhaps one day are testable; and giving rise to a general framework that provides an account of determinacy and indeterminacy, which yields a valuable guide to the different interpretations of QT, and their assignment of determinacy and indeterminacy to different features of their ontology. These potential payoffs aim to support the view that EnDQT is worth taking seriously, potentially circumventing the problems of relationalist views and with a series of benefits, not being just a view formulated to deal with the idealized Wigner's friend scenarios.

More details of the argument will be provided in the appendices. I will elaborate on how EnDQT gives a more realistic account of the Wigner's friend scenario and interference phenomena (Appendix 1). In the Appendix 2, I will explain how EnDQT and other interpretations that could adopt a strategy similar to EnDQT, would regard the phenomenal, mental, or cognitive states of an agent in an isolated lab and of a quantum AI agent running on a quantum computer. This allows EnDQT to answer some plausible objections. I will put this strategy into practice by explaining how EnDQT and these other interpretations interpret in a new way recent extended Wigner's friend scenario theorem[32] where the friend is a quantum AI agent, and clarify some of its assumptions. As I will also argue, contrary to EnDQT, this version of the theorem will be considered disanalogous by some relationalists with the typical extended Wigner's friend scenarios in some relevant aspects because there is no (large) decoherence inside the circuits of the quantum computer that process the experiment. So, according to this view, the authors of the above paper failed in their goals of replicating the original extended Wigner's friend experiment. However, according to EnDQT they didn't fail to achieve their goals (but due to different reasons than the ones expected). I will then present further objections to EnDQT, respond to them, and present some new challenges to relationalist views (Appendix 3). Afterward, I will further sketch how EnDQT provides a non-relationalist local explanation of Bell correlations (appendix 4). To simplify, I will assume a non-relativistic QT,[33] as well as the Schrödinger picture Hilbert space-based finite dimensional QT. I will also omit some unnecessary formal details.

## 2. EnDQT: the basics

In section 2.1 I will present an account of quantum determinate and indeterminate properties. To do that, I will explain what are quantum properties, value properties, value-determining quantum properties, stable and unstable differentiator properties, and unstable undifferentiator properties (section 2.1). Then, given the ontology presented in the first section, I will explain the most innovative features of EnDQT (section 2.2). More concretely, I will explain what the differentiation makeup, the stable differentiation chain, and the unstable differentiation chain are. These features will allow EnDQT to surpass the Wigner's friend dilemma, as will be argued in section 3.

### 2.1. Quantum properties and value properties

I will now show how via QT and decoherence we can represent, characterize and gain knowledge about the differentiation of quantum properties of systems and analyze their behavior under measurement-like interactions. So, in virtue of the ontology presented here, decoherence will acquire these roles. However, given the indirect predictor strategy, we will not reify the quantum states involved in the decoherence process. So, since I am adopting this strategy, I will consider that systems are localized in spacetime regions and change via local interactions.

---
[32] See Wiseman et al. (2022).
[33] Quantum field theory will be used to account for interference phenomena in an idealized way in the appendix 1.

As anticipated in the last section, there are two related types of properties: quantum properties, which have degrees of differentiation D*, and value properties (henceforward, values) which have degrees of determinacy D. More concretely, quantum systems have fundamentally and intrinsically quantum properties.[34] I will consider a system as an instance of a collection of quantum properties that occupy local regions of spacetime. It's on us to discover what those properties are and what systems instantiating them do. I will be very liberal about what constitutes a system. For example, some internal degrees of freedom of an atom could constitute a quantum system. However, when I say that a system occupies local regions of spacetimes, this excludes considering "holistic systems," such as entangled systems space-like separated from each other as constituting a system. We may call these later holistic systems "larger systems." Moreover, given the indirect predictor strategy, we shouldn't reify this quantum state. So, we shouldn't consider that it represents a system constituted by non-locally causally connected subsystems. Quantum states acquire representational roles beyond being simply indirect predictors via their relationship with other elements of the quantum framework, such as observables. P, M,… will stand for the typically called physical properties/magnitudes such as position, momentum, etc., which often observables are considered to refer to.

A system S (which could be a subsystem of a larger system S') instantiating a certain quantum property D*-P in the cases discussed here can be represented[35] via either one of the following triples: $\left(|\psi(t)>_S, \hat{O}_P, D^*\right)_{D^*-P\,(S)}$ or $\left(|\psi(t)>_{SE}, \hat{O}_P, D^*\right)_{D^*-P\,(S)}$ or $\left(\hat{\rho}_S, \hat{O}_P, D^*\right)_{D^*-P\,(S)}$, where

- $\hat{O}_P$ is a self-adjoint operator concerning a physical magnitude P (e.g., P=momentum, or P=spin projections, etc.) acting on the Hilbert space of S or on the quantum states concerning the "degrees of freedom" of S; and

- $|\psi(t)>_S$ (ignoring the global phase) is a quantum state that is equal to an eigenstate of $\hat{O}_P$ or superpositions of eigenstates of $\hat{O}_P$; or $|\psi(t)>_{SE}$ is the state of S entangled in

---

[34] Or again at least more fundamentally than values.
[35] This is just one way of representing quantum properties.

some degree with the state of system or systems E, where E may[36] give rise to the full or partial decoherence of S under a certain local interactions as we will see (I will omit for simplicity referring to rays in a Hilbert space and talk instead of quantum states throughout this article, leaving implicit I am referring to an equivalence class that includes these quantum states), or $\hat{\rho}_S$ is the reduced density operator obtained via the partial trace over E of density operator deduced from $|\psi(t)>_{SE}$ and

-D* is a variable representing the degree of differentiation of a quantum property, and that may be quantified in different ways. In our case, $D^*$ will take values $0 \leq D^* \leq 1$, and can be in certain cases inferred and quantified with the help of "the amount of decoherence" that a system S suffers. The quantum states that are superpositions of eigenstates of $\hat{O}_P$ may be entangled to some degree with the quantum states of a system E in the environment of S (i.e., $|\psi(t)>_{SE}$). The distinguishability between the quantum states of the environment interacting with the system, which will be related to this degree of entanglement between S and E, will be used to quantify the D*.

To simplify, I will write $D^* - P(S)$ instead of either $\left(|\psi(t)>_S, \hat{O}_P, D^*\right)_{D^*-P(S)}$ or $\left(|\psi(t)>_{SE}, \hat{O}_P, D^*\right)_{D^*-P(S)}$ or $\left(\hat{\rho}_S, \hat{O}_P, D^*\right)_{D^*-P(S)}$, but a more specific representation of $D^* - P(S)$ will be given by either one of these triples depending on the situation.[37] I will write $D^* - P$ when I am referring collectively to the uninstantiated quantum properties of any system that have $0 \leq D^* \leq 1$, and are further represented via one of the above triples. When D* takes a specific value, I will use apostrophes such as D*', D*'', etc. $D^{*\prime} - P$ is further represented by one of the above triples with a specific value D*'. So, for example, we have quantum properties that are designated as D*'-P, D*''-P, … or D*'-M, D*''-M, … where D*'< D*''< …. and $M \neq P$, etc. How to choose between the different representations of quantum properties will be specified below, as well as how to precisely quantify D*.

---

[36] See Appendix 4 for a more detailed explanation of how EnDQT interprets entangled states with systems that don't give rise to decoherence, but to Bell correlations. In the main body of the paper, I will focus more in this paper on systems that give rise to decoherence.

[37] I am assuming that all quantum properties can have arbitrary degrees of differentiation, and can be stably and unstably differentiated. So, the system can be placed in superposition of the eigenstates of observables related with those properties and in principle via decoherence, we can explain why those properties are always differentiated. This applies to quantum properties such as D*-electrical charge, which were never found to be in a superposition. The quantum states (and in a sense the related observables) concerning these properties are typically said to be subject to superselection rules (see, e.g., Bartlett et al., 2007). These rules can be regarded as prohibiting the preparation of quantum states in a superposition, which are eigenstates of a certain observable, and assume certain coherent behavior. Rather than postulate these rules (using certain symmetry considerations, for example), decoherence in a certain widespread environment in spacetime might be used to explain this superselection (see, e.g., Earman, 2008; Giulini et al., 1995). For reasons of ontological parsimony, this is the perspective taken here. However, one may object to this perspective and still want to maintain a EnDQT perspective. In that case, one may argue, for example, that there are certain quantum properties that are always stably differentiated independently of their interactions with an environment, or environmental systems can be determinators of certain quantum properties independently of their interactions with an SDC. We may call these special quantum properties, superselected quantum properties.

Values come with different degrees of determinacy D, where D is a variable taking values $0 \leq D \leq 1$. A system with a non-maximally determinate value (a value with a non-maximal degree of determinacy D, i.e., $D \neq 1$) is represented like we represent quantum properties, but instead of D*, we have D. A determinate value (i.e., a value with D=1) is represented by a certain value or value interval, which in the simple cases presented here, are eigenvalues of $\hat{O}_P$. Like quantum properties, determinate values will be represented in a simpler way via the label D-v-P, and where P is a physical magnitude (e.g., energy, position, etc.). When D=1, v assumes a value. So, for example, we have 1-+1/2-spin-z, or 0.6-v-spin-z.

A system S with a quantum property with a certain degree of differentiation D*' has a D-value with a certain degree of determinacy D=D*' with D>0 only while interacting with certain systems during a time interval. Outside these interactions, values are indeterminate (i.e., D=0). So, the phenomenon of quantum indeterminacy[38] occurs when a system has an indeterminate value (D=0), and this happens when it isn't interacting with other systems that I will specify now. The dependency between instantiating certain values and quantum properties will be established here via a certain kind of functionalism explained later in this section (although this view is not strictly committed to a functionalist account).

In the cases of decoherence that I will consider, the systems that we typically consider that "really decohere" other systems will instantiate quantum properties that are *value-determining or stable differentiators* of the quantum properties associated with the quantum states that decohere. So, I will consider that a system S with a D*'-P in interactions with systems E with a value-determining or stable differentiator properties of D*-P of S *determines* the degree of determinacy of the value of S. On the other hand, the system that "virtually decohere" other systems will still be represented

---

[38] Quantum indeterminacy is often regarded as a case of ontological indeterminacy: its source is in the world rather than our knowledge or representations (see, e.g., Barnes & Williams (2011), and Calosi & Wilson (2019)). Evidence for value indeterminacy comes from theorems such as Bell's and the Kochen-Specker theorem. See, e.g., Bell (1964), Bell (1976) and Kochen & Specker (1975). In Pipa (in preparation-b) I formulate different ontologies for quantum indeterminacy based on a more generalized account of the interpretation presented here. I will return to this topic in section 4.

by decoherence but will lead the system to have indeterminate values. Below, I will present these properties and interactions. So, let's consider that,

*A system E with a value-determining quantum property D\*''-M of quantum properties D\*-P is a system that doesn't change the degree of differentiation of a quantum property D\*=1-P of an S under interactions with S at time t. Moreover, it gives rise to S having a determinate value (D=1) under these interactions at time t (in black). I will call this measurement-like interaction the value-determining process of D\*-P of S or the determination of system S. I will call E a determinator of S or a determinator of D\*-P.*

To put it using the usual terms, a system value-determining a D\*-P of an S occurs, for example, when the quantum state of a system concerning a differentiated quantum property 1-P of S "is measured in the same basis in which it is prepared" by a certain measurement device, yielding a determinate outcome/determinate value. For example, a system with a differentiated quantum property spin-z *measured* in a Stern-Gerlach apparatus oriented in the z direction, yielding the value +1/2 spin-z is one example of value-determining D\*-spin-z of a system. The measurement device instantiates value-determining quantum properties of the property D\*-spin-z.

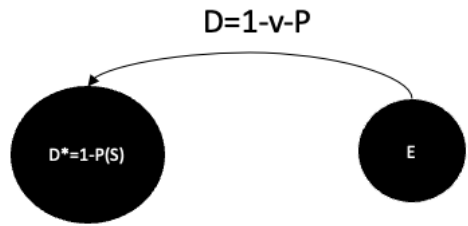

Figure 2: Value-determining process of a quantum properties D\*-P of S by E or determination of S by E (dark arrow)

The value-determining process and the process of stable differentiation examined here will be focused on the paradigmatic strong measurement-like interactions between systems S and E where the changes in system S due to interactions with E are much stronger than the changes due to its own dynamics. So, we can neglect the self-Hamiltonian of the system and just consider the interaction Hamiltonian in governing/describing the evolution of the system. A value-determining process also occurs in the case where the changes in the system due to interactions with E are much slower than the changes due to its own dynamics (see, e.g., Paz & Zurek, 1999). I won't analyze this situation here because the techniques used to model it are more complicated. The intermediary case where both changes are roughly the same doesn't relate so straightforwardly with the stable differentiation and value-determining process and will be examined in future work.[39]

Let's then see how to represent the above value-determining process. Let's consider $|s_i>$, being an eigenstate of the observable $\hat{O}_P$, which, together with $\hat{O}_P$ and the degree of differentiation D*=1, represents a differentiated property D*=1-P of S. The eigenvalue $s_i$ of $\hat{O}_P$ that results in the latter acting on $|s_i>$, represents a determinate $s_i$ −P property. Note that, for simplicity, throughout this article, I will assume that observables have non-degenerate eigenvalues.

The value-determining quantum property of D*-P is represented by the operator $\hat{O}_M$ concerning M, and a quantum state that is an eigenstate of $\hat{O}_M$. If I wanted to be more precise, I would distinguish between potential and actual value-determining properties. An actual value-determining property of D*-P is a property of system E that, from the beginning until the end of an interaction between S and E, will give rise to S having a determinate $s_i$ −P. A system E that may end up having an *actual value-determining property of D\*-P* while interacting with a certain system S has a *potential value-determining property of D\*-P* before interacting with S. So, consider the set of value-determining properties of D*-P. Subsets of this set will be sets of potential or actual value-determining properties of D*-P. I will refer to both as value-determining properties of D*-P.

Let's associate the quantum state $|E_0>$ with a D*'-M of E, being an eigenstate of $\hat{O}_{M_0}$, which will represent a potential value-determining property of D*-P. Let's also associate the quantum state $|E_i(t)>$ with another D*'-M of E at t, being an eigenstate of $\hat{O}_M$, which will represent an actual value-determining property of D*-P.[40] Given the appropriate Hamiltonian for this situation, EnDQT represents the value-determining process occurring in spacetime region *ST* in the following way,

$$|s_i>_S \otimes |E_0>_E = |s_i>_S|E_0>_E \rightarrow |s_i>_S|E_i(t)>_E.$$

---

[39] More complex techniques, such as the predictability sieve, are typically used to find the states to predict and represent the system in more complex regimes, namely, when the self-Hamiltonian of the system (i.e., the intrinsic dynamics of the system) is roughly the same order of magnitude as the interaction Hamiltonian (see, e.g., Zurek, 2003, Zurek et al., 1993) . The techniques and the selected minimum uncertainty phase space localized coherent states in this intermediary regime will have a different interpretation than the ones provided here. The analysis of this case is more complicated and will be left for future work. However, the interpretation given here will serve as a basis for this analysis. Sketching briefly and roughly an interpretation of these states, they will represent the system's position and momentum quantum properties as not being both extremely differentiated or undifferentiated on average over time (hence being states of minimum uncertainty in both the position and the momentum). So, in this situation we don't get full determinacy and indeterminacy on average over time.
[40] We will see below that determinators have determinate values v-M in interaction with the determinators of D*-M.

So, in the situation where S already has a completely differentiated quantum property (i.e., D*=1-P) over *ST*, whose degree of differentiation is represented by $D^*(P, S, ST, t) = 1$,[41] the associated value $D' - P$ of S has a degree of determinacy $D(v - P, S, ST, t) = D^*(P, S, ST, t) = 1$.[42] To simplify, I will not talk in terms of actual and potential differentiating quantum properties anymore, and leave that implicit.

I will turn to another related measurement-like interaction called the process of stable differentiation, where this process might culminate in a value-determinating process, and which will be represented via decoherence.[43] In order to characterize and understand this process, let's consider that

*a system S with certain properties D\*'-P in interaction with system E with a differentiator property of D\*-P at t will lead system S to have quantum properties D\*''-P at t' where D\*''>D\*' and t' > t. I will call E, a differentiator of D\*-P.*

Consider a

system E *with a differentiator property of D\*-P that also gives rise to system S having a value with a non-minimal degree of determinacy D=D\*' (i.e., D≠0) at t when S has a D\*'-P during this interaction at t. I will consider that E has a stable differentiator property of D\*-P.*

---

[41] As we will see, the measure of differentiation will be related with the distinguishability of certain quantum states of the environment. Here the state of E is maximally distinguishable.

[42] Quick value-determining interactions of a certain kind give rise to the quantum Zeno effect [QZE]. As Wallace (2012) summarized: "[it] consists of the observation that if a system, whose state would otherwise evolve from out of some subspace S of its Hilbert space H, is subject to some measurement process which (a) determines whether or not it is still in that subspace and (b) occurs on timescales fast compared to the evolution, then the system's evolution out of the subspace will be dramatically suppressed." Perhaps the QZE maintain some parts of SDCs stable. Note that value-determining interactions that give rise to QZE need not to occur in order for a given system S* to belong to SDCs for a given quantum property D*-P during most of the time in an interval of time. For instance, there might be times during that interval that D*-P of S* is not being value-determined by systems belonging to an SDC or not so strongly (in such a way that the Hamiltonian of interaction totally dominates its evolution during that time), which allows the system "to evolve from out of some subspace S of its Hilbert space H."

[43] This process is more general than the value-determining process, where the later should be regarded as an idealization since a system being in an eigenstate of some observable in order to have determinate values of that observable should be regarded as an idealization (see, e.g., Wallace, 2019).

Now, the process of stable differentiation is the following (see Figure 3): a system S with a D*'-P interacts with a system E with a stable differentiator property of D*-P, becoming stably differentiated over time. S, due to its interaction with system E, will have quantum properties with an increasing degree of differentiation D*' until it has a quantum property that is maximally stably differentiated (i.e., D*=1). During this interaction, it will also have value properties D-v-P with an increasing degree of determinacy D=D*', until D=1. So, if this process goes on long enough, S will have a completely stably differentiated quantum property and a completely determinate v-P. System E has stable differentiator properties of the property D*-P of S until it doesn't increase or decrease D*=1 of a D*-P of S. When that happens, E has a value-determining quantum property of properties D*-P of S, and S has a 1-v-P (i.e., a determinate value). I will also say that E is value-determining a quantum property of S or value-determines S (during that interaction). Note that system E could be composed of many subsystems, each with stable differentiator properties of a D*-P.

This is the process that occurs, for example, when the not-completely differentiated quantum property D*'-spin-x of a system is measured via a Stern-Gerlach apparatus with a non-homogeneous magnetic field oriented in the x direction, yielding the value +1/2 spin-x with probability ½ or -1/2 spin-x with probability ½, where D*-spin-x ends up being completely differentiated. The measurement device "will have" stable differentiator properties of spin-x, and a value-determining property of spin-x if D*-spin-x of S becomes completely differentiated.

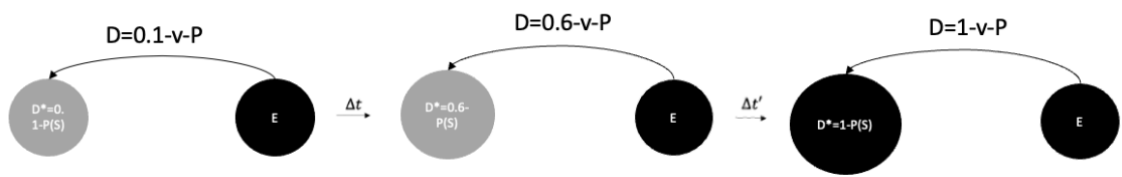

Figure 3: Process of stable differentiation of quantum properties D*-P, which becomes a value-determining process when D=1-v-P (both represented with dark arrows).

Decoherence models represent the stable process of differentiation of a quantum property of a system and its degree of differentiation, and so they represent how quantum properties of systems are affected in these interactions. I will use the expression "decoherence of systems" only to facilitate communication since decoherence will be "a tool" to represent the phenomenon of differentiation and quantum properties. To see how I will return to our simple example above and assume instead that D*-P is initially completely undifferentiated. Let's represent that undifferentiated D*=0-P by

$$|\psi>_S = \sum_{i=1}^{N} \alpha_i |s_i>_S$$

and the operator $\hat{O}_P$ concerning P, where $|s_i>_S$ corresponds to the different eigenstates $i$ of $\hat{O}_P$.

S will interact with a system E that has a stable differentiator property of D*-P in one of the regions ST. Like in the case of value-determining properties, the distinction between potential and actual differentiators will apply, and again for simplicity further below I will mostly neglect this distinction. System E will be the same system with the same property as above, represented initially in a similar way to systems with value-determining properties, but now it's going to act as a differentiator of D*-P. We then have the standard von Neumann interaction given by the Hamiltonian that describes the interactions between S and E,

$$(\sum_{i=1}^{N} \alpha_i |s_i>_S)|E_0>_E \xrightarrow{\hat{U}} \sum_{i=1}^{N} \alpha_i |s_i>_S |E_i(t)>_E = |\Psi>_{SE}.$$

The amount of overlap between two quantum states $<E_i(t)|E_j(t)>_E$ for the different quantum states in the superposition above (let's label the different terms with the natural numbers i and j, with i different from j) measures the amount of distinguishability between $|E_i(t)>_E$ and $|E_j(t)>_E$. Let's make the plausible assumption that the amount of distinguishability between all these different quantum states depends on the distinguishability of all the possible actual stable differentiator or value-determining properties of D*-P that E instantiates in interactions with S. The distinguishability of these quantum properties of E concerns the distinguishability of the possible behaviors of S while interacting with E (or with respect to E) in the physical situation that is being modeled, where E affects and is affected by S. Moreover, since $|E_i(t)>_E$ and $|E_j(t)>_E$ for all i and j with i different from j are correlated with the states that are eigenstates of the operator $\hat{O}_P$ giving rise to eigenvalues $s_i$, and given that E has stable differentiator properties of D*-P, it's plausible to consider that the above amount of distinguishability and how it changes over time allow us to probe and represent the degree of differentiation of the properties D*-P, the degree of determinacy of v-P that S has during these interactions, and how they change. Thus, we should study how $<E_i(t)|E_j(t)>_E$ for all i and j with i different from j behaves and evolves to analyze these features. Note that although "distinguishability" refers to an extrinsic feature, it allows us to probe intrinsic features of systems, which is the differentiation of the quantum properties that they instantiate (more on this below).

The reduced density operator $\hat{\rho}_S(t)$ obtained from partial tracing the degrees of freedom of the environment, which roughly corresponds to "marginalizing over" the degrees of freedom of the environment, allows us to measure such overlap or distinguishability, and represent and study further the impact that the interactions with E have on the behavior of S,

$$\hat{\rho}_S(t) = \sum_{i,j=1}^{N} \alpha_i \alpha_j^* |s_i>_S <s_i| <E_j(t)|E_i(t)>_E$$

$$= \sum_{i=1}^{N} |\alpha_i|^2 |s_i>_S <s_i| + \sum_{i \neq j}^{N} \alpha_i \alpha_j^* |s_i>_S <s_j| <E_j(t)|E_i(t)>_E,$$

where $<E_i(t)|E_j(t)>_E$ for each i and j with i different from j are called coherences. A *possible* measure of the degree of differentiation of the different D*-P of S in ST over time t for the above simple scenario will be given by the von Neumann entropy[44] $S(\hat{\rho}_S(t))$ of $\hat{\rho}_S(t)$ over $lnN$, where $N$ is the number of eigenvalues of $\hat{\rho}_S(t)$,

$$D^*(P,S,ST,t) = \frac{S(\hat{\rho}_S(t))}{lnN}.$$

Thus, we can measure and represent the degree of differentiation D*' of a quantum property D*'-P of S at a certain time t, and how the differentiation of quantum properties of S change over t, and how much time it takes differentiation to arise (which is equal to the decoherence timescale[45]), with $0 \leq D^*(P,S,ST,t) \leq 1$, in *the possible*

---

[44] Given a density operator $\hat{\rho}_S$ for quantum system S, the von Neumann entropy is $S(\hat{\rho}_S) = -tr(\hat{\rho}_S ln\hat{\rho}_S)$. $S(\hat{\rho}_S)$ is zero for pure states and is equal to $ln\ N$ for maximally mixed states in this finite-dimensional case.
[45] I.e., it is equal to how much time it takes the overlap terms to be approximately zero quasi-irreversibly.

*set of spacetime regions* ST where they are differentiated via interactions with other systems E. Or after those interactions in other STs in the absence of further interactions with differentiators, so-called undifferentiators (see below), or determinators of properties D*-P. The differentiation is maximal when $D^*(P,S,ST,t) = 1$. The above $D^*(P,S,ST,t)$ is also interpreted as representing the evolution over time of degree of determinacy of $v_i - P$ of S when S is interacting with stable differentiators of properties D*-P. When $D^*(P,S,ST,t) = 1$ and this process is effectively irreversible because "we cannot control the many states of the systems of the environment that got entangled with S to reverse this process," taking an astronomical amount of time to reverse spontaneously, it is typically said that decoherence has occurred. However, for EnDQT this will mean that S has a completely differentiated quantum property. So, together with $\hat{O}_M$ after a sufficient amount of time $t$, $|E_i(t)>_E$ and $|E_j(t)>_E$ and a certain D*' will represent the value-determining properties of D*-P.[46] Whereas before $t$, together with the projectors onto these states and a certain D*', they will represent the differentiator properties of D*-P.

In models of decoherence, we never get $D^*(P,S,ST,t) = 1$, but $D^*(P,S,ST,t) \approx 1$ is enough to approximately represent the typically complex open systems situation where S has D*=1-P. This is because $D^*(P,S,ST,t) \approx 1$ stays like that over time, and for not very large systems the recurrence time of this quantity may even exceed the age of the universe. So, this value doesn't have a meaningful difference from one. Note then that we can also use the above density operator to represent quantum properties or non-maximally determinate values.

Note also that all potential stable differentiator properties of quantum properties D*-P are potential value-determining properties of D*-P. Moreover, all processes of stable differentiation end up being value-determining/determination processes. So, from now on, I will mostly talk in terms of value-determining properties since these are the properties that matter to give rise to determinacy, and value-determining/determination interactions.

It's plausible to consider that the distinguishability of S instantiating D*'-P by another E under interactions with S presupposes that the E has determinate values in interaction with other systems so that *we* can access and determinately represent such distinguishability. Also, it often presupposes that the local system E is interacting with many other inaccessible systems in such a way that it's harder to have control over this process and reverse it (i.e., "reversing decoherence"). As we will see below, these two features will be related in a certain way. More concretely, the determinate values of the environment arise under interactions with a great number of members of the "stable differentiation chain" mentioned in the introduction, giving rise to E having determinate values and allowing E ultimately to value-determinate a quantum property of S. The (at least hypothetical) interactions with SDCs and the distinguishability qua extrinsic feature will provide an epistemological anchoring (in the sense of reference point) that we use to help us know about and represent all the quantum properties, as we will see throughout this article.[47]

---

[46] Or more realistically, they will be quantum states whose projectors onto these states end up commuting at least approximately with $\hat{O}_M$ over time $t$. I will ignore here these precisifications.

A few features and consequences of determinate values become manifest with this ontology as represented by decoherence, allowing us to characterize systems instantiating determinate values and systems with less determinate or indeterminate values. As we can see, when the system has determinate values, we can assign determinate values to all their possible behaviors in a physical situation PS (such as in the double-slit case with monitoring of the slits); their possible behaviors are distinguishable from one another; and we can establish a quantity that assumes determinate values in PS in certain spatiotemporal regions. Then, we can assign a probability measure[48] to these different possibilities in PS and make inferences about them based on the probability calculus. So, we have that $\hat{\rho}_S(t) = \sum_{i=1}^{N} |\alpha_i|^2 |s_i>_S<s_i|$, for the different determinate values of S represented by $s_i$.

On the other hand, when a system hasn't determinate values v-P, in general, we cannot assign determinate values to their possible behaviors in a physical situation PS* and we cannot establish a quantity related to P that assumes determinate values in that situation. Their possible behaviors are indistinguishable or not completely indistinguishable (but not completely distinguishable either), and we cannot assign a probability measure to these different possibilities in that situation, making inferences about them based on the probability calculus.

---

[47] This may include assuming certain superselection rules, even "when those rules dynamically arise" via interaction of the system with its environment. See footnote 36.
[48] A probability measure is typically defined via the following Kolmogorov axioms: a Kolmogorovian probability space is a triple (X, Σ ,μ), where one has that σ-algebra is an algebra for the nonempty set X and the measure μ: Σ → [0, 1] is such that μ(X) = 1, μ(E)≥ 0 for all E ∈ Σ, and for all countable sequence of disjoint sets $E_1, E_2, ... ∈ Σ$, $μ(\cup_{i=1}^{\infty} E_i) = \sum_{i=1}^{\infty} μ(E_i)$. A σ-algebra on a set X is a collection of subsets of X that is closed under the complement and is also closed under a countable union and intersection.

It's thus unlike the first case where the system has determinate values, where for example the probability of having a certain determinate value $s_1$ or having a certain determinate value $s_2$ is equal to the probability of having a certain determinate value $s_1$ plus the probability of having a certain determinate value $s_2$ (where having $s_1$ and $s_2$ are the only possible events that can occur in this situation and are mutually exclusive). We have in general the extra term $\sum_{i \neq j}^{N} \alpha_i \alpha_j^* |s_i>_S <s_j| <E_j(t)|E_i(t)>_E$ in $\hat{\rho}_S(t)$ that shows that this cannot be the case.

As mentioned above, decoherence also models situations where systems have values with a degree of determinacy between the two extremes. The extra term quantifies this, becoming smaller with the amount of distinguishability and the related amount of determinacy. As we can see, in these cases, the behavior of these systems also doesn't offer an exact probabilistic treatment, but the probabilistic treatment becomes closer to being accurate as the degree of determinacy of the values increases.[49]

As we will see more clearly in the next section, not all processes of differentiation give rise to non-indeterminate values. Some of them give rise to the system having indeterminate values. Moreover, some processes that we would consider to be a value-determining process aren't and don't give rise to determinate values at all. I will thus consider that

*a system S' has unstable differentiator properties of D\*-P if, under interactions with S, differentiates to any degree the properties D\*-P of S or maintains the degree differentiation of D\*-P of S during time t but gives rise to S having indeterminate values during this process.*

I will say that S' *unstably differentiates D\*-P of S or gives rise to an unstable process of differentiation where D\*-P is unstably differentiated* to a degree D\*' (so, completely undifferentiated properties are by default unstably minimally differentiated). Unstable differentiation will be represented like stable differentiation or determination but gives rise to systems having indeterminate values. The process of unstable differentiation is the process represented by the so-called virtual or reversible decoherence (which is typically considered not to be decoherence at all). This process may happen, for example, because the environmental system may not be composed of many subsystems that give rise to the irreversible process of decoherence, being "virtually decohered" (however, we will see a better justification for this in the next section). Interactions that unstably differentiate properties (which are represented via decoherence) shouldn't license a state update, given that no determinate values arise, and we can't associate states to the system related to these determinate values (like in the friend's interactions with her system in an isolated lab, as we will see).

---

[49] One may wonder about the case when it is assigned to the non-interacting system an eigenstate of a certain observable. It seems that in this case its behavior is distinguishable, and it behaves as if it has a determinate value. Note, however, that in so far that a system being in an eigenstate of an observable is realistic, for EnDQT such determinacy and distinguishability just occurs upon value-determining interactions.

In the case of the unstable differentiation of a D*-P of S by an E instantiating non-value-determined (by other systems) quantum properties D*-M and hence having indeterminate values v-M, the distinguishability that arises via interactions between S and E (represented through quantum state overlaps such as $<E_j(t)|E_i(t)>_E$) can be justified via hypothetical interactions. More concretely, it will be justified by hypothetical interactions between E and determinators of D*-M of E (which have to belong to a stable differentiation chain as we will see).

Just like there are systems that increase the degree of differentiation of the quantum properties that a system has, there are systems that decrease their degree of differentiation. Let's consider that

*a system S with certain properties* D*'-P *in interaction with system S'' with undifferentiator property of D*-P at time t, will lead system S to have properties* D*''-P *at t' where D*''< D*' and t' > t. I will call S'' an undifferentiator of D*-P.*

This happens when a system is unstably differentiated ("being so-called virtually decohered") by S and "we can reverse its decoherence." Ideally, we could reverse this virtual decoherence by, for example, applying a unitary (i.e., a countertransformation to the unitary one that entangled the system and the environment) to the system-environment that reverses that state of the system to the previously non-decohered state. The system that performs these operations is an undifferentiator of S.[50] As we will see more clearly, all systems S' with undifferentiator quantum properties of a D*-P are unstable since systems S will have indeterminate values v-P in interactions with them (like the friend and her "measured" system when the lab is isolated and this whole

---

[50] See, e.g., Raimond et al. (1997) and Myatt et al. (2000) for experimental proposal and signatures concerning the reversal of decoherence. See also Schlosshauer (2007).

interaction is reversed by Wigner's device qua system that has an undifferentiator property).

## 2.2. The differentiation makeup, the stable differentiation chain, and the unstable differentiation chain

We have seen what are quantum properties, value properties (which I am calling values), stable and unstable differentiators, value-determining and unstable undifferentiator properties, as well as the related systems. I haven't characterized how to values depend on quantum properties. One possible way is via functionalism.[51] Very roughly, functionalism is the position that a property P* is the property of having some other property P that plays a certain role or function R.[52] Using the above interactions, I can provide a more precise account of the dependence relation between value and quantum properties through a certain functionalist account:

*The non-minimally determinate value property, i.e., D-v-P where $0 < D \leq 1$, instantiated by system S at time t is the property of having a quantum property D\*'-P with D\*'=D≠ 0, when S is interacting at time t with systems with a stable differentiator or value-determining properties of D\*-P.*

*The indeterminate value property, i.e., D-v-P where $D = 0$, instantiated by system S at time t is the property of having a quantum property D\*'-P with an arbitrary degree of differentiation when S is not interacting at time t with systems with a stable differentiator or value-determining quantum properties of D\*-P.*

Adopting this view, we look at the representations of quantum properties (via decoherence, etc.) and see what theoretical roles they have in terms of accounting for such interactions. Through them, they will represent values with a certain degree of determinacy.

What we haven't seen so far is how determinators turn into unstable differentiators and vice-versa. In other words, how can some systems give rise to determinate values under interactions and others not? Answering that questions will allow EnDQT to surpass the Wigner's friend dilemma. To do that, we have to add other elements to our ontology. In this section, I will introduce the differentiation makeup, the stable differentiation chain, and the unstable differentiation chain.

---

[51] I have adopted a functionalist perspective because it is a well-studied case of a dependence relation. However, other accounts are possible (Pipa, in preparation-b).
[52] There is certainly more to say about how to characterize the kind of functionalism I am appealing. I will leave that for future work.

So far, it might seem that EnDQT just follows the decoherence technique under a different dressing to assign determinate values to observables of the systems under interactions. However, as explained in Section 1, decoherence is insufficient to address the Wigner's friend dilemma. We need to complement standard QT with a criterion that allows us to distinguish between the processes occurring inside the isolated lab that may be represented through "virtual" decoherence plus the features of EnDQT and which doesn't give rise to values with a non-minimal degree of determinacy; from the other processes occurring in non-isolated labs that do give rise to determinate values, and which may be represented through ("the real") decoherence plus the features of EnDQT.

To achieve that, let's introduce *the differentiation makeup* (DM) of the world:

*The DM is constituted by systems, their evolution, and local interactions. The latter is a particular kind of evolution[53], which sometimes gives rise to determinate values. Such systems (in the simple non-quantum gravity case presented here) occupy regions of spacetime.[54] A part of the DM involves stable differentiation chains (SDCs) and another part involves unstable differentiation chains (UDCs).*

Chains can be composed of systems that influence other systems, which influence other systems and so on under interactions in space and over time (see Figure 1 and 4 for simplified examples of chains). Each SDC will be characterized by a 5-tuple,

$$SDC = ((initiator\ systems, non-initiator\ systems), laws, stability\ conditions, structure).$$

The initiator systems or *initiators* of that chain are the systems that begin a chain, being its first elements, and can be represented like I have explained in the previous.[55] And in at least the simple initiators' picture set up here, they ultimately maintain the existence of a chain.[56] The *non-initiators* are all other systems that constitute a chain and can be represented as I have explained in the previous section.

---

[53] See the final section for a discussion of how EnDQT can provide local explanations of phenomena such as Bell correlations, and further support this assumption.
[54] Note that systems can determinately occupy certain regions of spacetime. Moreover, we will see in section 4 and Appendix 4 ways of representing how system evolve between regions of spacetime through quantum causal models. See also appendix 1 for an example of how to infer and represent the spatiotemporal localization of systems.

Another feature of the SDCs is the laws given by QT and a Hamiltonian that allow us to describe the evolution and the interaction between systems that belong or may belong to an SDC.[57]

An SDC is characterized as well by certain *stability conditions*. The stability conditions specify what it takes for a system to belong to an SDC. More adequate stability conditions might exist than the ones presented, but the ones explained below are the simplest ones that I could think of. As we shall see, they will be enough to block relationalism in the Wigner's friend scenarios. I will divide these conditions into two subconditions, one for the so-called non-initiator systems and one for the initiator systems.

Regarding the stability conditions for a non-initiator system, I will assume that:

*-for a system to become a non-initiator member of an SDC' it has to have a quantum property value-determined by other members of SDC'.*

*-For a non-initiator system to continue being a member of a certain SDC' during a time interval Δt whose structure is stable over that time, the system has to have a quantum property constantly or frequently value-determined*[58] *by members of SDC' during Δt.*[59]

*-Moreover, only if a non-initiator system belongs to an SDC', having its quantum property D\*-P value-determined by members of SDC' can it have a stable differentiator or value-determining properties of other quantum properties.*

---

[55] In the case presented here, prime initiators don't have determinate values (since no system value-determines their quantum properties). Alternatively, we may consider that they always have determinate values. I will stay with the first option because it assumes less things.

[56] Note that in another version we might not be committed to the existence of systems like initiators, but certain initial conditions connected with regions of spacetime (or other arenas) that systems find themselves into and that lead to the propagation of determinacy through spacetime (or through that arena); or simply not consider systems as fundamental but SDCs as more fundamental where such SDC propagates over spacetime. Another possibility is to endow initiators with certain hidden variables (see deterministic EnDQT below).

[57] Laws may be seen as mere summaries of the evolution and interactions between initiators and non-initiators and thus dispensable as describing SDCs. However, I will keep them there.

[58] These constant interactions can be represented by a system interacting with other system, where one decoheres the other, and the Hamiltonian of interaction that describes the interaction between the two is non-zero.

[59] There is a weaker possible stability condition that would claim that to belong to an SDC, the properties of the system must be only stably differentiated to some non-minimal degree. I will adopt the above stronger condition because it's less vague and fits with the evidence that the information about determinate values spreads to the environment through entangled systems.

To avoid falling into a regress, we need stability conditions for the initiators that are different from the ones for non-initiators. I will adopt certain simple conditions concerning certain initiator members of an SDC that don't need to be value-determined to be part of it starting an SDC. These stability conditions are the following:

*Some of the initiators of an SDC will already have value-determining properties, giving rise to determinate values in certain interactions without the need to interact with other systems that value-determine them. The rest of the initiators, which interact with these initiators, follow the stability conditions for non-initiators.*

Initiators can be regarded as forming a set composed of 2-tuples. The initiators in these sets that don't need to be value-determined by other systems to stably differentiate or value-determine others will be called *prime initiators*. The other systems will be *subordinate initiators*. So, we have the ordered pair ($prime\ initiators, subordinate\ initiators$) representing systems (which might be composed by many subsystems), and that give rise to selective interactions. Such selective interaction can be described by certain initial conditions and laws (including a certain Hamiltonian). More concretely, the prime initiators, such as S, will have so-called *initiator quantum properties*, specified via the Hamiltonian, which are value-determining properties D*'-P, allowing S to give rise to determinate values in interaction with the subordinate initiators without D*'-P of S needing to be value-determined by other systems. So, S will be a determinator of the subordinate initiator S' or the subsystems that constitute S'. This then allows S' to be a determinator of other systems to perpetuate a chain. I am not considering that the subordinate initiators have any special quantum properties, just the prime initiators. There might exist many more 2-tuples that belong to the set of initiators (more on this below), and so we have that

$$Initiators = (prime\ initiators, subordinate\ initiators),$$
$$(prime\ initiators', subordinate\ initiators'), \dots$$

A prime initiator and a subordinate initiator will give rise to what I will call *blueprint interactions*, which will be certain interactions between the initiators in a certain region, giving rise to determinate values and a certain chain SDC'. Moreover, they are interactions that maintain SDC'. They need to keep going in order for SDC' to persist.

An analogy to provide an intuition about what are prime initiators, subordinate initiators, and blueprint interactions is the following. Consider prime initiators as the fuel of determinate values and subordinate initiators as the piston (and all the associated parts) that helps to give rise to those properties. They need to interact constantly between them to give rise to determinate values. Also, the subordinate initiator (i.e., the piston) has to interact constantly with other systems (i.e., the other mechanisms of the car) to stably give rise to determinate values over time and to allow other systems to give rise to determinate values (i.e., to allow the car to move).

Why am I considering that interaction blueprints give rise to determinate values instead of initiators taken individually? Because, in this way, we aren't pressed so much to assume hidden variables and a preferred observable. Systems in isolation have quantum properties; thus, they have indeterminate values. Systems give rise to certain determinate values only while interacting with certain systems. Moreover, SDCs can be represented further, for example, by a certain network together with a related relative-state formalism, as I will explain. So, it will be describable via QT without necessarily requiring the typically called hidden variables.

Note that more adequate and realistic stability conditions for initiators and non-initiators might exist. I consider what I am posing here as roughly or approximately true, and to be further developed by future models. Let's call the above simple stability conditions for initiators and non-initiators the *standard stability conditions*.

A fifth feature of an SDC is its *structure* (see Figures 1 and 4). Given the features of initiators and that the interactions in the structure are modeled via decoherence, it will be represented by a directed acyclic graph (i.e., a directed graph with no cycles, DAG)[60] in space that evolves over time at least approximatelly. The nodes will represent systems instantiating certain quantum properties, and the arrows will represent value-determining relations, some of them constant or frequent, which allow an SDC with a certain structure to be maintained over time $\Delta t$. So, $X \rightarrow Y$ represents system X instantiating certain quantum properties that allow it to value-determine certain quantum properties of Y over time $\Delta t$ in a region of space. $X \rightarrow Y \rightarrow Z$ represents X value-determining a certain quantum property of Y, which allows Y to value-determining a certain quantum property of Z over a certain period of time in a region of space. These systems may be composed of many subsystems. The DAG will obey at least two other constraints, which are also encoded in the laws plus the standard stability conditions. First, no system can value-determine the prime initiator. Second, every system has to connect directly or indirectly to the prime initiator.

---

[60] Why acyclic and not cyclic in space and over time? I use decoherence as a guide for this assumption, as well as considerations of simplicity. It is not typically considered the case that, for example, a system A decohere/renders distinguishable B, which decoheres/renders distinguishable C, which decoheres/render distinguishable A over time, forming a highly organized correlated structure. On top of that, decoherence at a moment in time represents an asymmetric process between systems that are uncorrelated at first. The above considerations can be justified mainly due to the fact that decoherence occurs in uncontrollable environments. More concretely, such asymmetry can be justified by the many interactions between systems of the environment that decorrelate it from any interaction that it had with the system whose quantum property is or will be value-determined, as well as due to the initial low entropy conditions (that can be interpreted as establishing that system are uncorrelated initially at the beginning). Thus, such cycles of value-determination become unlikely if not impossible. Note that EnDQT in principle is not tied fundamentally to a notion of acyclicity in this process, and this acyclicity might rather considered as an approximation. The important part is that systems ultimately connect directly or indirectly to initiators. Much more to say here, but I will leave it like this.

For simplicity, I will not distinguish between type and token networks, where type networks represent the regular features of concrete/token networks. I will also not distinguish between more fine-grained details, leaving some ambiguity in the interpretation of the networks. For instance, if we have a system $S_1$ with systems $S_2, \dots, S_n$ value-determining $S_1$, it might be that $S_1$ is composed of many subsystems and $S_2, \dots, S_n$ are value-determining each one of its subsystems. Or it might be that $S_1$ is a single system, and all of these other systems are value-determining it. Moreover, also for simplicity, I will not represent the quantum properties of systems in the network with their different roles.

Systems in the DM that are not part of an SDC with respect to some quantum properties are part of Unstable Differentiation Chain (UDCs) with respect to som properties. Members of an UDC:

-don't have completely determinate values.

-Systems in the UDCs don't form a chain that can be tracked down to the initiators of the chain or aren't necessarily part of any structure at all, such as isolated systems. That's why I say that they form an unstable chain.

-members of a UDC that have differentiator quantum properties are *necessarily* unstable differentiators. Remember that systems S' with these quantum properties give rise to another system S having indeterminate values, even if S' is "decohering" S. So, a system S' might have unstable differentiator properties because S' with those properties isn't connected with an SDC.

Let's finally make clearer what an SDC and a UDC are through a toy example, as well as what are the differences between the probabilistic and the deterministic EnDQT. In the examples that I will provide, I am going to adopt the following conventions. When I place a subscript SDC in the quantum states related to system S having properties D'-P (i.e., being eigenstates of the operator concerning P) this will mean one of the following two things:

- If S is a prime initiator, S can give rise to an SDC, or

-If S is not a prime initiator, S is connected or will connect with a SDC via frequent value-determining interactions between S and members of that SDC, where such interactions involve D'-P.[61]

Now, if a system S' having some quantum states concerning a certain quantum property becomes locally strongly correlated or entangled with the states of S concerning D'-P due to frequent interactions between S and S' and those interactions continue (i.e., in the cases I am analyzing this means that the Hamiltonian of interaction between S and S' is non-zero), sharing an index with these quantum states of S, it can mean either two things:

-If S is a prime initiator and S' is a subordinate initiator belonging to the same ordered set of initiators as S, it gives rise to the first bit of a SDC, or

-If S is not a prime initiator, S' will start belonging to a SDC or will belong to a SDC during an instantaneous or a longer period of time, depending on the duration of the interaction.

---

[61] Certain models of decoherence treat at least the initial environmental system explicitly as being a tensor product of subsystems of that system that are in a superpositions of states, such as superposition of qubits in the central spin model (see, e.g., Cucchietti et al., 2005). So, they treat the environment as being constituted by certain subsystems, represented via these qubits. In this case, if we put a subscript in each of these quantum states, each term in that superposition plus operators should be interpreted as different quantum properties that will be connected with an SDC. Moreover, if each subsystem connects with an SDC, we should interpret that the overall environmental system constituted by these subsystems as being connected with an SDC. Each subsystem has value-determining properties of the relevant quantum property of the target system, and the overall environment value-determines this quantum property.

Given these conventions, let's consider the non-space-like separated systems A, B, C, and D, and let's consider how the formation of the following SDC in a region of space over time,

$$SDC' = ((A, B); C, D; laws; standard\ stability\ conditions; A \rightarrow B \rightarrow C \rightarrow D),$$

could be represented in a certain toy universe. So, A and B belong to the 2-tuple of initiators for this toy universe, where A is a prime initiator, and B is a subordinate initiator, i.e., $(A, B)$. $A \rightarrow B \rightarrow C \rightarrow D$ means that A value-determines B which allows B to value-determine C, and which allows C to value-determine D concerning certain quantum properties. For simplicity, I will omit those properties talking about those properties.

I will first describe the above interaction from the point of view of the probabilistic EnDQT, then I will turn to the deterministic EnDQT. It's an open problem whether the deterministic or the probabilistic EnDQT are the right EnDQT views.

The interaction between A and B could be represented in the following standard way using the above conventions,

$$|E_0>_{A\ SDC'} (\alpha_1|s_i>_B + \beta_1|s_j>_B) \rightarrow_{t=1} \alpha_1|E_i>_{A\ SDC'}|s_i>_B + \beta_1|E_j>_{A\ SDC}|s_j>_B.$$

Above I have assumed that the quantum property D*'-P of B is represented by the above quantum state of B (plus the respective observable and D*'). So, A will have value-determining properties of D*-P and is going to value-determine D*-P of B from $t = 0$ to $t = 1$. The SDC initially has the following structure at $t = 1$, $A \rightarrow B$, and where the arrows mean A value-determines B. This interaction will be a blueprint interaction, giving rise to certain determinate values probabilistically, where, as I have said in section 1, these probabilities are meant to be always objective. Given that A is a prime initiator, it doesn't need to be value-determined by other systems to be a stable differentiator or determinator. Also, note that according to the Hamiltonian implicit here, A cannot give rise to determinate values by interacting with other systems, only via (this rather selective) interaction with the subordinate initiator B.[62]

---

[62] More realistically, we could have many more systems as initiators.

Also, let's assume that from $t = 0$ to $t = 1$ C unstably differentiated D. It's an unstable differentiation because C and D are not connected to an SDC, so they form a UDC. Note therefore that D has an indeterminate value in interactions with C. We can represent their interactions in the following way given our conventions and given the initial states of these systems described below,

$$|E'_0>_C (\alpha_2|v_k>_D + \beta_2|v_l>_D) \to_{t=1} \alpha_2|E'_k>_C |v_k>_D + \beta_2|E'_l>_C|v_l>_D.$$

Some chains involving blueprint interactions can develop into longer chains if, for example, other systems interact with subordinate initiators while the latter is being value-determined by prime initiators. This is the case in this example.

Let's assume that in this toy universe, the interaction between A and B gives rise to B having a determinate value $s_i$ through their continuous interactions. Also, for simplicity, let's ignore the prime initiator (although the interaction Hamiltonian that describes their interactions is still non-zero and so the interactions are still going on). Updating the state concerning the result of the interaction between A and B, and given a certain Hamiltonian, we can now have the following interactions B and C from $t = 1$ to $t = 2$ represented in the following way given our conventions,

$$|s_i>_{B\ SDC'} (\alpha_2|E'_k>_C |v_k>_D +$$
$$\beta_2|E'_l>_C|v_l>_D) \to_{t=2} \alpha_2|s_k>_{B\ SDC'} |E'_k>_C |v_k>_D + \beta_2|s_l>_{B\ SDC} |E'_l>_C|v_l>_D.$$

Only when B interacts with C, C will have a determinate value. Also only then, C gives rise to D having a determinate value in interaction with D. So, C and D share an index with B and thus will now form the SDC'. We arrived at the following structure, $A \to B \to C \to D$. Note that perhaps for this situation to be more realistic, A would have to value-determine multiple systems $B_1, \ldots, B_n$ that would go on to value-determine C, and so on.

Thinking about this example and, by induction extending it to more complex examples, we can see another important role of Born probabilities mentioned in the introduction. They allow us to predict how SDCs evolve over time, giving rise to determinate values under the interactions that constitute an SDC. The Born rule will thus depend on the chances of determinate values arising in the circumstances of interactions with members of the SDCs.[63]

The deterministic EnDQT differs from the probabilistic one. According to the deterministic EnDQT, there is an underlying deterministic dynamics of the SDCs that gives rise to determinate values, and a deterministic dynamics of the UDCs that give rise to indeterminate values (except perhaps at the beginning of universe where a chancy process occurs). There are certain initial conditions that will determine the different possible histories of the evolution of the SDCs as it evolves and interacts with members of the UDCs.

The deterministic EnDQT view will therefore consider the above scenario in a different way from the probabilistic EnDQT when regarding it from a global point of view. It adopts certain global quantum states to represent and predict the dynamics of the SDCs from a global point of view. I will now sketch very roughly one possible way of how the evolution of a small and simple SDC may go in a universe U. The universe U has a certain prime initiator A, and many systems {B, C, D, E, F, G, H,…} that initially belong to a UDC, and that may or may not end up forming an SDC by directly or indirectly interacting with A. They evolve deterministically, and, given a certain Hamiltonian, they may end up interacting with an SDC. Let's assume that in the universe U, given the Hamiltonian for the SDC and the UDC we will end up with an SDC formed by systems A, B, C, and D.

The different possibilities of evolution of this possible configuration of the SDC (not of UDCs) given by systems A, B, C, and D and how it evolves deterministically can be represented via the (initial) global state below and each term of the state,

$$|\Psi(t)>_{ABCD} = \alpha |E_0(t)>_{A\,SDC'} |s_i>_B |E_0'(t)>_C |v_k>_D$$
$$+ \beta |E_0(t)>_{A\,SDC'} |s_j>_B |E_0'(t)>_C |v_k>_D$$
$$+ \gamma |E_0(t)>_{A\,SDC'} |s_i>_B |E_0'(t)>_C |v_l>_D$$
$$+ \delta |E_0(t)>_{A\,SDC'} |s_j>_B |E_0'(t)>_C |v_l>_D.$$

This state and its evolution allow us to predict and represent how the chain $A \rightarrow B \rightarrow C \rightarrow D$, as well as our uncertainty about its evolution. Since the state $|\Psi(t)>_{ABCD}$ will represent all physical possibilities of evolution of this SDC, it occurs with probability 1. For predictive and explanatory purposes, the above quantum states in a superposition are already written on the basis concerning the values that arise from the value-determining process (i.e., "via decoherence"). Since the evolution of the different possible SDCs and UDCs is deterministic, we can write the above global state.

---

[63] Note that even if this process is typically chancy, it takes a finite but non-zero time to arise of the order of the decoherence timescales.

I will explore here three possible kinds of simple options regarding the types of initial conditions, and I will illustrate them through the above toy model. Future work will develop these options further, making them more adequate, and explore other options that are perhaps better.

One simple option called *the chancy option*, involves a chancy event at the beginning of the universe that establishes how A will value-determine B in a certain way from $t_0$ to $t_1$ corresponding to each term in the superposition. The evolution of the different possible ways where A value-determines B is given by $\hat{U}_{AB}(t_0, t_1)$. Such different possible ways determine the way B will value-determine C, which in turn determines how C will value-determine D. The probabilities for the different consequences of this chancy event are given by the Born rule and are $|\alpha|^2$, $|\beta|^2$, $|\gamma|^2$, and $|\delta|^2$ for a certain possible SDC.

We also consider from $t_0$ to $t_1$ C unstably differentiates D, and this is accounted via a unitary (deterministic) evolution acting on C and D, $\hat{U}_{CD}(t_0, t_1)$. Note that this doesn't result in D having determinate values; the different options in each term of the superposition are about the different possible ways the SDCs could develop when it interacts with C, influencing how C interacts with D. Then, from $t_1$ to $t_2$, B will value-determine C, where such evolution is given by $\hat{U}_{BC}(t_1, t_2)$, and C will value-determine D. So, given the "Hamiltonian of the SDCs," (which should include terms that include the interactions with members of SDC and UDCs over time) and some "initial conditions" (which should include the possible different initial states of members of the UDCs when they meet the SDC) we can represent and predict the different possible evolutions of the SDC of our world and how it expands as it meets members of the UDCs. Note that UDCs can evolve independently of the SDC.

According to the chancy option, we can never go beyond the Born rule because there are some objective chancy events that arise from the interaction between A and B, whose probabilities are predicted via the Born rule. Moreover, we need to study how SDCs possibly evolve by looking globally at the different possibilities. Note that if we wanted to represent the evolution of UDCs independently of the evolution of SDCs, we would just proceed to use the standard QT way to analyze the dynamical evolution and interaction between systems without collapse because UDCs don't give rise to determinate values.

The chancy option spelled in this way might be considered rather ad-hoc and unsatisfactory because it doesn't give us an explanation of why certain initial conditions are selected instead of others. Perhaps more satisfactorily, instead of assuming the existence of this chancy process at the beginning of the universe in this way, an alternative is to consider certain hidden variables at the beginning of the universe that by chance the prime initiator or prime initiators have. I will call this option *the early universe hidden variables option*.

A toy example is the following, prime initiators have certain possible determinate values D=1-P arising from stably differentiated initiator quantum properties D*=1-P (which due to their specialness don't need to be value-determined to give rise to determinate values). These determinate values are represented by the local hidden variables $\lambda_\alpha$ or $\lambda_\beta$ or $\lambda_\gamma$ or $\lambda_\delta$. A chancy process assigns either $\lambda_\alpha$ or $\lambda_\beta$ or $\lambda_\gamma$ or $\lambda_\delta$ to A, where the probability is given by the Born rule.[64] So, we have an unsurpassable uncertainty about which one they have, where this uncertainty is given by the Born rule. The nature of such chancy process might be law-like or due to the nature of other physical processes. I will remain neutral about this here.

Each one of those determinate values will determine which one of the above four possible SDCs also involving B, C, and D will actually arise. $|\Psi(t)>_{ABCD}$ would then also represent our uncertainty about the evolution of this chain, with this particular configuration, which dates back to our uncertainty about these initial conditions determined by some chance event that gave rise to the system having certain hidden variables and not others.[65] These conditions and the adequate Hamiltonian will give rise to a particular history where A interacting with B gives rise to certain determinate values. This allows B to interact with C to give rise to certain determinate values, and where C will interact with D giving rise to certain determinate values over space and time, and so on. The chancy process at the beginning of the universe could have given rise to different possible chains with different sizes with certain probabilities, and this probability obeys Born rule, and one can justify the Born rule in part via this chancy event. To predict the chain that we belong to, we should use the Born rule because we are uncertain about which one of the possibilities the chancy process gave rise to, and how the SDCs evolved. There is much more to say here regarding how to justify Born probabilities, but I will leave it like this and further develop this approach in future work.[66]

---

[64] Or by some rule beyond the Born rule that would evolve into the Born rule, but let's not complicate the picture.
[65] EnDQT, may provide resources for a different notion of hidden variables to account for these dynamics. These non-standard notion of hidden variables (or "ontic states", Harrigan & Spekkens (2010) assumes determinate values only when certain systems interact in a certain way, such as in the case of value-determining interactions, and depend on a certain network of interactions to arise. We might call them environmental-based determinacist ontic states. On the other hand, relationalists appeal to a simpler account of ontic states that doesn't depend on this network, which we might call relationalist ontic states, and which make the values private. Future work should formalize the different relationalists and environmental-based determinacist ontic states and explore the possibilities of the latter more systematically. New predictions might be generated by exploring further these states states.

As I will argue in the next section, representing the determinate values of systems like A via the above local hidden variables isn't problematic. First, system A can initially have a non-entangled state for various reasons, such as being the first system, and thus cannot be entangled with anything else or some low entropy conditions that may guarantee some non-entangled states at the beginning of the universe. Second, A cannot be used to perform any Bell-like experiment since prime initiators will be inaccessible to manipulation. Also, A being the prime initiator is not value-determined by other systems, and so it cannot be measured. The further specification of the above hidden variables will be left for future work and more realistic models.

I want to make clear three points about the deterministic EnDQT. First, note that the above analysis doesn't commit deterministic EnDQT to parallel evolving worlds. Although it might be tempting to assume that the whole global state deterministically evolves, the important thing is that each possibility (where one of them is the actually occurring one) deterministically evolves, mathematically represented via the action of the unitary evolution that acts on each term of the superposition. Contrary to MWI-like views, this is also not merely selecting "one branch of the wavefunction." The interactions with the initiators over spacetime matter and the overall structure of SDCs.

Second, the above systems in $|\Psi(t)>_{ABCD}$ should be considered as occupying spacetime regions and interacting locally, and as having indeterminate values when not connected/interacting with the members of the SDC (and hence not describable via hidden variables and treated via superposition of quantum states in general). Moreover, the above state has no operational significance. No agent can perform operations in it since that would presuppose that there was an entity belonging to an SDC that wasn't represented by this state.

---

[66] Also, future work should use the decoherent-histories approach (see, e.g., Griffiths, 2019 and references therein) to represent the different possible histories of the evolutions of the SDCs.

So, according to the deterministic EnDQT, what the probabilistic QT (and standard QT) is doing with amplitudes, probabilities, state superpositions, etc., it's to predict and represent the evolution of UDCs, and what occurs when these elements interact with elements of SDCs, as well as predicting and representing the evolution of SDCs. The origin of quantum uncertainties can be understood, for example, via one of the ways mentioned above. From our condition immersed in this universe, we can't ever know in which SDC we are from the possible ones (e.g., in the case of the toy universe above, we can't know in which SDCs we are, where these SDCs are represented by $|\Psi(t)>_{ABCD}$).

Third, as I will discuss further in section 4 and in the Appendix 4, EnDQT in principle doesn't involve non-locality or action at a distance or superdeterminism or retrocausality.[67] So, the above-mentioned hidden variables aren't of that sort. We can start seeing that by noticing that systems that belong to the UDCs have indeterminate values before interacting with the elements of the SDCs (they are not describable via hidden variables).

Moreover, we should consider that the influence between systems is local, given by the Hamiltonian that represents the local interaction between systems; and the speed of this influence is lower than the speed of light, being bounded by the accepted speed of signaling between systems (which is lower than the speed of light).

For instance, let's see how we could estimate how much time it takes for A to influence D in the interactions between A, B, C, and D or what would be the speed of this influence. To estimate that, note that from the EnDQT perspective, A is not "passively receiving a signal/mark from D" when it interacts with B (which has marks of D via the interactions of B with C, and the interactions of C with D). In an indirect sense, A is also value-determining D via B. Although A initially receives a mark from D via its initial value-determining interaction with B, if this interaction continues and assuming that all the systems have determinate values in these interactions (i.e., that they continue to form an SDC),[68] the time it takes for A to continue influencing D to have a determinate value is the time it takes for A to be influenced qua receiving a signal/mark from D. This is because A has to interact with B to both value-determine B (which ends allowing C to value-determine D) but also to receive marks from D via B and C, when it interacts with B. So, the time it takes to influence the determinate value of D and the time it takes to receive a signal/mark from D are identical. Thus, plausibly, the time of influence between A and D is also the time it takes for D to become indeterminate if A stops interacting with B.

---

[67] Note also that this view doesn't go against the Kochen-Specker theorem because not all observables of B will have determinate values. Just the ones selected via the Hamiltonian, i.e., via decoherence.
[68] And that the systems with these quantum states don't change (or at least these macrosystems, allowing for their subsystems to change).

Moreover, since nothing basically changes from the first interaction between A and the rest of the systems to the later interactions between A and the other systems (in terms of how we represent this situation via unitary interactions except that there aren't decoherence times anymore), it is plausible to consider that the time of influence between A and D in the first interaction is similar to the time of influence of the next interactions, except that we need to add the decoherence times[69] to the estimated time of influence of the first interaction.

So, the process of signaling between systems in the same SDC and value-determination, and the associated time and speed of signaling (given the distance between systems, such as the distance between A and D), are related, allowing us to estimate the speed and time of the local value determination (which at least in these simplified scenarios can be done easily).[70]

Much more needs to be said about the deterministic view and how to justify Born probabilities in this view. I will leave that for future work.

Given the above conceptual apparatus, two features of EnDQT become clearer regarding the conditions for determinacy.

First, and dropping the functionalist account for simplicity,

*A system S has a determinate value v-P at time t if and only if it is interacting at t with a system E having a quantum property D\*=1-M that is a value-determining quantum property of D\*-P. In order for E to instantiate 1-M that is a value-determining quantum property, E has to interact with other members of a stable differentiation chain (SDC) that value-determine this quantum property. If S is not interacting with an E, S has an indeterminate value v-P (i.e., D=0).*

---

[69] The decoherence times would be roughly the time it takes for the coherences (mentioned in section 2.1) to go quasi-irreversibly to zero (i.e., becoming much smaller than the diagonal terms of the density operator).
[70] What happens if we close the lab when the friend is about to measure her system? Does it mean that the friend instantaneously/at infinite speeds will have an indeterminate value that she cannot give rise to determinate values? Given what I have said above, of course not. Does this mean that Wigner can then reverse the results of the friend, and we can't surpass the Wigner's friend dilemma because determinate outcomes arose inside the lab? Also no. In order for the Wigner's friend dilemma to occur (and in order for to lead to the violation of the related inequalities), the lab of the friend has to be isolated from the beginning until the end of the interaction, so that Wigner is able to treat the whole interaction as given by pure states that unitarily evolve and there isn't any uncontrollable "information leakage." This is not what happens in this scenario.

As I will explain in the next section, the friend's isolated lab is a case where a system has an indeterminate value, not interacting with an E.

Second, both directions of the Eigenstate-Eigenvalue link (EEL) are rejected completely as a criterion to assign determinate values to the observable concerning P, where the EEL is typically adopted by the "orthodox QT"[71]:

*A system S has a determinate value q of a property O represented by the self-adjoint operator $\hat{O}$ if and only if the system S is in eigenstate $|q>_S$ of the $\hat{O}$ with an eigenvalue q.*

A system can be in an eigenstate of $\hat{O}$, but that doesn't per se imply that it has a determinate value of $\hat{O}$ since that system being connected with an SDC matters. Moreover, a system can have a determinate value of some observable $\hat{O}$, but that doesn't imply that it is in an eigenstate of $\hat{O}$. It can be attributed to that system a certain reduced density operator or entangled state like the ones attributed in the case of interactions involving differentiation, which aren't eigenstates of $\hat{O}$.

One might consider strange that a system in an eigenstate of some observable doesn't have a determinate value of that observable if it's not interacting with the relevant SDC. However, note that systems are never prepared in a pure state in realistic situations, and when examining in more detail this is rather an artifact of an idealization. Using standard quantum theory, what actually happens is that the system whose state is getting prepared gets entangled with the preparation device degrees of freedom or to some other relevant degrees of freedom. This gives rise to all the coefficients in the reduced density operator of the system whose state is being prepared (tracing out the degrees of freedom of the preparation device or the relevant environment) as being approximately zero, except the coefficient that concerns the "pure state" being prepared. So, the system in a sense is still in an entangled state with some other degrees of freedom of some other system or systems. This prepared state doesn't correspond what we can assign in general determinate values of some observable precisely (only "unstably" so, hence the name unstably differentiated).[72] We might assign more generally determinate values to "larger systems," but these are not necessarily local systems. The system and the preparation device might be entangled but separated by arbitrary distances. As I have explained in section 2.1, EnDQT is focusing on local systems. Note that, given the indirect predictor strategy, we shouldn't reify this quantum state. So, we shouldn't consider that it represents a system constituted by non-locally causally connected systems.[73]

---

[71] See Gilton (2016).
[72] See Wessels (1997) for more details on this issue.

## 3. Surpassing the Wigner's friend dilemma

Which kind of SDCs whose existence in our universe we should hypothesize and where to locate their beginning is a matter of speculation, explanatory power, ability to circumvent relationalism, and other virtues. So, I think the kind of SDCs to choose should fulfill at least three natural desiderata.

The first desideratum is, of course, to circumvent the Wigner's friend dilemma. As I have explained in the introduction, the strategy adopted will be the environmental-based indeterminacist one, i.e., one which appeals to a system having absolutely indeterminate values (i.e. not varying depending on the system, agent, etc.) when the system is in the Wigner's friend situation (i.e., isolated from an external environment), while holding a unitary non-hidden variable universal QT view (see introduction). In order for this desideratum to be fulfilled two other desiderata need to be fulfilled.

The second desideratum is the following: like decoherence, SDCs should help explain the widespread success of certain aspects of classical physics in accounting for diverse phenomena in certain contexts, where importantly, these aspects involve classical physics being based on systems represented by variables that assume determinate values. According to our best science, such success apparently, (at least for now) goes back to the beginning of the universe. Even models of inflationary cosmology appeal to classical physics.

The third desideratum follows from EnDQT's use of decoherence, which should be supported by SDCs: the SDCs should justify the success of decoherence. More concretely, it should justify how real decoherence can help to account for determinate values. Real decoherence, as opposed to virtual decoherence, is tied to effective irreversibility as mentioned in the previous subsection. For instance, if the lab is open, "the information encoded via quantum states" about the initial state of the friend and/or the system or their interaction quickly and uncontrollably becomes "delocalized" due to the constant "interactions/entanglement" of the system and the friend with *many* other systems becoming inaccessible to Wigner in such a way that he cannot unitarily reverse the process via local operations.

---

[73] See also section 4 and appendix 4 for a characterization of these kinds of states.

To fulfill these desiderata, I will hypothesize the existence of certain conditions a)-e) that the elements of the DM (differentiation makeup, which includes SDCs and UDCs) in our universe should satisfy. I will call these conditions *actual universe conditions*. These conditions impose constraints in both the initial conditions (involving certain systems with certain quantum properties), on systems. and the laws of our universe. The SDCs that satisfy these conditions will be called *early universe robust SDCs (eurSDCs)*. There is a lot of flexibility concerning how these conditions should be implemented. Thus, they should be subject to certain precisifications and improvements involving more concrete, realistic, and scientifically sound models. Again, I will take something like what I hypothesize here to be approximately or roughly true.

The first type of conditions concerns the origin, persistence, and expansion of SDCs. a) eurSDCs are constituted by initiators that started interacting and value-determining other systems or getting value-determined at the beginning of the universe; where b) to maintain locality, systems that belong to eurSDCs and UDCs, aren't space-like separated from each other when they interact (thus, the Hamiltonians of interaction represent local interactions). Also, in agreement with standard QT, there is no retrocausality. So, no systems in the future can influence the past as the network evolves over time. Moreover, c) eurSDCs should expand over time to include elements of UDCs.

So, as a consequence of a)-c) plus the standard stability conditions, the interactions between initiators maintain the existence of a chain. Any systems have to connect locally (directly or indirectly) with the interacting initiators in order to give rise to systems having determinate values. These interactions started in the early universe, giving rise to chains, and some of them (i.e., the ones of the SDCs that persist[74]) at least continue until now and can continue in the future.

Other conditions concern the pervasiveness and robustness of the structure of eurSDCs over time and space and their relationship with UDCs. The d) structure of the eurSDC or eurSDCs, as opposed to the structure of UDCs, of our universe[75] is robust in the sense that the elements belonging to them are densely present throughout our universe in such a way that the value-determining process, which is represented via (the effectively irreversible) well-supported decoherence models,[76] involves systems that are stably connected with a eurSDC.[77]

---

[74] It would be interesting to see what happened with the systems that belong to the SDCs that didn't persist.
[75] One weaker hypothesis would just refer to our local region of the universe or just certain regions. This possibility deserves further exploration. I have adopted the one above because it is the simplest.

However, e) these SDCs should be such that they leave space for the independent existence of a sufficient number of systems belonging to UDCs, allowing them to evolve and persist over time without being influenced by SDCs, where members of UDCs account for the quantum phenomena in our universe. So, SDCs cannot be completely robust and pervasive. Otherwise, we wouldn't explain why we see interference and other quantum effects throughout the universe. For instance, imagine that in the toy scenario above from $t = 0$ to $t = 1$, A, B, C, and D immediately form an SDC, which stays like that forever. In that case, we wouldn't be capable of seeing certain quantum effects anymore in this toy universe. Universes like this toy universe shouldn't exist. Note that conditions like e) are implicitly assumed by other interpretations of QT and should be rather seen as natural, i.e., non-ad-hoc. Imagine that every system was decohered in a MWI multiverse, given certain initial conditions and laws. We wouldn't be able to see interference anymore. EnDQT via e) renders what is implicit by many interpretations of QT, explicit and accounts for what is left implicit via the DM. I will argue for two other conditions that eurSDCs should fulfil below, but for now, let's see how they fulfill some of the desiderata.

Given the existence of eurSDCs, we fulfill the third desideratum, justifying the success of theoretically or empirically well-supported models of "real decoherence." The many systems of the environment "that give rise to decoherence" should be connected with SDCs to explain how decoherence can give rise to determinate values during interactions.[78] So, according to this hypothesis, theoretical and empirically well-supported decoherence phenomena will be clear empirical evidence for the existence of a eurSDC.

---

[76] When the system is in an eigenstate of the measured observable, its state formally is at least locally reversible, but as mentioned above, those situations should be seen as idealized.

[77] Seeing what could happen if this hypothesis doesn't hold and what sort of predictions arrive from it deserves future exploration. See also section 4.

[78] Additionally, typically the time it takes for the effects of decoherence to be reversed in measurement-like situations is longer than the age of the universe. This means that determination can be reversed via unitary evolution, although it is a remote possibility or even an impossibility. If this is a possibility in measurement-like situations, it is a very remote one. If it turns out to occur, this (remote) possibility of reversing the determination of a quantum property in such a way that it becomes unstably differentiated is harder to justify assuming the probabilistic EnDQT view. However, it can be justified by assuming that in certain circumstances the system can probabilistically reverse back to the previous state after a long time under interactions, where differentiators become undifferentiators.

eurSDCs also justify other features of decoherence, such as its possibility of its reversal in certain conditions. Given eurSDCs, an SDC becoming a UDC is accompanied by the greater easiness of reversing the differentiation of the quantum properties of the systems that were previously part of such SDC via certain operations. This is because the systems that belong to that UDC won't be interacting anymore with the dense number of elements that constitute the SDCs, which have differentiator properties. The reversal of differentiation is plausibly modeled by the reversal of decoherence.

One moral to draw from the above discussion about the reversibility of unstably differentiated quantum properties, and that was mentioned briefly before, is the following. Typically, just when a property of a system is unstably differentiated, it can be undifferentiated, i.e., becoming less unstably differentiated. Since unstably differentiated properties have indeterminate values, keeping the terminology adopted here, the undifferentiators of quantum properties of systems should be said to typically have unstable (rather than stable) undifferentiator properties of quantum properties of systems.

Since determinate values arose in the early universe and continue arising, we can fulfill the second desiderata, which is explaining the success of classical physics dating back to the early universe and its underlying systems with determinate values. We thus can build plausible explanations about the determinate values of our world, where these explanations appeal to the origin and persistence of determinate values. They go roughly and abstractly like this: the first systems with determinate values arose in the early stages of the universe through certain blueprint interactions.[79] Moreover, these early universe initiators, due to their privileged position in the history of the universe, connect with many systems in the rest of the universe, forming widespread robust chains already in the past that tend to persist in the future and further give rise to systems with determinate values. So, determinate values persist over time because of value-determining interactions between systems. Like the origin of matter or spacetime at the big-bang, there was also the origin of determinacy. Also, like the expansion of the universe, "there is also the expansion of determinacy."[80]

---

[79] If it helps the intuition, we can conceive an SDC in the following way. Imagine a large train pushed around by hand from one car in the tip of the train. This car influences other cars to move, and so on. The hand is analogous to the blueprint interactions (i.e., the energy of the hand that gives rise to a force is generated via these interactions), the cars are analogous to the rest of the systems of the chain. The different relations of influence, i.e., between the hand and the car, or between the cars, is analogous to giving rise to determinate values, as well as the movement of the hand. This train can attach to more cars (which before interacting with the train, it's analogous to these cars belonging to a UDC), which is analogous to the expansion of the chain.

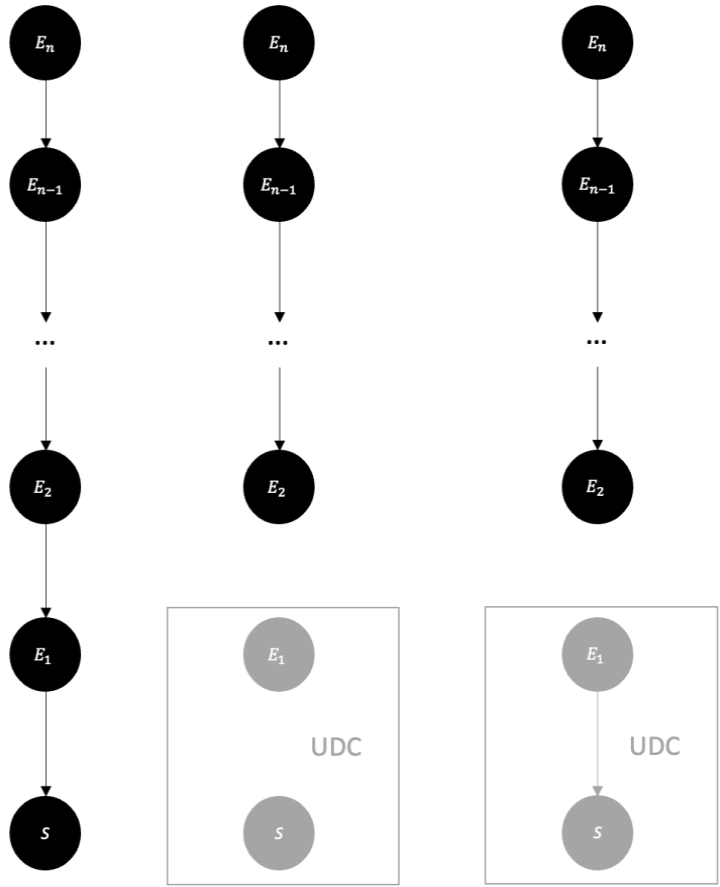

Figure 5: Representation of a Wigner's friend scenario toy example in three different situations in a region of space in a temporal interval $\Delta t$. Systems in black belong to an SDC and systems in grey belong to UDCs. The network on the left represents the open lab situation where we have an SDC with respect to certain quantum properties that leads to the value-determining of the quantum property D*-spin-z of S by the friend $E_1$ at at least one instant within $\Delta$t in a spatial. The directed dark arrow $X \rightarrow Y$ denotes a system X value-determining a quantum property of system Y. $X \rightarrow Y \rightarrow Z$ represents X value-determining a certain quantum property of Y, which allows Y to value-determine a certain quantum property of Z. If this chain represented a toy universe, $E_n$ is the prime initiator and $E_{n-1}$ is the subordinate initiator. The images in the center and right, represent the isolated lab situation, where we have only UDCs inside the lab. The grey arrow represents an unstable differentiation of the D*-spin-z of S by the friend $E_1$.

---

[80] If these analogies are physically related or are just analogies deserves future research.

The first desideratum, which is surpassing the Wigner's dilemma, is also achieved. To start seeing how it's achieved, let's look at Figure 5 on the left. This figure depicts a toy universe obeying the early universe conditions, and resembling our actual universe in the relevant aspects that I need for my argument. Let's suppose that the friend (or her measurement device) is system $E_1$, and her target system is system $S$. Let's assume that the Wigner's friend scenario occurs in the toy universe whose SDC' that exists when the lab is open and the friend is measuring her system is characterized in the following way,

$$SDC' = \Big((E_n, E_{n-1}); \ldots, E_2, E_1, S; laws;\ standard\ stability\ conditions; E_n \to E_{n-1} \to \cdots \to E_2 \to E_1 \to S\Big).$$

So, SDC' is represented by the network above, representing the local interactions between systems in spatial regions over the time interval $\Delta t$. Let's assume that given the laws and the systems above, the friend is value-determining the D*-spin-z of her system $S$, giving rise to $S$ having a determinate value in a time interval $\Delta t$. Note that $E_1$ is connected to SDC' by being connected with its elements that are outside the open lab. So, the frequent interactions with $E_2$[81] maintain $E_1$ as a member of SDC' over $\Delta t$. Keeping with our conventions, just focusing on $E_1$ and $S$ and neglecting the other systems, we would have, for example, the following quantum state representing system S instantiating D*=1-spin-z and a determinate value of spin-z,

$$\alpha_1 |E_\uparrow>_{E_1\ SDC} |\uparrow>_S + \beta_1 |E_\downarrow>_{E_1\ SDC} |\downarrow>_S.$$

In Figure 5, at the center, we have the standard initial situation in an extended Wigner's friend scenario, where we have an isolated lab with the friend and her system inside of it. The friend $E_1$ "is preparing herself" to measure $S$, however; she is disconnected from the SDC. When the lab is isolated, the SDC that previously existed inside the lab is now a UDC. Thus, $E_1$ will not be able to value-determine $S$. The interaction between $E_1$ and S in this case will result in $E_1$ unstably differentiating the spin-z of *S*. In Figure 5 on the right, I have represented that relation through a dashed arrow. In this case, we would have that the following state represents *S* instantiating D*=1-spin-z,

$$\alpha_1 |E_\uparrow>_{E_1} |\uparrow>_S + \beta_1 |E_\downarrow>_{E_1} |\downarrow>_S,$$

and S would have indeterminate values during this interaction.

---

[81] These interactions could involve interactions with the subsystems that constitute $E_1$.

Note that since in the EWFS, there is a process of unstable differentiation inside the lab of the spin-z of the target system, Wigner can reverse with more easiness the contents of the lab to the previous states involving a non-decohered system *S*.

I will now turn to the last two conditions that eurSDC should fulfill, and which allow us to circumvent the Wigner's friend dilemma. These conditions are based on the other conditions a)-e), and they are f) *inaccessibility of initiators* and g) *certain natural initial conditions at the beginning of the universe that allows for initiators with certain characteristics that circumvent extended Wigner's friend-like conditions.*

Regarding f), this structure of the SDCs can be achieved, for example, via a layered structure where the initiators are in the first layer of many layers, and that makes initiators impossible to manipulate from the other layers. It is as if there was a barrier to interact with them. More concretely, initiators in the early stages of the universe began interacting continuously. Given that plausibly the number of interactions increases from the past to the future, in the meantime many other systems interacted with the initiators, and therefore a direct interaction with them is impossible. Thus, they become inaccessible. So, I will assume that this robust chain started in the past and already has many layers constituted by many systems. As a consequence, initiators become inaccessible and cannot be manipulated because we cannot have access to them.

The reason I am assuming f) is that if we could have access to those initiators, that would be problematic for EnDQT goals of circumventing the Wigner's friend dilemma. To see why, let's consider that instead of having a universe where blueprint interactions started developing in the early inaccessible universe, we had a universe where blueprint interactions started developing in the current stage of evolution of the universe. Let's call them currentSDCs. Plausibly, in this universe, we would more likely have scenarios with labs isolated with those initiators inside, where such initiators are manipulated to give rise to extended Wigner's friend scenarios. For instance, the prime initiator could be the friend, and the target system of the friend could be a subordinate initiator. This universe would perhaps press us to assume relationalism in order to circumvent the Wigner's friend dilemma. Condition f) also allows for an unproblematic *early universe hidden variables option* and other options of that kind (see previous section). Even if prime initiators assume certain determinate values (associated with certain observables), no one can manipulate and measure them to violate Bell inequalities.

Assuming EnDQT with the standard stability conditions and conditions a)-e), there are good theoretical and empirical reasons to reject the existence of currentSDCs and accept the existence of eurSDCs obeying f). First, the SDCs that would form would render the determinacy of values of systems very fragile because one could manipulate the initiators (by for example reversing their state to a state that doesn't value-determinate the other systems that such initiators connect with and therefore "breaking" the SDC) and change the determinacy of values elsewhere. We don't have empirical evidence that that's occurring. Second, it would render decoherence as interpreted by EnDQT and QT (since decoherence can be seen as a consequence of QT) as a deficient tool to account for determinate values. A fully decohered system wouldn't be enough to guarantee determinacy since the associated SDC could be manipulated by manipulating its initiators. Given the at least apparent success of decoherence to yield determinacy, in many situations, we don't seem to live in such a universe. So, I think that condition f) inaccessibility of initiators is naturally supported by our experience and the success of QT.

Let's turn to g) *certain natural initial conditions at the beginning of the universe that allows for initiators with certain characteristics that circumvent extended Wigner's friend-like conditions.* The motivation for g) is to postulate the existence of a certain kind of initiators that circumvent certain skeptical scenarios involving the existence of extended Wigner's friend scenarios (EWFS) with these initiators. A skeptic of EnDQT could say that, although initiators are inaccessible, an EWFS could be established at the beginning of the universe when initiators started interacting before becoming inaccessible. More concretely, we could suppose we had a friend plus a system at the beginning of the universe that are initiators in an EWFS-like setup and that established blueprint interactions. Let's consider that the friends are prime initiators, and their systems are subordinate initiators. Moreover, we could have (at least) two friends and two systems, spacelike separated from each other, each pair is isolated with Wigners outside manipulating them. In that case, the friends could give rise to determinate values inside the isolated lab, which could press the adoption of relationalism to deal with the Wigner's friend dilemma.

There are multiple ways of ruling out that initiators constitute the kind of systems that could give rise to EWFS, responding to a skeptic. These ways rely on the assumption that these initiators started interacting at the beginning of the universe. One way is by the brute physical necessity of the structure of the eurSDCs. There cannot exist universes where EWFSs exist with their isolated lab features right in the beginning. If there is a possible universe like that, it would be a universe without true initiator systems and hence without SDCs.

Another way of ruling out that we can have these early-universe EWFS is by appealing to certain other familiar phenomena or features or hypotheses. For instance, consider the past hypothesis (Albert, 2000), i.e., roughly, the hypothesis that our universe started in a low entropy state. At this low entropy beginning, the higher (Boltzmanian) entropy contraptions associated with the EWFS aren't expected to be present. So, blueprint interactions should not involve EWFS-like scenarios or isolated system scenarios.

However, note that we don't need to recur to this hypothesis and the notion of entropy, as I will also discuss in section 4. As I will also discuss in that section, EnDQT with the eurSDCs even provides another way of justifying the past hypothesis (flipping the arrow). If we assume that the universe had a beginning and that the entanglement relevant to the EWFS arises through interactions, we could assume that necessarily there weren't interactions before the universe began and no system was in an entangled state. So, none of the initiators prior to interacting with each other can be described via an entangled state.

Condition g) can also provide the extra benefit of allowing for an unproblematic *early universe hidden variables option* and other options of that kind (see previous section). Even if prime initiators assume certain determinate values (associated with certain observables) prior to interacting with other systems, they are not in an entangled state, and so they don't give rise to Bell inequality violations.

As we can see, there are different plausible conditions to assume to rule out these scenarios, and thus condition g) can involve multiple possible conditions. It is an open question to establish which one is the most adequate.

From now on, when I refer to eurSDCs, I will refer to eurSDCs whose initiators are inaccessible and are systems that exist in conditions that don't give rise to EWFS at the beginning of the universe, and thus obeying actual universe conditions a)-g). The fact that doesn't give rise to EWFS can be justified, for example, via one of the ways mentioned above.[82] What if the actual universe conditions don't hold? As I will discuss in the next section, this would give rise to the exciting prospects of disconfirm EnDQT and confirm other interpretations of QT that use decoherence as a sufficient criterion for the destruction of interference or vice-versa. However, it's unclear if these conditions could be disconfirmed and how they could be disconfirmed.

## 4. Conclusion and future directions

EnDQT surpasses the Wigner's friend dilemma by considering that systems absolutely have determinate values only while interacting with other systems of an SDC. I thus have shown that we can have a coherent non-relationalist non-hidden variable unitary universal quantum theory and that the received view is misleading. Consequently, I have shown that relationalist interpretations potentially add avoidable complications and that a less costly alternative may exist. EnDQT should be taken seriously if we don't want to incur the potential costs of relationalist interpretations of QT.

As we can now see more clearly, contrary to relationalist interpretations, for EnDQT there are "non-relative/absolute facts" regarding what's happening inside the isolated lab, disagreeing with what is often recently claimed about universal unitary non-hidden variables interpretations of QT.[83] The state of Alice (or any friend inside the isolated lab) will always be absolutely described by the states that the Wigners ascribe to them. So, by adopting EnDQT, contrary to relationalist interpretations, no complications regarding different hyperplanes or structures of that kind are needed, no private or varying facts or conditions to explain the consistency between such facts are needed, nor assuming the existence of a myriad of splitting worlds. Moreover, scientific objectivity is clearly justified through an account of how systems instantiating quantum properties behave and the related D determinate values presented by EnDQT.

I haven't addressed what is like to be a friend in an isolated lab or an agent running on a quantum computer. EnDQT provides several accounts of that, some of them involving a new (quantum) extended mind hypothesis, and in appendix 2 I will begin to sketch them. I also show that some of the challenges that EnDQT faces regarding what is like to be a friend-like system are also shared with other interpretations of QT (see also appendix 3).

---

[82] At the beginning of time or in the early universe, plausibly, no system was influenced by other systems before, and prime initiators are not differentiated (or influenced) by other systems. So, a minor benefit of eurSDCs is that they ground this assumption regarding prime initiators.
[83] See, e.g., Healey (2022), Di Biagio & Rovelli (2021) and Brukner (2020, 2022).

It might seem that EnDQT was built just to deal with the extremely idealized Wigner's friend scenarios. In this section, I will provide six other examples of good reasons to take EnDQT seriously, which are still a work in progress, and that are independent of circumventing the Wigner's friend dilemma. These reasons are linked to the promising resources that EnDQT offers, and it's unclear whether relationalist interpretations can offer similar good reasons.

The first reason is that EnDQT may provide a local explanation of Bell correlations, and that can lead to a potentially appealing, clear, peaceful coexistence with relativity. For simplicity, I didn't go into detail about how EnDQT can do that. However, the tools are available. Bell's factorizability condition (assumed in certain important versions of Bell's theorem[84]) is the following,

$$P(AB|XY) = \sum_\Lambda P(\Lambda)P(A|X\Lambda)P(B|Y\Lambda).$$

It's unclear what the above expression tells us about the Bell experiment and whatever gives rise to those correlations. It's just probabilistic relations devoid of any causal meaning. A clearer way of finding the causal meaning of this condition is by deriving it from the classical Markov condition (CMC). Roughly, the CMC connects the causal structure provided by some theory represented by a DAG with certain probabilistic statements. The factorization CMC (which I will still call CMC) is the following. Let's assume that we have a DAG G, representing a certain causal structure over the variables $V = \{X_1, \ldots, X_n\}$. A joint probability distribution $P(X_1, \ldots, X_n)$ is Markov with respect to G if and only if it satisfies the following condition: For all distinct variables in $V = \{X_1, \ldots, X_n\}$, $P$ over these variables factorizes as $P(X_1, \ldots, X_n) = \prod_j P\left(X_j \middle| Pa(X_j)\right)$, where $Pa(X_j)$ are the "parent nodes" of $X_j$, i.e., the nodes whose arrows point to the "child nodes," i.e., $X_j$. The CMC together with the assumption of faithfulness[85] applied to the DAG in Figure 6, which respects relativistic causality, can be used to derive the factorizability condition.

---

[84] See, e.g., Bell (1995).

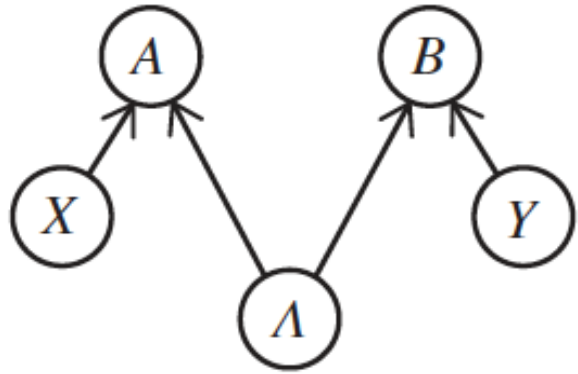

Figure 6: DAG of the common cause structure of Bell correlations, which respects relativity and is represented by classical causal models.

A and B are the variables that represent the different measurement results of Alice and Bob, X and Y are the different possible choices of measurement settings for Alice and Bob. Λ represents some "hidden" variable in the past lightcone of A and B, which like the other variables, assumes determinate values. Λ serves as a complete common cause, together with A and B, of the correlations between X and Y. This causal structure respects the so-called relativistic causality because X or A doesn't influence Y or B, and vice-versa; where A, B, Λ, X and Y concern events embedded in a Minkowski spacetime. Moreover, no other variables influence the variables A, B, X, or Y, or they don't influence anything else. So, there are no so-called retrocausal or superdeterministic causal relations.[86]

---

[85] As Shrapnel (2016) explained "[f]aithfulness (alternatively stability or no fine-tuning) states that the only conditional independencies in the distribution P are the independencies that hold for any set of causal parameters D. Another way to put this is that all the independence relations in the probability distribution over the variables in V [see text] must be a consequence of the Markov condition. The idea here is that one does not wish to allow for "accidental" independencies that are created when causal paths cancel. Faithfulness therefore licenses the inference from unconnectedness in the graph to causal independence. (…) Two variables X and Y are conditionally independent given a third Z, if the joint conditional factorizes: P(X, Y |Z) = P(X|Z)P(Y |Z)."

[86] The factorizability condition can be derived in the following way from faithfulness and the classical causal Markov conditions. From these conditions applied to the above DAG, we arrive at the following expression $P(ABXY\Lambda) = P(A|X\Lambda)P(B|Y\Lambda)P(X)P(Y)P(\Lambda)$. Then, assuming that the choices of measurement settings by Alice and Bob are independent P(XY)=P(X)P(Y), we arrive at the factorizability condition, i.e., $P(AB|XY) = \sum_\Lambda P(\Lambda)P(A|X\Lambda)P(B|Y\Lambda)$. This condition and the related inequalities are violated by certain quantum correlations, given certain non-aligned measurement settings.

But why should we believe in the CMC? We better have a good reason to believe in it, and it's unclear we should. For EnDQT, this belief arises from the metaphysical bias toward believing in a world constituted by systems that have in general determinate values. Structural equations allow for a rigorous justification and derivation of the CMC based on deterministic relationships between variables that are related to each other via these equations (Pearl, 2009; Pearl & Verma, 1995; Hitchcock, 2022). Roughly, in acyclic structural equations we have a set of endogenous variables $V_j$ (i.e., variables whose [*determinate*] values are determined by other variables in the model) that depend on their endogenous parent variables $Pa(V_j)$ (which correspond to parent nodes in the DAGs and that also assume determinate values), plus a probability distribution P' over exogenous variables $U_j$ (i.e., variables whose [*determinate*] values are determined from outside the model), establishing a deterministic relationship between them $V_j = f(Pa(V_j), U_j)$. Pearl and Verma (1995) proved that if we have a DAG G' that represents the causal structure on $V_j$, the probability distribution P on $V_j$ that results from the marginalization of the noise sources if error variables $U_i$ are probabilistically independent in P, will respect the (classical) Markov condition with respect to the DAG G'.

So, a rigorous justification of the classical Markov condition is based on systems that assume determinate values in general. However, by adopting EnDQT, we deny that those common causes in Bell scenarios have determinate values but rather hold that they have indeterminate values (and the associated undifferentiated quantum properties).[87] So we can deny the justification for classical CMC and the classical CMC to account for causal relations in general in QT, including in the Bell scenarios. More concretely, the assumptions qua rigorous justification underlying this condition amount to assuming that (isolated or non-interacting) quantum systems have determinate values that contribute to giving rise to outcomes A and B (in the DAG above), and of course, EnDQT denies that. Local common causes in the Bell scenario have undifferentiated quantum properties (and indeterminate values),[88] and thus they are represented by quantum states plus operators that represent those properties.

---

[87] See Glymour (2006) and Wood & Spekkens (2015).
[88] Note that the systems in the EPR-Bell scenarios are prepared with certain undifferentiated quantum properties that help giving rise to those correlations, such as spins in a certain direction. Systems with such quantum properties are represented via certain quantum states involving entangled systems (see Appendix 4 for an account of how EnDQT interprets those states).

In a paper in preparation (Pipa, in preparation-a, see also Appendix 4 for a sketch of the interpretation), I will argue through a realist interpretation of Quantum Causal Models (QCM, e.g., Costa & Shrapnel, 2016; Allen et al., 2017; Barrett et al., 2019) that EnDQT provides a local causal (non-relationalist) explanation of Bell correlations. Local here is in the sense of not leading to the choice of a preferred relativistic reference frame, no action at a distance (respecting relativistic causality, i.e., the causes of events are always in their past lightcone), and no non-local influence between system at each wing in the EPR-Bell scenario (even at a speed lower than the speed of light).[89] The systems prepared at the source act as common causes for Bell correlations, having indeterminate values/undifferentiated quantum properties until each system interacts with Alice and Bob measurement devices. The latter have value-determining properties of the quantum properties of these systems, giving rise to the correlated outcomes.

Instead of appealing to the CMC, quantum causal models appeal to a Quantum Markov Condition, which from EnDQT's point of view, expresses common causes of the correlations between both systems in terms of quantum states representing systems with indeterminate values. These common causes plus the determinators at each wing will be enough to explain these correlations. It perhaps should be reminded that like any correlations, Bell correlations don't imply causation, and don't imply that the common causes have to be given by systems with determinate values (or with stably differentiated quantum properties).

Note that analogously to the classical case, the Quantum Markov Condition can be derived via quantum structural models (QSM, see Barrett et al., 2019 and Suresh et al., 2023). But contrary to the classical case, according to EnDQT the systems treated by these models don't fundamentally have determinate values but rather quantum properties, being represented by quantum states evolving unitarily and self-adjoint operators. The justification for the QMC, and the denial of the CMC, comes naturally from EnDQT ontology and it's another benefit of adopting it. Note also that, as I said in the introduction, by adopting the informed indirect predictor strategy, it is not considered that the measurement of Alice affects the system of Bob and Bob, and vice-versa by looking at the quantum states that represent each system that Alice and Bob have. I am not reifying quantum states. What is important is to keep track of the determinator systems in each wing of the EPR-Bell scenario, which act locally on each system of Alice and Bob.

---

[89] Note that, since EnDQT doesn't invoke collapse or modifies the basis of QT, it can be extended to quantum field theory and respect Lorentz symmetry.

Relationalists can also reject the CMC and assume the quantum one, but the problem is that many (or all?) of them such as the MWI or RQM will have to reject that non-relative outcomes occur in the EPR-Bell scenarios, and they definitely reject that in the extended Wigner's friend scenario. Assuming EnDQT, we should think that they reject more than is needed to deal satisfactorily with the EPR-Bell scenarios. Moreover, as will also be seen in future work, the simplicity in the application of QCM by EnDQT seems to contrast with the complexity of relationalist interpretations. They are pressed to take into account the multiple perspectives/worlds when QCMs are applied to extended Wigner's friend scenarios and sometimes they also have to do that in the case of EPR-Bell's scenario (in the case of the MWI, for example). Note that these extended Wigner's friend scenarios can become very complex, involving multiple friends and Wigner's in nested labs, sharing entangled systems with other Wigners and other friends giving rise to complex QCMs. In my view, this will be another reason for wanting to adopt EnDQT, which derives from the first one.

A second good reason to take EnDQT seriously is that, as I sketched briefly in the previous section, EnDQT via SDCs and UDCs has the resources to provide a deterministic interpretation of QT that is different from the MWI, Bohmian mechanics, retrocausal, and superdeterministic views, and in principle without their issues.

One might think that in order for deterministic EnDQT to reproduce the Bell experiments, it will need to invoke some sort of superdeterminism (since EnDQT is clearly not an MWI view nor a retrocausal one). However, note that this is not the case. Superdeterministic theories assume that the following equality doesn't hold,

$$P(\Lambda|X,Y) = P(\Lambda),$$

where $X$ and $Y$ are the different measurement settings of Alice and Bob, and $\Lambda$ are the hidden variables that determine the measurement outcomes in the Bell correlation. So, it rejects that the hidden variables that determine the measurement outcomes and the measurement settings are uncorrelated.

First of all, if we consider that $\Lambda$ represents the common causes at the source of the quantum systems that affect their latter evolution, it's important to note that, as I have argued, for EnDQT it cannot be assigned in general a probability measure $P(\Lambda)$ to these common causes or an histogram (concerning certain frequencies) because they have indeterminate values (so $P(\Lambda|X,Y)$ cannot be different from something that is not well-defined). Now, we might insist that such $\Lambda$ can stand for something else beyond the common causes at the source. Perhaps it stands for the hidden variables at the beginning of the universe and the values that arise in interactions between the members of an SDC, given such variables. However, even those $\Lambda$ can be independent of the measurement settings $X,Y$ since we can base our choices of measurement settings on interactions with members of a UDC, which aren't causally connected with any SDC or the initiators, and it can exist a myriad of them (given certain initial conditions). We could even use some quantum systems belonging to UDCs that never interacted with any elements of an SDC as the basis for the choices of measurement settings. Or alternatively, use systems whose potential past interactions with members of SDCs don't matter anymore due to the interactions of these systems with many other systems that don't belong to SDCs, which screen off past influences. So, such systems belonging to UDCs would be clearly causally disconnected from the SDCs that value-determine the target systems of Alice and Bob.

The existence of these UDCs can be seen as a consequence of the actual universe conditions, more specifically, the natural condition that says that "e) these SDCs should be such that they leave space for the independent existence of a sufficient number of systems belonging to UDCs, allowing them to evolve and persist over time without being influenced by SDCs, where members of UDCs account for the quantum phenomena in our universe."

On top of the above aspects, given the size of UDCs, the elements of the UDCs that act as a common cause at the source (or that determine the preparation of the systems at the source) and the (myriad possible) elements of the UDCs that act to decide the measurement settings can be causally unrelated (in a careful Bell experiment). In other words, given the size of UDC and the long history of interactions between its elements, we don't expect any correlation between the systems that interact with the measurement device over time to "decide" the choice of measurement settings and the systems that act as a common cause are washed out. So, they cannot conspire to determine the measurement results, and no hidden variables can be attributed to them to determine the measurement outcomes. Thus, there are many factors that can guarantee the statistical independence mentioned above, and therefore we aren't facing a superdeterministic view.

A third reason is the following: a single-world non-relationalist interpretation of QT seems to provide a good basis to explain the probabilities provided by the Born rule, likely being in a better position than the MWI. In the case of the deterministic EnDQT, as I have mentioned above via the toy example, the Born rule and probabilities can be seen at least as arising, for example, from the uncertainty regarding the initial conditions that gave rise to the SDCs. However, more needs to be said, and this view needs to be further developed.

A fourth reason is that the existence of eurSDCs and the time-asymmetric direction of their evolution may at least contribute to providing unificatory explanations for the diverse temporal asymmetries (explaining them together with different features of QT).[90] For instance, it could be used to explain the temporal asymmetry of thermodynamics, i.e., the tendency of entropy to increase in an isolated system over time.[91] As briefly mentioned, the initial conditions of eurSDCs may provide support to the past hypothesis (i.e., very roughly, the assumption regarding lower entropy initial conditions of the universe), grounding it on something even more basic than the typical considerations about microstates in statistical mechanics. Moreover, with the evolution and expansion of eurSDCs (represented via the irreversible process of decoherence), more and more determinate values arise (associated with traces or records), distributed towards spacetime. Such determinate values are hard or even impossible to reverse back to being indeterminate. This evolution can provide the basis behind the entropy gradient (towards the overall increasing entropy in our universe) and therefore help explain the thermodynamic arrow.[92]

However, the initial conditions of eurSDCs may be stated independently of the past hypothesis, i.e., without appealing to a lower (Boltzmannian) entropy in the past, and so it might satisfy the skeptics of using entropy to characterize the initial conditions of the universe and explain the time asymmetries. So, EnDQT can develop its own program of (partial or total) explanation of various temporal asymmetries, independently of entropy consideration. For instance, it can help to explain the mutability arrow, i.e., "[w]e feel the future is "open" or indeterminate in a way the past is not. The past is closed, fixed for all eternity." (Callender, 2021) Following the direction of the evolution of SDCs, indeterminate values of systems become determinate as time evolves, helping to explain our psychological sense that the "indeterminate future becomes determinate."[93] Note that this unificatory potential of EnDQT stems from the actual universe hypothesis, which as I have said above, grounds quantum theory and the irreversible process of decoherence. So, EnDQT may explain both the different features of QT and the different temporal asymmetries.

---

[90] These explanations will divergence at some point depending on adoption a probabilistic EnDQT or a deterministic EnDQT.
[91] See, e.g., Callender (2021) for a review of these problems.
[92] To see this very roughly, note that the von Neumann entropy allows us to calculate the degree of determinacy of a value under interactions when a SDC expands. Moreover, in so far that the von Neumann entropy corresponds to the Boltzmannian entropy and vice-versa (see Chua, 2021 and references therein), when we have more and more systems with a high von Neumann entropy, we have more and more systems with high Boltzmannian entropy. Since entropy is an extensive quantity, we have an overall increase in entropy in space over time as SDCs evolve and expand in spacetime, starting from the lower entropy conditions.

Fifth, EnDQT with the SDCs clearly offers novel empirical posits. This distinguishes EnDQT from other views in so far as the other views don't rely on those SDCs. However, perhaps one day, under future developments of this view, EnDQT will have the advantage of being able to offer some novel predictions that other interpretations of QT don't provide. More concretely, an investigation of how different SDCs with their different structures and features impact the determinacy of values would render certain features of certain SDCs testable. More concretely, we could investigate whether or not and how certain SDCs could impact the determinacy of values and what are the most robust SDCs (in the sense of the ones that are less destructible under perturbations[94]) at the cosmological scale since it seems that those are the structures that more likely exist, assuming the eurSDCs. Remember that eurSDCs lead to widespread determinacy starting at the beginning of the universe. Then, we could make experiments or do observations to find out those structures or if they even exist.

Unfortunately, if the precise structure of eurSDCs of our world is found, it will be very hard (or perhaps impossible?) to definitely distinguish EnDQT empirically from most of the other unitary interpretations of QT because in practice all of them appeal to (irreversible) decoherence connected with certain environments.[95] Despite all this,

---

[93] Relatedly, the SDCs expansion may one day be able to explain the determinate ordering associated with temporal asymmetric processes. It is unclear that other programs that aim to explain temporal asymmetries can do the same. More concretely, throughout this paper, I have been assuming that the temporal order has "determinate values" and an associated "causal order" between events. However, in quantum gravity regimes where we may have superpositions of massive bodies, that assumption might no longer be valid due to the effects of such superposition in the resultant causal order between events or temporal order. We may have to accept an indeterminate temporal or causal order of events (see, for example, Hardy, 2009, Zych et al., 2019, Oreshkov et al., 2012). Relatedly, the order of differentiation of some quantum properties might even become indeterminate. So, we will end up with scenarios where the causal/temporal order of events is indeterminate. The determinators of this spatiotemporal/causal order will be the "measurers" of such masses, which give rise to determinate value causal/temporal orders.
[94] I.e., under arbitrary transformations on the elements of the SDC.
[95] One may also try to find deviations from the Born rule by studying how SDCs form and persist. For instance, the uncertainty regarding which initiators started the chain could deviate from the Born rule in certain scenarios. This would also make EnDQT in principle empirically distinguishable from other interpretations.

EnDQT offers a finer account of how determinacy propagates than other views since for EnDQT these connections become important, possibly offering novel predictions in the future. If this finer account ends up being further developed and empirically confirmed, it provides good support for EnDQT since the other interpretations of QT don't require that.

It would also be interesting if the actual universe hypothesis was somehow proven to be false, although it's unclear how. If it was disproven, there wouldn't exist eurSDCs. In other words, we wouldn't find SDCs that match the standard decoherence predictions in our universe. Even more interesting would be if we couldn't find potentially existing satisfactory SDCs that would substitute eurSDCs to save EnDQT. For instance, if we tried to substitute eurSDCs with other SDCs*, we find that these SDCs* are very fragile ones for EnDQT to be satisfactory, work, licensing the existence of systems that we typically associated with having determinate values. For instance, it would license the existence of planets or other macrostructures that are far removed from any possible SDCs. Unless we are willing to accept actual universes with large macroscopic systems that exist and existed having indeterminate values during long periods of time, this would render the existence of these SDCs hard to believe.

Alternatively, to save EnDQT while maintaining the actual universe hypothesis, we would perhaps need to adopt other kinds of SDCs with other stability conditions and/or initiators. However, it would be interesting if such initiators and/or stability conditions would be unsatisfactory for some reasons.

In practice, the above issues, if found, would amount to clear empirical discrepancies between EnDQT and other interpretations that rely on decoherence as a sufficient criterion for the destruction of interference; and could lead one to prefer these interpretations of QT. Note that, for instance, contrary to EnDQT, for these interpretations, there would exist regions of spacetime where interference was destroyed, and determinate values would arise, but without any existing connection between the detectors and the SDCs (such as the planets above).

EnDQT seems to have the potential to offer new predictions, and it definitely offers new empirical posits. Future work remains to be done on whether we can develop EnDQT further to offer new predictions and potentially disconfirm the existence of eurSDCs and even disconfirm EnDQT or render it not very credible.

Sixth, the interpretation underlying EnDQT can be generalized, giving rise to a framework that provides an account of ontological indeterminacy. Ontological indeterminacy is a kind of indeterminacy whose source is in the world rather than in our knowledge or representations of it. Quantum indeterminacy is an example of ontological indeterminacy.[96] This framework will allow us to understand EnDQT interpretation more broadly, fitting it into a general pattern of assignment of ontological determinacy and indeterminacy to entities in the world. Moreover, it will provide a comprehensive guide to understanding the different interpretations/interpretations of quantum theory and a new account of their ontology. In Pipa (in preparation-b) I will propose a *generative theory* of quantum indeterminacy and other kinds of ontological indeterminacy that cuts across the different interpretations/interpretations of QT. Let's start by the quantum case. Each interpretation of QT has its own *generators* of determinacy. The generators are a generalization of the initiators and the features of SDCs, applicable to other interpretations of QT. Generators include certain ontological elements and other kinds of elements (nomological, etc.) postulated by the theory that generate determinacy, giving rise to entities with determinate values or to the persistence of such entities.

In the case of EnDQT, we can consider that the generators are initiators, certain initial conditions, and/or the laws that perpetuate SDCs in interactions with other systems, generating systems having certain determinate values over time and space. In the case of Bohmian mechanics, the generators are, roughly, an initial configuration of particles with determinate value positions and the laws that evolve those particles across spacetime (or the dynamics of the wavefunction), giving further rise to particles over spacetime with determinate values of position. In the case of a certain interpretation of the MWI (i.e., one that assumes the existence of indeterminate state of affairs when there isn't branching), the generators are the initial wavefunction of the universe and certain laws that give rise to decoherence, giving rise to determinate values within each branch of the wavefunction. In the case of certain collapse theories, the generators are, for example, the wavefunction and certain laws, giving rise to flashes that have determinate values of position over spacetime, etc.

Value properties that aren't generated by these generators are indeterminate or less determinate. Different interpretations can be divided, for example, on how widespread they consider determinacy to be in spacetime and how many value properties may end up being determinate. Bohmian mechanics is one of the most extreme ones in terms of widespread determinacy in spacetime and less extreme in terms of value properties (just the position has determinate values). EnDQT is perhaps one of the less extreme ones in terms of much it considers determinacy to be widespread in spacetime. However, it is an extreme one in terms of value properties (any value property can be determinate). So, we can map the different interpretations of QT in different ways via the generator theory of quantum indeterminacy.

---

[96] See, e.g., Barnes & Williams (2011), and Calosi & Wilson (2019).

The ontological indeterminacy of other features of other *levels* can be captured via this view. On top of that, the generative theory allows for interlevel generations in the sense that the determinacy or indeterminacy at one level, such as the determinacy or indeterminacy about the future, may depend, for example, on the more fundamental quantum indeterminacy, as I have sketched above. More concretely and very roughly, in a generative theory of time, the claim would be that, as time unfolds, certain conditions generate determinacy in the present with respect to certain features observed by certain agents (which belong to an SDC), and these generators depend on certain more fundamental quantum generators.

The generative theory of indeterminacy gives rise to various underexplored questions, such as how much quantum indeterminacy and determinacy do we need to explain satisfactorily diverse phenomena in our world? What is the interpretation of QT that has sufficient resources that allow it to explain the other kinds of indeterminacy better?[97] If my argument is correct, EnDQT is a promising interpretation, potentially providing the right *amount* of determinacy and indeterminacy to understand various phenomena.

The above reasons involve just examples of potential applications of EnDQT that need to be fully developed. I hope that I have conveyed the idea that the further development of EnDQT is worth pursuing.

I should mention that in this article, I didn't provide an account of more general measurements than projective measurements, represented via Positive-Operator Valued Measurements (POVMs).[98] Future work should provide this account. As I have mentioned in section 4, it should also provide a more general account of decoherence

---

[97] In this framework, questions such as the following arise: does a fundamentally determinate picture of the world, such as Bohmian mechanics, has all the resources to explain the other ontological indeterminacies (assuming they exist) in a parsimonious inter-level way?

[98] There are different ways of accounting for them from an EnDQT point of view. For instance, such measurements typically involve the coupling of the target system S with an ancilla system A. We can roughly consider that what's happening at least in some of these measurements is that A unstably differentiates to some non-maximal degree D*' the quantum properties D*-P of S via a quantum property D*''-P' of A. Then, this quantum property of of A is value-determined to obtain information about D*'-P of S.

beyond the one provided here, i.e., where the environment dominates the evolution of the system.

## Acknowledgments

I want to thank Peter Lewis and John Symons for their moral support and valuable feedback on multiple earlier drafts. I want to thank Harvey Brown, Claudio Calosi, Ricardo Z. Ferreira, Sam Fletcher, Richard Healey, Noel Swanson, and Chris Timpson for helpful discussions or valuable feedback on earlier drafts.

# Appendix 1: More realistic Wigner's friend scenario and interference

In this appendix, I put into practice the above features of EnDQT in a more realistic interferometer-based setting. It will also show how EnDQT can account for interference phenomena through a simple example. Let's consider the situation where the friend Alice is doing an experiment with an interferometer with her lab open.[99]

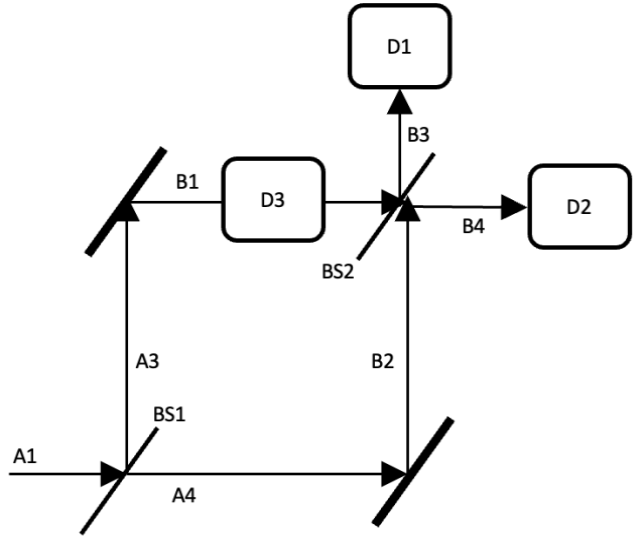

Figure 7

The electromagnetic field can be quantized, where such quantization proceeds by associating to each radiation mode a quantum harmonic oscillator and the corresponding so-called creation and annihilation operators, allowing us to express the particle number operator $\hat{N}_{Ch}$. Each of the channels of the beam splitters in the interferometer is associated to a number $Ch$. Let's consider that the eigenvalues $n_{Ch}$ of the operator $\hat{N}_{Ch}$ obtained from

$$\hat{N}_{Ch}|n>_{Ch} = n_{Ch}|n>_{Ch}$$

---

[99] This model is based on von der Linde (2021). See Cohen-Tannoudji et al. (1997) for a more extensive discussion of this framework.

represent the number of photons (the particle number) in the channel $Ch$, where each channel is associated with a radiation mode.

Now, let's consider the following states of the channels whose numbers appear in Figure 7, $|1000> = |1>_1 \otimes |0>_2 \otimes |0>_3 \otimes |0>_4$, the same in the case of $|0100>$, $|0010>$, and $|0001>$. The context will make clear whether, for example, 1 refers to A1 or B1, and so on. $|0>$ is the vacuum state.

Let's start with the case where detector D3 is not present and consider that the initial state of the quantum system inserted into channel A1 is given by

$$|Input> = |1000>.$$

This system has an unstably differentiated particle number since it won't be interacting with systems that belong to an SDC after being prepared, and so it will belong to a UDC, having an indeterminate particle number.

After the interaction with the first beam-splitter, we obtain two systems with an undifferentiated particle number each whose state is

$$|Final>_{BS1} = 1/\sqrt{2}|0010> + i/\sqrt{2}|0001>.$$

BS1 is a second-order *unstable undifferentiator of D\*-P*, where P here is the particle number:

*Second-order unstable undifferentiators of D\*-P are systems with certain quantum properties that, given certain systems with an unstably differentiated D\*-P at t, will give rise to systems with a less unstably differentiated D\*''-P and D\*'''-P' than D\*-P at t', where t' > t.*

BS1 takes systems with an unstably differentiated property, giving rise to multiple systems with unstably undifferentiated properties. How do we know that it undifferentiates them? Because they behave like the undifferentiated properties analyzed above after interacting with the beam splitter if we place certain measurement devices immediately afterward. It is unstable because we have seen above that undifferentiators in these circumstances only undifferentiates quantum properties of systems with unstably differentiated quantum properties, not stable ones. Note that although the beam splitter is typically a macroscopic object plausibly belonging to SDCs for some of its properties, a "good" beam splitter doesn't have a stable differentiator property of the particle number. Using the typical way of talking, "it doesn't decohere S."

Afterward, the two systems meet at the beamsplitter BS2, which gives rise to a system with the following state:

$$|Final>_{BS2} = |0001>.$$

BS2 has certain properties that make it a *second-order unstable differentiators of D\*-P,* where P, in this case, is the particle numbe*r property:*

*Second-order unstable differentiators of D\*-P are systems that have certain properties which, given certain systems such as S' having a not completely unstably differentiated D\*'-P and S'' having a not completely unstably differentiated D\*''-P at t, give rise to a system having an unstably completely differentiated (i.e., D\*'''=1-P') at t', where t' > t.*

We find that BS2 is an unstable differentiator because the system with the resultant quantum property behaves as an unstably differentiated property after interacting with it. One way of seeing this is that we can afterward unstably undifferentiate this property. However, if the resultant quantum property of this system was value-determined, we couldn't.

After BS2, the system will interact with the detector D2, giving rise to this system having a 1 particle-particle number property during the interaction. This later property is an example of a determinate value. Note that since the lab is open, D2 is connected to an SDC. In this article, I only mention two kinds of differentiator properties, the second-order ones and those also mentioned in previous sections, but other ones might exist.

Let me now clarify how the process of stable differentiation works by examining what happens when detector D3 is placed at B1 (see Figure 7). This detector interacts with the quantum system, annihilating the above interference phenomenon. I am going to adopt the same SDC subscripts convention that I have adopted in the last sections.

Let's consider that detector D3 is like system E in the network in Figure 5. However, now we have a situation with S having a particle number instead of a spin-z. The interactions after time t involving D3 (and omitting the interactions with D1 and D2) lead to the following state,

$$|Final(t)> = \frac{1}{\sqrt{2}}|1000>|E_1(t)>_{SDC} - \frac{1}{2}|0010>|E_0(t)>_{SDC} + \frac{i}{2}|0001>|E_0(t)>_{SDC},$$

where these interactions can be represented via decoherence.[100] How does EnDQT interpret the above phenomenon? First, note that contrary to $|E_1(t)>$, $|E_0(t)>$ concerns the inexistence of the measurement signal. It will also mean that the measurement device interacted with a quantum system, giving rise to a 0 particle-particle number property (qua the vacuum). Upon the placement of D3, there is a probability of ½ of finding a photon there. According to EnDQT, there is a probability ½ of one of the systems, which has at first an undifferentiated particle number, to interact with D3, giving rise to this system having a stable differentiated particle number and a 1 particle-particle number. The system that goes through the other channel interacts with D1 or D2, giving rise to this system having a 0 particle-particle number at D1 or D2.

Moreover, there is a ¼ probability of one of the systems with undifferentiated particle number interacting with D1 or D2 and having a 1 particle-particle number. The system that goes through the other channel interacts with D3, giving rise to this system having a 0 particle-particle number at D3.

Importantly since D3 (as well as D1 and D2, but let's not analyze them) belongs to an SDC, their value-determining properties of the particle number need to be value-determined by other determinators E'.

Let's now consider the following Wigner's friend example. Let's assume that S and E are already interacting before E', which is connected with an SDC, interacts with

---

[100] For simplicity, I will not analyze in detail decoherence in the Fock basis and assume that the Schrödinger picture is applicable. See Walls and Milburn (1995), Mcclung et al. (2010) and Myatt et al. (2000) for a detailed account. I will also assume that a notion of spatiotemporal localization of particles arises during these interactions. See Fraser (2022) for a survey of different options that consider particles as non-fundamental, but emergent.

E, making S also belong to that SDC. The interaction between S, E, and E', focusing on D*-particle number of S becoming value-determinate, is, for example, the following,

$$(\frac{1}{\sqrt{2}}|1000> |E_1(t)> -\frac{1}{2}|0010> |E_0(t)>$$
$$+\frac{i}{2}|0001> |E_0(t)>)|E'_{ready}>_{SDC}$$
$$\rightarrow \frac{1}{\sqrt{2}}|1000> |E_1(t')> |E'_1(t')>_{SDC}$$
$$-\frac{1}{2}|0010> |E_0(t')> |E'_0(t')>_{SDC}$$
$$+\frac{i}{2}|0001> |E_0(t')> |E'_0(t')>_{SDC} = |Final'(t')>$$

where $t$ is the time the particle number of S is completely unstably differentiated E. t' is the time when E' becomes entangled with E, and thus E value-determines the particle number of S. Hence, S will have a determinate value particle number when E' value-determines the relevant quantum property of E.

Note that, instead of the situation above, we could have a more realistic situation where E' is already interacting with E before t. In that case, at least non-minimally determinate values would already arise when E is interacting with S before t. Or a situation where E interacts with E' only after interacting with S. In that case, the determinate values that would arise in this interaction between E and E' would contain marks of the interaction with between E and S.

Let's now consider instead the situation where the lab of the friend is completely isolated, leaving E' outside the lab and S and E inside of it. Moreover, E still didn't interact with S. By isolating the lab, E' won't be able to continuously interact with E and value-determine D*-P of E. Thus, in this case, the systems inside the lab would be part of a UDC instead (Figure 5, center and right DAGs). The quantum state $|Final'(t')>$ isn't anymore applicable to correctly represent the situation inside the lab. We would rather have

$$|Final(t)> = \frac{1}{\sqrt{2}}|1000> |E_1(t)> -\frac{1}{2}|0010> |E_0(t)> +\frac{i}{2}|0001> |E_0(t)>,$$

but S will not have a determinate value particle number. Note that the properties of E inside the lab could be already differentiated; however, to be a stable differentiator/determinator during a time interval, they need to be frequently value-determined by determinators during that time interval.

# Appendix 2: The indeterminacist approach to the friend's phenomenal, mental, or cognitive states and AI agents running on a quantum computer

In this section, I will explain the *indeterminacist approach to the friend's phenomenal, mental, or cognitive states*. This approach involves the adoption of some hypotheses concerning the friend's phenomenal, mental, or cognitive states. This approach answers some plausible objections against EnDQT. Moreover, I will put the lessons learned in this section into practice by showing how it clarifies the results of a recently extended Wigner's friend theorem by Wiseman et al. (2023) with friends as artificially intelligent (AI) agents running on quantum computer. As I will argue, for many relationalists and other interpretations, these theorems will be considered disanalogous to the original Wigner's friend theorems because there isn't decoherence or collapse inside the quantum computer. However, for EnDQT they are analogous because for EnDQT decoherence is not a sufficient criterion for the friend to give rise to determinate values in interactions with their system. However, the indeterminacist interpretation will reframe the assumptions of this theorem.

I will now turn to three objections regarding the agent's experiences, which motivate this approach. The first objection is the following: EnDQT contrary to other popular interpretations of QT, doesn't solve what should be considered the main issue in quantum theory, which is to tell us what it is like to be the friend inside the isolated lab or what the friend *experiences* (some may even call this main issue, the measurement problem). So, EnDQT shouldn't be taken seriously as a viable interpretation of QT.

I will assume that answering this objection demands that QT helps explain what it is like to be a friend inside an isolated lab and what it is like to be a friend in a non-isolated lab. I will talk briefly about how EnDQT satisfies these two demands. Future work should go into more detail on these issues.

Let's start with the second demand. Real decoherence is associated with the expansion of the eurSDCs and their dynamics. The experience of someone outside the lab with its determinate features can be physically explainable with the help of the dynamics of eurSDCs bringing about determinacy, in so far we consider that decoherence as always happening. So, the experience of being outside the lab is the experience of being in an SDC with its dynamics and can plausibly be explained via that (more on this below).

Let's turn to the first demand, which is that EnDQT should help explain what is like to be a friend inside the lab.

Regarding this demand, in fact, EnDQT with its ontology generates certain predictions regarding the experiences of a quantum agent, which I have called the indeterminacist approach to the mental, phenomenal, or cognitive states. So, the above objection has no significant power as I will explain. The approach can be used by other interpretations of QT.

The above objection could apply as well, for example, to the MWI and collapse theories in realistic circumstances, as well as other relationalist or non-relationalist interpretations.[101] So, this objection is applicable to other interpretations. To see why, note that as it is recognized by many MWI proponents,[102] we can have branching into worlds when there is decoherence, but inside some quantum computers, we shouldn't have such branching because there isn't decoherence (at least ideally). The systems associated with an AI agent running on a quantum computer shouldn't be decohered by the internal environment (at least massively because otherwise we couldn't quantum error correct the issues caused by decoherence). Something similar could be said for the case of collapse theories. We shouldn't (ideally) have collapses inside a quantum computer (at least massively).

Let's assume the plausible and often-held hypothesis that AI agents are capable of experiencing the world or have certain cognitive or mental states that stand for it. Can the MWI or other interpretations that need decoherence for the determinacy of outcomes and collapse theories also describe or help explain what it is like *to be* agents in a superposition inside a quantum computer? It seems that these interpretations would be obliged to adopt an *indeterminacist approach to the friend's states* that I will soon present, like EnDQT.[103]

---

[101] Such as Healey (2021) (see Appendix 3) or Kent (2015), or RQM if we have multiple AI agents inside the isolated lab.
[102] See most prominently, Wallace (2012, section 10.3).

There are at least four possible hypotheses that EnDQT and other interpretations can provide regarding the experience of friend-like systems in isolated labs or that have some *parts* of the brain in a similar situation. Whether one adopts some of the hypotheses depends on considering true or false the following thesis:[104]

(*) Let's consider situations where we have a friend-like entity S (e.g., agents in isolated labs or AI agents running on quantum computers) typically, but not necessarily, capable of having certain phenomenal, mental, or cognitive states PMCS (outside an isolated lab), but
-S is instead inside an isolated lab-like environment, or
-certain quantum properties of certain systems S' responsible for PMCS (whose bearers belong to the internal body or processor) of S are not value-determined by certain appropriate systems of the environment of S (for example, systems S' are isolated where S', where S' are a part of the brain of human agent).
In these situations, S won't have these PMCS because its mental constitution, dependent on S', will have indeterminate or less determinate value properties, and these are not the kind of properties that give rise to phenomenal, or certain mental, or cognitive states. Note that these mental or cognitive states are the kind of states that functionally stand for phenomenal states in case we deny that S lacks phenomenal states (such as perhaps in the case of certain quantum AI agents running on a quantum computer or quantum robots).

---

[103] However, at least typically the MWI proponents don't claim that non-decohered systems give rise to indeterminate values or appeal to degrees of determinacy. They rather consider that there is always a determinate entity, which is the wavefunction. So, it's unclear whether they would or could adopt the quantum experience hypothesis. Thus, contrary to EndQT, they generate less precise predictions about what it would be like to be an agent running on a quantum computer experiencing things if it does experience things. Of course, MWI could transform their view to include that hypothesis, although their view is not typically built on indeterminacy. Perhaps they would consider indeterminacy and degrees of determinacy as, in some sense, emergent or just an epiphenomenon, but it's unclear if this move would render indeterminacy too weak to play any role in our experiences. Anyway, I will assume that that can be done satisfactorily. See Calosi & Wilson (2022) and Wilson (2020) for examples of adoption of indeterminacy in a MWI setting.

[104] In fulfilling these demands, we are setting aside more exotic dualist replies where the mental doesn't supervene on the physical (e.g., Albert and Lower's single mind theory, see Albert (1992) and references therein) or replies where the mental strongly emerges from the physical. In that case, we could have determinate mental, cognitive, or phenomenal states, or states of this kind that aren't in a superposition, existing in parallel to or arising from physical states that are in a superposition.

Note that (*) assumes that there is such a thing as not having determinate values, i.e., having less determinate or indeterminate values, and value-determination. However, we will see that other interpretations of QT that don't have the same ontology as EnDQT can adopt the hypotheses below. So, we could formulate another assumption (**) appropriate to the MWI where instead of value-determination, we would have decoherence and branching. So, we would consider that systems instantiate determinate values when there is enough branching and different versions of those systems are part of these branches. We could also formulate a general assumption (***), adapted and also applicable to relationalist views and other views, such as collapse theories, that use, for example, decoherence or collapse as a criterion for determinate values.

The first hypothesis, which I will call the *absent experience hypothesis* assumes that (*) or (**) or more generally (***) is true. So, according to this hypothesis, a quantum AI agent, the friend, or systems maybe capable of certain phenomenal, mental, or cognitive states when in an isolated environment or in an environment that doesn't value determine or decoheres or collapses, etc. the systems with features that determine PMCS, will lack those PMCS. So, the friend may be like a philosophical zombie, i.e., an entity lacking phenomenal states, or lacking the mental or cognitive states that functionally stand for such experience (in the case we deny that that entity has phenomenal consciousness).

Instead of the *absent experience hypothesis*, we can deny that friend-like systems *internally* lack phenomenal, or the related mental or cognitive states. Rather, these states arise through the interactions of the outputs of the friend-like system with the relevant environment (e.g., the environment external to the friend's isolated lab or the environment where there is a relevant SDC or where is decoherence in the external boundaries of the system responsible for the PMCS, etc.). So, it allows for a kind of what I will call *extended quantum mind hypothesis*. Extended mind theses are familiar in philosophy of mind,[105] and QT presents us novel versions of such theses. Extended mind theses roughly consider that not all mental states are exclusively located inside the entity that has those mental states (e.g., beliefs, desires, hopes, etc.). So, the *bearers* of those mental states can be external to the entity. The extended quantum mind hypotheses can be extended to all mental states, but here I will consider phenomenal states (as mental states), cognitive states (that in the case of a computer might not be considered mental states), and the mental states that play the functional role of phenomenal states (in the case we deny the existence of consciousness, considering it as an illusion, or that there are genuine phenomenal states).

---

[105] See, e.g., Rowlands et al. (2020) for a review and references therein. See also Clark & Chalmers (1998).

The extended quantum mind hypothesis considers that the bearers of mental states of a certain quantum agent QA running on a quantum computer QC, with a certain architecture, are the output systems/outputs[106] of QC due to QA in interactions with the external systems to QC that value-determine/decohere/branch/collapse, etc. these output systems/outputs.[107] Note that this hypothesis can be modified to admit QC with other bearers.

In the more general case, let's consider a certain entity E that is capable of having mental states but whose internal states IS are not part of a branched world or depend on systems with value-determined quantum properties or don't involve systems "concerning collapsed quantum states"[108] (or some other states analogous to these ones). The extended quantum mind hypothesis considers that the bearers of mental states of E are the output systems (or outputs) of E in interactions with the external systems E that value-determine/decohere/branch/collapse, etc. these output systems/outputs.

The third hypothesis is the *quantum experience hypothesis*, and it denies (***). The idea of this hypothesis is that cognitive, mental, or phenomenal states can depend on systems with indeterminate or not completely determinate value properties (or systems with non-decohered or non-collapsed quantum states), making them indeterminate or not completely determinate states. On the other hand, there are states that are determinate, which depend on systems with determinate values.

According to this hypothesis, friend-like systems will have experiences or whatever mental or cognitive states play the functional role of those experiences. However, such experiences or mental or cognitive states will differ from the typical (or perhaps not so typical?) ones we experience outside the lab, which arise from systems with determinate values.

A possible kind of quantum experience hypothesis (call it QE-1) would be the following: agents (like friend-like systems) would experience (or have certain mental or cognitive states concerning) quantum properties (or what arises from them) without experiencing the determinate values associated with those properties. For example, they would experience the energy of a system without experiencing the determinate values of that energy, the same for position, momentum, higher-level features (that arise from physical features), etc. (more precisely, the experience of these values will be indeterminate). An analogy might help. I can sometimes see some objects with a certain color, but it's not determinate which color the object has. According to the QE-1, the phenomenal, or relevant cognitive, or mental states of a friend-like system in an isolated lab would be affected by quantum effects in a certain way.

---

[106] Note that these outputs take into account the inputs into QC.
[107] More precisely, in a MWI perspective, the bearers of those states will be the branches that arise due to the decoherence of the output quantum system by its external environment. The content of those states will be determined by those branches that respond to certain external stimuli.
[108] E.g., mass densities or flashes.

Moreover, in non-isolated environments, these states would vary with the degree of determinacy of the values of the systems that determine them, giving rise to more or less determinate states. Using the analogy above, we could perceive the color of the object with more determinacy, but not necessarily complete determinacy. However, in this case, we would, for example, experience the position or energy of an object but with more (but not necessarily complete) determinacy. Here, I am assuming that other interpretations of QT adopt an account of degrees of determinacy or something analogous since many of them use decoherence (although, in the case of collapse theories, things are more subtle, and I am not sure they would adopt that degree-based account).

Another possible kind of quantum experience hypothesis (QE-2) would be simpler than QE-1, not considering that the above states could come in terms of degrees of determinacy. So, one would experience either a physical feature as determinate or indeterminate, and nothing in between. If QE-2 turns out to be true, this will cast doubt on the existence of degrees of determinacy, and EnDQT would perhaps need to be revised just to include only a binary determinacy.

The quantum experience hypothesis naturally leads to another extended quantum mind hypothesis. The *extended determinate quantum mind hypothesis* considers that the bearers of determinate mental states of a certain quantum agent QA running on a quantum computer QC, with a certain architecture, can be the output systems (or outputs) of QC due to QA in interactions with the external systems to QC that value-determine/decohere/branch/collapse, etc. these output systems/outputs.[109] The internal mental, cognitive, or phenomenal states are rather indeterminate, but like in the quantum experience hypothesis, it doesn't mean that it lacks them.

---

[109] In a MWI perspective, for example, the bearer of those states will be the branches that arise due to the decoherence of the output quantum system by its external environment. The content of those states will be determined by those branches that respond to certain external stimuli.

In the more general case of any entity, let's consider a certain entity E that is capable of having determinate mental states but whose internal states are not part of a branched world or depend on systems with value-determined quantum properties or don't involve systems "concerning collapsed quantum states" (or some other states analogous to these ones). So, its internal states are rather indeterminate. The extended quantum mind hypothesis considers that the bearers of determinate mental states of E are the output systems (or outputs) of E in interactions with the external systems E that value-determine/decohere/branch/collapse, etc. these output systems/outputs. There is much more to say about both the extended mind theses, but the important point is that friend-like systems may have mental states or determinate mental states.

The quantum extended mind hypotheses circumvent traditional objections against extended mind theses. One of them is the so-called cognitive bloat objection. In the traditional theses, sentences in the handbook of an agent qualify as the bearers of their beliefs. The question is where the boundaries lie for something to be a bearer of mental states. Why not the dictionary of the agent or something that they check on their smartphone?[110] However, we don't have the same worry in the case of the extended quantum mind hypotheses. The output quantum systems have to come from the internal states of the entity E and its bearers.[111] Another objection is the coupling-constitution confusion:

"On the state-oriented interpretation, a vehicle of content is extended if it is identical with, or constituted by, a structure that lies outside the biological boundaries of an individual. But a vehicle of cognition is (merely) embedded if it depends on, or is coupled with (causally or otherwise) structures that lie outside the biological boundaries of the individual without being identical with those structures. According to this second objection, arguments for extended mind only establish that vehicles of content are embedded, or coupled, with the environment rather than being extended or constituted, by the environment." (Rowlands et al., 2020).

However, note that the output interacting systems are the only bearers of determinate values or the only systems that give rise to branching structures. There is nothing like those systems or structures inside the entity E, or at least typically it shouldn't massively exit. To see this more clearly, we cannot simply couple them with the internal environment of a quantum computer at the risk of opening such quantum computer to the external environment and destroying the superposition of qubits internal to the quantum computers.[112]

---

[110] See, e.g., Rowlands et al. (2020).
[111] Even if the output systems travel large distances before interacting with, for example, an SDC.

More research needs to be done on the above hypotheses, how to develop further and perhaps test them, as well as on other possible kinds of hypotheses that EnDQT allows for (which derive from the above ones). An interesting topic to further study is whether we might have quantum experiences in situations outside Wigner's friend-like environments (such as our current everyday environment) or gaps in our experience, but we are not aware of them, not associating their source (for example) to systems with indeterminate or less determinate value properties. Finding empirical markers that would allow us to test whether the absent consciousness or the quantum consciousness hypothesis is the right one, or neither, would test these hypotheses.

Moreover, for example, there could exist hybrids between the absent experience and quantum experience theses, where certain internal cognitive or mental states X involve or depend on indeterminate values or non-decohered/non-collapsed quantum states of systems (e.g., memory, which might not be directly linked to consciousness), others not and so we might lack them at least inside the friend's lab-like environments (e.g., perhaps certain kind of phenomenal consciousness). We can also have hybrids that involve also the extended quantum mind hypotheses. A lot more could be said here. The main point is that EnDQT does generate hypotheses about what it is like to be a friend inside the isolated lab, addressing the above concerns and the above issue.

As a side note, the challenges of accounting for experiences of quantum AI agents that interpretations of QT have, such as collapse theories and the MWI, add to the well-known "lost in a huge or unfamiliar space" challenge that these interpretations need to address. This is the challenge of accounting for our experience in the everyday spatiotemporal world if we live in a world constituted by a multidimensional and/or unfamiliar wavefunction. I don't seem to live in a 3N dimensional configuration space, and I have never seen an object made of wavefunctions, which even interferes with itself. How to get the manifest image out of this ontology? Since EnDQT doesn't reify the wavefunction and systems occupy spatiotemporal regions, EnDQT has the benefit of not having to respond to this challenge. On top of these challenges, interpretations such as Bohmian mechanics and collapse theories, which privilege a certain basis (or determinate value properties) and consider them as fundamental, have the challenge of explaining how from that poor qua minimal basis (e.g., position, energy, etc.) we can have all the qualities we experience in the macroscopic world (call this, the "poor basis" challenge). Again, this challenge can't be posed to EnDQT because it doesn't privilege any basis.

---

[112] Note that the extended quantum mind theses differ from the traditional extended quantum mind theses by considering that even phenomenal states can have extended bearers. I don't see any problem with considering that. However, the extended mind thesis might be justified via individuating mental states via its functional roles (Clark & Chalmers, 1998). Moreover, some may reject the claim that phenomenal states can be individuated by their functional roles (e.g., Chalmers, 1996). There is a lot to say about this. First, it's unclear that his account requires a functionalist account of phenomenal states. It is also unclear that we can't appeal to a functionalist account. I will leave the investigation of this topic for future work.

Of course, there are all sorts of ways of responding to the challenges that I am explaining here, and it's an open question whether they are satisfactory. The point is that EnDQT doesn't even have to address them. So, taking into account all the other benefits of EnDQT, one could in fact make the argument that EnDQT is less problematic and thus preferable to the currently popular interpretations of QT.

Relationalists and other interpretations of QT might respond by arguing that EnDQT, contrary to them, has trouble explaining why if we opened the door of the lab with a human agent inside and asked them what they observed (or "apply a projective measurement on the lab onto the basis" they observed the outcome), they will provide us the answer of what they observed (e.g., spin up or spin down with 50 % of probability). Was the friend mistaken in having observed anything?

First, note that, as I have argued above, other interpretations of QT will have the same issue if the friend is a quantum computer. Moreover, the reply that EnDQT and other interpretations can provide (in the case the friend is a quantum computer) will depend on which hypotheses we adopt from the ones that are involved in the indeterminacist interpretation of the friend's states.

So, what to say about the answers of the friend or the quantum computer to our questions, including her response about the (determinate) outcomes that she observed? If we adopt the quantum experience or the absent experience hypothesis, we would consider that the friend lacks determinate phenomenal, cognitive, or mental states or lacks those states entirely. Plausibly, we would be inclined to consider that the friend is delusional about having observed an outcome, let's call this the *delusional thesis*.

An alternative is to argue that the friend-like systems are aware that they are deceiving us or were designed to deceive, and they are deceiving us that they saw a determinate outcome. Let's call this the *deception thesis*. In the case of someone that adopts the quantum experience thesis or the absent experience thesis, the quantum agent running on a quantum computer could have been designed to pretend that it has determinate experiences or any experiences at all by some designer. The deception thesis could also be adopted by someone that assumes one of the extended quantum mind hypotheses. In that case, roughly, the states where the friend says yes or no to our question regarding if they obtained an outcome would be correlated with an eigenstate of the operator that represents the "deception mental state," with eigenvalue "deception mental state." So, any answer that the friend gives could give rise to self-awareness of deception (although I would have to complicate the example a bit more to make it more realistic). Note that by assuming the extended quantum mind hypotheses, one can claim that the (determinate or not) phenomenal or mental or cognitive states of the friend that arise through interactions are their own phenomenal or mental or cognitive states. The bearers of those states arise through interactions with their environment.

Another alternative is what I will call *the translation thesis*. This thesis holds that the friend is translating what occurred inside the lab, which involves systems with indeterminate values or indeterminate phenomenal, mental, or cognitive states that depend on these systems, to what in our classical world plays a similar role in terms of determinate values. So, when we talk with the quantum friend-like system, and we ask them if they observed an outcome, we should actually take their answers as translating their indeterminate phenomenal or mental or cognitive states to determinate value-based answers that we can relate with. Moreover, the agents can learn or be programmed to be aware of that in their interactions with us (external to the lab). The replies to our questions would be correlated, for example (and simplifying a lot), with an eigenstate of the operator representing the mental state "translation." The eigenstates of this operator would have (simplifying again) eigenvalues "translating my world to your worldviews" and "ready to translate." The eigenstates whose eigenvalues are "translating my world to your worldviews" would always be correlated with the replies that the friend-like agent gives to us.

As it is clear that this thesis is more amenable to be adopted by someone that assumes the extended determinate quantum mind hypothesis since it would be plausible to consider that the friend-like systems would have some cognitive, mental, or phenomenal states before interacting with us (where we are outside the systems). Then, they would be just translating qua inferring whatever happened inside the lab to our determinate-value-based worldview. Knowing that we are talking with a quantum agent inside a quantum computer or someone trapped inside an isolated lab, what we have to do in these cases is to infer their mental, cognitive, or phenomenal states. This would be analogous to some quantum state tomography tasks since in practice we would be reconstructing the quantum states of the friend-like system via a series of measurements. In this case, we are inferring their mental, cognitive, or phenomenal states. Indeed, this is a hard communication task and interaction, but not one that is impossible or inconceivable. Future work will further develop this thesis taking into account the different architectures of quantum computers.

As a side comment, note that the extended determinate quantum mind hypothesis and the associated translation thesis have the advantage of allowing for a more friendly and ethical attitude towards any kind of friend-like or quantum agents, paving the way to quantum AI ethics and hypothetical quantum-based life (e.g., see the quantum alien example in Appendix 3). This is because consciousness or awareness frequently is an important criteria in assigning moral status (i.e., the degree in which an entity deserves ethical consideration) to an entity. We might even argue that, since, for all we know, the extended quantum mind hypotheses might be true, in the future we should grant some rights to sufficiently developed quantum AI agents, as well as friend-like agents (which could be certain quantum-based aliens, etc.). Future work should develop ethical theories which consider these quantum entities.

Some may object that EnDQT, with the above extended quantum mind hypotheses, allows for the following skeptical scenario and question: how do we know that the extended quantum mind hypothesis doesn't apply to us? Or how do we know that we don't live in a lab-like world and communicate with others as if our internal reality is determinate when it's not?

The first thing to notice is that we know from the testimony that our world is not the same as the world of a friend in an isolated lab due to the value-determination (or decoherence or massive collapse) processes occurring in it (only a skeptic would deny this). Second, if one of the extended quantum mind hypotheses is correct, we could in principle be made aware (even if over time) that our internal reality is different from the external one.

Note that relationalists in typical Wigner's friend scenarios also allow for skeptical scenarios and questions: how do we know that we are not inside a huge lab and someone is erasing our memory again and again, having the illusion that nothing happened in our lives, but in fact, something happened? This skeptical scenario doesn't apply to EnDQT because there is nothing determinate inside to be illuded about. So, relationalist views allow for other similar kinds of illusions.

The fourth objection, which is similar to the previous one, is the following: can we ever know that we are connected with an SDC, or how do we know that we are not living inside a friend's lab-like environment? EnDQT might not seem to provide a clear answer to that.

Given what I have said above, there are at least two related responses to the above objection. First, if we weren't connected with SDCs, or our measurement devices, classical physics wouldn't work. Our experiences would be very different from our classical physics-based world. We could then use QT if we (or, more rigorously, someone that we know) had access to some SDC so that we could test whether we were embedded in some UDC and what are the features of that UDC (of course, we would have to step away from this UDC to talk to that person). Second and more directly, either we would lack consciousness if the absent experience hypothesis is true, or we have some kind of quantum consciousness if the quantum experience hypothesis is true. Or we would have one of those experiences explained by one of the above extended quantum mind theses. So, we would be aware of that.

Let's now apply the lessons learned in this section to a concrete case. In principle, we could attempt to mimic the extended Wigner's friend scenario in a quantum computer. We could program an artificially intelligent (AI) agent on the computer that talks and provide us "determinate" verbal reports. Moreover, these AI agents could have experiences or thoughts, which seem to be determinate, reporting them to us. This agent could be an AI "friend" doing experiments with a quantum system inside a quantum computer. We could talk to them before erasing their memory and obtaining an outcome like in the extended Wigner's friend scenario. And like in those scenarios, such virtual agent would share a quantum system entangled with another spacelike separated (virtual or non-virtual) agent. Note that someone could again object that it seems that we have here evidence for some determinate value properties or phenomena occurring inside the quantum computer since the friend communicates with us its determinate thoughts or experiences. The above scenario was proposed and made precise through a recent theorem (Wiseman et al., 2022).

However, we can see that EnDQT has the resources to respond to it and analyze it. As we will see, this theorem fails to be analogous to the extended Wigner's friend theorem with human agents, at least for the relationalist views, since, as I have pointed out, there is no decoherence occurring inside a quantum computer. On the other hand, for EnDQT both scenarios are analogous in the sense that both human-like and virtual friend give rise to indeterminate value properties when interacting with their target systems.

Let's look more closely at the relevant assumptions of this theorem for us and interpret them via the lessons learned.[113] Besides the assumptions associated with locality and statistical independence (the so-called, local agency), one of them is that the agent in a quantum computer displays Human-Level Artificial Intelligence (HLAI).

Another one is (Wiseman et al., 2022)

"Physical Supervenience. Any thought supervenes upon some physical event(s) in the brain (or other information-processing unit as appropriate) which can thus be located within a bounded region in space-time."

Assuming one of the extended quantum mind hypotheses, we would reject this assumption without rejecting physical supervenience. So, it seems that this assumption has baked in two assumptions. One about physical supervenience and another one about internalism.

However, the authors say the following,

"The confinement of the physical events upon which thoughts supervene in space-time is necessary to be able to apply the first assumption, Local Agency, to prove the theorem. The reader may wonder whether this is in conflict with certain types of mind externalism [28] which hold that the physical processes on which a thought supervenes are not restricted to the brain, or even body, of a particular party, but may include that party's environment and even other parties with which they are interacting. In fact we think there is no conflict, because of the types of thoughts that are relevant for the [no-go theorem]. (…) That is because the thoughts we will consider are tied to observations of inputs, and are assumed to take place on a time scale of a second or less. Moreover, for the case of the "friend" in the [theorem] we will consider, this party is prevented from interacting with the physical environment external to its information-processing unit for the time during which the relevant thought would exist, were it to exist."

---

[113] The assumptions of the theorem are the following:
"1. Local Agency
2. Physicalism
3. Ego Absolutism
4. Friendliness
(which together constitute LF#); and these two technological assumptions:
5. Human-Level Artificial Intelligence (HLAI)
6. Universal Quantum Computing (UQC)."
I will explain ego absolutism afterwards. Local agency is similar to the locality assumption in Bell's theorem.

It's unclear that is on us to decide when and where the thoughts of a quantum AI occur, and which kinds of thoughts it has (this later issue is already a problem with the current more basic classic AI agents). The thoughts might occur in the interaction with the external environment, and given that the quantum agent displays HLAI, it's plausible to assume either the deception thesis or the translation thesis. So, the quantum agent could be aware that is lying to us or (without us noticing) they are translating their cognitive, mental, or phenomenal states into our worldview. One could deny that the above theses are possible, but then one would be (in my view) putting restrictions on the kind of HLAI and adding further assumptions to the theorem. One could also argue that in this theorem, there is a "getting to know" phase and that we would find out whether one of the above extended quantum mind hypotheses is correct. Is that so if the system has HLAI? If we have a HLAI, it is conceivable that they could deceive us, or we could misunderstand them in all sorts of ways. So, to make this theorem as general as possible, I suggest the following modification, which EnDQT and other interpretations (such as possibly certain relationalist interpretations) that assume one of the extended quantum mind hypotheses could accept,

Physical Supervenience II. Any exclusively internally determined thought supervenes upon some physical event(s) in the brain (or other information-processing unit as appropriate) which can thus be located within a bounded region in space-time.

Another assumption of the theorem is the following,

"3. Ego Absolutism. My communicable thoughts are absolutely real."

Where the meaning of the absolute above is the following

"For my thoughts to be real in an absolute sense means that I do not have to qualify any statements about my thoughts as being relative to anyone or anything. For example, when I have a thought "it's alive!" it is not the case that this is my thought only in this world, or that this thought's existence is a fact only for me, or a fact only relative to certain other thoughts or states or facts or systems. Rather, my thought exists unconditionally."

EnDQT would accept this assumption. Relationalists are inclined to reject it in general at least in the typical extended Wigner's friend scenario, where decoherence occurs inside the lab (more on this below). Another assumption is the Friendliness assumption:

"4. Friendliness. If a system displays independent cognitive ability at least on par with my own, then they are a party with cognition at least on par with my own, and any thought they communicate is as real as any communicable thought of my own."

What does this assumption amount to? It is not clear what is meant by "reality," and how thoughts can be compared to reality. It could mean many things. To see what it amounts to more clearly, let's see what amounts to mathematically. Imagine that a quantum system (that will be delivered to the friend) is initially in the state $\sum_{i=1}^{N=2}|s_i>_S$ (I am omitting here that this quantum system is in an entangled state with another quantum system), the virtual friend is in the initial state $|V-Friend_0>_F$ and the message that he could communicate to Wigner is in the state $|M_0>_M$. After interacting with its system, the friend sends a message with the measurement outcome to Wigner. The evolution of this process is the following,

$$\sum_{i=1}^{N=2} c_i |s_i>_S |V-Friend_0>_F |M_0>_M \rightarrow \sum_{i=1}^{N=2} |s_i>_S |V-Friend_i>_F |M_i>_M.$$

Then, a random number generator determines whether one of the following two things happen. Either Wigner doesn't read the message and erases the memory of the friend like in the above Wigner's friend scenario. Or it reads the message of the friend.[114] According to the message of the friend, Wigner infers the thoughts of the friend. If the outcome of the friend is (for example) $a$, it infers that the friend had the thought $c$. Notice that the thoughts are labeled through determinate values. I will call it a *determinate thought assumption*. Importantly, Wigner only runs the experiment if, after talking with the virtual friend, he acknowledges that the friend has determinate thoughts.[115] However, note that in this "getting to know stage," there is no large-scale entanglement occurring inside the quantum computer. Only the classical logical states of the quantum processor are being used to run a classical AI algorithm. So, it's a "normal conversation," where after the friend obtained an outcome $a$ (with 100% of probability), we may infer that the friend had the (internal) thought $c$. Note that I think that in this "getting to know stage," any of the above hypotheses can still hold, and we might be unaware of certain features of the quantum AI agent mental, cognitive, or phenomenal states.

---

[114] This would correspond to a measurement in the same basis as the state of the friend.

Using the concepts that I have been developing, the above assumption amounts to considering that the

i) The virtual friend running on a quantum computer has determinate thoughts (i.e., it has determinate internal phenomenal, cognitive, or mental states) where these thoughts are exclusively internally determined,

And that

ii) Wigner or humans or anyone that participates in this experiment has only determinate phenomenal, cognitive or mental states in general.

Both assumptions can be denied by at least collapse theorists, relationalists, or EnDQT, adopting an indeterminacist interpretation of the friend's mental, cognitive, or phenomenal states. ii) may be denied because we don't know whether humans have determinate phenomenal, cognitive, or mental states exclusively. We may have (perhaps subliminally) a kind of quantum experience like the one assumed by the quantum experience hypothesis. However, instead of focusing on the more speculative denial of ii), let's focus on i), which clearly can be denied by both EnDQT and many relationalists.

By adopting any of the hypotheses regarding the mental, cognitive, or phenomenal states of the quantum AI agent EnDQT denies i) by denying that *internally*

---

[115] Note that "independent party" above means that the thought of the virtual agent are not previously known to Wigner. This is a reasonable assumption to hold since this is an AI system that has a certain autonomy.

they have determinate cognitive, mental, or phenomenal states, or they lack such states completely. EnDQT, for example, can further add that only the output of the quantum computer needs to give rise to determinate values when it interacts with members of an SDC, which are external to the QT. The quantum computer internally doesn't need to involve processes involving determinate values. The authors of the paper allude to this idea when they explain how collapse theories deal with their theorem:

"Thus Quall-E can be put in a two-component superposition (as in our proposed experiment) and nothing in $\psi$, or the collapses, will correspond to one thought process or the other. Quall-E's thoughts are thus not real in the way that my thoughts as a human are real, and so Friendliness is rejected."

In fact, besides collapse theories, other relationalist interpretations like some versions of the many-worlds interpretation could also reject i) and the friendliness assumption.[116] For instance, as I have discussed, at least the Oxford version of MWI (e.g., Wallace, 2010) would deny that there is enough branching structure qua decoherence to have an agent that instantiates determinate thoughts. Healey (2021) would perhaps deny it because the interaction between the quantum agent and their system doesn't constitute an event where

"A necessary condition for the occurrence of a quantum event involving a system is that there be a process in the system's environment that can be modeled by the robust decoherence of states in a system's 'pointer basis', each associated with a different value of that magnitude."

No robust decoherence occurs inside of a quantum computer occurs, and so no event qua thoughts (like our thoughts) occurs.

As we can see, at least for some relationalist views that rely primarily on decoherence, we will never be able to replicate the Wigner's friend scenario in quantum computers due to their "double standards" (see the double standards challenge in the previous section). More concretely, a quantum friend-like agent is different from a human agent because there isn't sufficiently robust decoherence occurring inside of the relevant circuits of the quantum computer. So, according to this view, the authors of the above paper didn't achieve their goals because they failed this replication. However, for EnDQT, we are able to replicate the relevant aspects of the workings of a human agent via a (quasi-) isolated quantum AI agent running on a quantum computer. This perhaps provides another advantage of EnDQT because allows for the possibility of that investigation. Moreover, according to EnDQT, the authors achieved their goals of replication of the relevant aspects of the extended Wigner's friend scenario, although in a different way from what perhaps they expected.

---

[116] It's unclear how Bohmian mechanics, which also appeals to decoherence (although it doesn't provide a fundamental role for determinacy) responds to these scenarios.

# Appendix 3: Further objections, replies, and the double-standards objection to relationalist views

In this appendix, I will respond to three plausible objections to EnDQT. I will also pose a new challenge to relationalist views with the third objection called the double-standards objection.

Let's turn to the first objection. First, one may object that in considering determinacy as being dependent on eurSDCs, EnDQT renders the determinacy of values too fragile. Sometimes in the past certain interactions between initiators occurred and the persistence of those interactions, or further interactions related to those, explain the determinacy that experience now. This renders determinacy prone to fluctuations since those interactions could fluctuate.

This objection misses the point that the value-determination of quantum properties associated with SDCs is represented via real decoherence. The fact that decoherence is widespread, plausibly even affecting the systems that give rise to decoherence, constitutes good evidence that value-determination is also widespread according to the actual universe hypothesis (see section 3). It might be possible that there are fluctuations of determinacy; if that occurs, like in traditional large-scale systems in traditional decoherence scenarios such fluctuations will be barely noticeable given the eurSDCs. Notice that there can exist many connections to initiators and many systems involved in this process, so EnDQT has sufficient resources to provide an account of a stable structure. It's thus unclear that EnDQT will inevitably lead to massive fluctuations of determinacy. On the other hand, if those fluctuations occur, this would be a good opportunity to test certain consequences of EnDQT. As I have mentioned in section 4, the possibility that the determinacy of values can fluctuate and change depending on certain SDCs, or be unstable, may render EnDQT testable, as well as the different SDCs. So, it might end up being a positive scientific outcome if this happens in the sense of leading to new predictions.

Second and relatedly, someone may object that the existence of eurSDCs with X special features seems to make EnDQT rely on certain special conditions to be satisfactory, and this is undesirable. To put this objection more strongly, it seems that the fact that we don't have a relationalist view is an accident, relying on some contingent conditions assumed by EnDQT. We would be pressed to assume a relationalist view in order to deal with the Wigner's friend dilemma if we didn't have these special conditions.

The problem with this objection is that it is framed from the point of view of assuming that QT or relationalist views provide a true/empirically adequate account of the world. However, standard QT and relationalist views are problematic and might provide a false or empirically inadequate account of the world. If we consider instead that EnDQT is the right theory or could be the right theory to account for the quantum domain, the objection has no power due to its assumption. First of all, the conditions involved in the actual universe hypothesis shouldn't be considered arbitrary. Assuming EnDQT, the existence of eurSDCs with their initiators in the early universe that give rise to determinate values and that allow them to propagate shouldn't be seen as something special. They are *the source* of the existence of determinate values, and so they are intimately connected with their existence. No initiators, no determinate values.

Moreover, eurSDCs can be regarded as explaining the stability of determinacy and why standard QT and decoherence end up working on accounting for determinate outcomes. Arguably many physicists think that decoherence can deliver that account. So, we can consider that QT implicitly assumes stability of determinacy accounted by the eurSDC. We can even consider the features of eurSDCs, and the actual universe hypothesis, as a law[117] and the laws of QT depend on these features. Or we can regard it as primitive facts that are necessary for QT to work. So, in this perspective, EnDQT offers the following reading of the EWFS scenario. If we assume absolutely determinate values inside of the friend's lab, we can get a joint probability distribution for the multiple outcomes of the friends and the Wigners, and certain inequalities that QT violates. This is evidence that we shouldn't assume determinate outcomes. So, EnDQT with its eurSDCs is instead regarded as the interpretation we need to justify why QT works to give us determinate values only in certain circumstances and explain the meaning of the violation of EWFS inequalities. Thus, the X special features of EnDQT are special in so far as QT has X special features.

---

[117] Regarding the actual universe conditions as laws has some similarities with regarding the past-hypothesis as a law.

Third, being an interpretation of QT that gets rid of many worlds, modifications of unitary evolution, undesirable hidden variables, etc., EnDQT arguably provides a simpler interpretation of QT than the other interpretations. However, one may still object that EnDQT provides a far more complicated picture of the world with its actual universe conditions and standard stability conditions. Moreover, one might still complain about EnDQT radicalness, "surely the human agent Alice inside the isolated lab obtained a determinate outcome," and consider that adoption of EnDQT is not worth these costs.

However, I think that under a more careful inspection of certain scenarios, we will rather see as advantageous the extremism of EnDQT in considering that isolated labs don't allow systems to give rise to determinate values inside of it, and regard even more evident the simplicity of EnDQT. As I have said above, it is recognized by many MWI proponents that we can have branching into worlds when there is decoherence, but inside some quantum computers, we shouldn't have such branching because there isn't decoherence (at least ideally). Now, imagine that we have a human inside a large box 1, and we have a quantum computer/robot inside a large box 2.[118] Let's assume that the quantum robot is a copy of the human (we live in the world of perfect synthetic robots). Also, we will have a large box 3 where there are certain aliens with human-like capacities, which can even transform themselves into copies of humans[119] but can live in environments where there is no decoherence. Note that we don't know if these quantum alien-like entities really exist, but they are at least conceivable.

Moreover, the agent in box 1 lives in an environment where there is decoherence, and the quantum robot and alien in box 2 and 3 live in an environment where there isn't decoherence. Let's further assume that if the quantum robot or alien step outside the box, their operations will be like the ones of a classical computer or a human. All these scenarios would be basically indistinguishable for someone outside the box talking with these agents or humans or aliens, making interference experiments with them plus their systems, etc. However, in box 1 we would have decoherence and branching of the agent into two versions upon measurement of a spin-1/2 system. Whereas in the case of the environment of the quantum robot or robot, there won't be any branching into worlds (somehow) for some MWI proponents. Moreover, if the humans and the aliens went outside the box, we basically could not distinguish them from each other (unless we had extra information about the history of the aliens).

---

[118] I could formulate a fourth scenario where there is a human that is a copy of the first human but suffers no decoherence whatsoever, but I am not even sure if such scenario would constitute a remote physical possibility.
[119] Modulo the limitations imposed by the no-cloning theorem.

What is the role of decoherence in explaining what we can empirically verify and the commonalities between the robot and the human and the alien when we consider their behavior? It doesn't seem to play any role. It seems that it would be simpler to do away with the distinction between the core features of these three scenarios and consider that both the agent and robot and the alien are constituted by systems that have indeterminate values or no determinate values as EnDQT does.[120] Moreover, importantly, indeterminate values of systems can be used to explain their common behavior. This behavior involves an entity that can talk with us, but we can also implement quantum operations on this entity. Somehow mysteriously, the MWI treats them as distinct.

Something similar could be said about other relationalist views, such as RQM (or any other views that adopt decoherence). For Di Biagio & Rovelli (2021) or Adlam & Rovelli (2022), very roughly there are shared relational facts (i.e., stable facts) between systems when there is decoherence or at least it is explained via decoherence. However, we could conceivably simulate a situation with a group of quantum robots where (apparently) there are stable facts between the member of the group but without decoherence. We could also have a group of quantum aliens in an isolated box sharing apparently stable facts. Let's consider that the robots are similar to a group of human agents G, and the aliens are identical to them. It would be difficult to distinguish the situation with the human agents from the above two situations.

On the other hand, EnDQT doesn't treat these basically indistinguishable scenarios as ontologically of a distinct kind but rather establishes a commonality between them that explain their features. The "brain" apparatus of both the human agent, the robot, and the alien have indeterminate values inside the lab, and this accounts for the common behavior of both. So, applying Occam's razor, it seems that EnDQT gives a better qua more parsimonious explanation than the other interpretations about how a human agent inside an isolated lab and the quantum robot and the alien can be identical in what we do with them or how they interact with us. There are no elements that play a futile role, contrary to relationalists. Let's call this objection to relationalist views concerning their treatment of similar scenarios, *the double-standards objection*.

---

[120] We may assume that there is branching inside a quantum computer in these situations, and decoherence doesn't account for the branching inside a quantum computer, but something else. This somehow diminishes the role of decoherence to give us a criterion for branching, a natural preferred basis (without having to add it to QT), and make us wonder if the MWI preferred basis problem is solved in these situations. One of the roles of decoherence in the MWI was to solve this problem, singling out approximately a basis, and in this case there would be no decoherence. Of course, we may still regard that there is branching occurring inside of a quantum computer, and regard it as not being justifiable via decoherence but some real patterns in the underlying microphysics. See Wallace (2012) for further elaboration on this criterion. However, the justification of the existence of this non decoherence based pattern looks much more as an addition to unitary QT than something justified almost directly by it. So, many MWI proponents might not be satisfied with this strategy because it adds explicitly something to unitary QT.

Recent work from Baumann & Brukner (2023), in my view, supports the claim that the above scenarios or other kinds of scenarios are empirically indistinguishable and should be theoretically indistinguishable; as well as my broader claim that EnDQT overall provides a better explanation, in the sense of more parsimonious explanation, about what's happening inside the lab. Assuming, for example, relative facts just adds unnecessary complications (when considering the benefits of not assuming relative facts). Very roughly, they have shown that if the friend has even a very limited awareness of the change of her memory by the operations of Wigner outside the lab, an observer (which shares an entangled pair with the friend) could use this change in awareness to send superluminal signals to both the friend and Wigner. So, what's the use of postulating friends that have experiences or have a memory qua awareness of determinate outcomes? Moreover, as we have seen in the previous appendix, EnDQT is able to account for such experiences (or lack thereof).

# Appendix 4: Sketch of a non-relationalist local explanation of Bell correlations

As I have said in section 4,[121] instead of appealing to a classical Markov condition, quantum causal models (QCM) appeal to a quantum Markov condition (QMC), rejecting the classical one. According to EnDQT, this rejection is due to the appeal of the classical Markov condition to systems that participate in causal relations (or relations of influence) having determinate values/stable differentiated quantum properties. On the other hand, the QMC expresses (correctly) common causes of the correlations between both systems in terms of quantum states representing systems having indeterminate values/undifferentiated quantum properties. These common causes plus the determinators at each wing (i.e., the measurement devices of Alice and Bob) will be enough to explain these correlations.

---

[121] This appendix is a brief sketch of what is argued for in more detail in Pipa (in preparation-a). See also section 4.

Let's sketch briefly how EnDQT provides a non-relationalist local causal explanation of Bell correlations via QCM. Consider below how via the quantum Markov condition, we can give a causal explanation of Bell correlations,

$$P(x,y|s,t) = Tr_{\Lambda AB} \left( \rho_\Lambda^{UDC} \rho_{A|\Lambda}^{UDC} \rho_{B|\Lambda}^{UDC} \tau_A^{x|s\,SDC} \otimes \tau_B^{y|t\,SDC} \right).$$

Compare this with the analogous factorizability condition (see section 4). Let's interpret the above expression. Let's assume that $\rho_\Lambda^{UDC}$ is a singlet state (see introduction) assigned to systems S and to S'. It represents, together with the appropriate operators and D*, S and S' prepared at the source with certain 0-spin-p (where p is a certain axis i.e., x, y, z, in a $\theta$ angle to z, etc.) quantum properties (more on this state below). The subscript UDC is to draw attention to the fact that these systems belong to a UDC, having each undifferentiated quantum properties spin-p. The system of Alice evolves to the region A, where Alice is, via the unitary evolution (or more precisely channel) $\rho_{A|\Lambda}^{UDC}$, and the same in the case of Bob via $\rho_{B|\Lambda}^{UDC}$.[122][123] Then,

---

[122] A quantum channel is a linear map ε that is a completely positive trace preserving (CPTP) map. A map is a CPTP map if: a) it is trace preserving, i.e., $Tr(\rho) = Tr(\varepsilon(\rho))$ for all density operators ρ, b) positive, i.e., $\varepsilon(\rho) \geq 0$ whenever the density operator $\rho \geq 0$, and c) completely positive, i.e., roughly, positive even when acting on a part of a larger system. When only b) and c) are fulfilled, it is a completely positive (CP) map rather than a CPTP. A CP-map can be associated with a positive operator-valued measure (POVM). A special type of POVM is a projector operator. Instead of random variables, as in the classical causal models' case, each node X of a DAG of a QCM concerns a possible locus of interventions on a system, and it is associated with certain input and output quantum states of systems given by CP-maps. An element of a set of CP maps acts on the system, where such maps are called quantum instruments. This set gives the "possibility space" that can be associated with the different ways the properties of a system with its associated quantum state can change under local interventions or interactions, which correspond to the preparation of quantum systems, transformations, measurements on them, and so on. We represent this map as a positive semi-definite operator (i.e., density operator) via the so-called Choi-Jamiolkowski isomorphism (CJ isomorphism), representing an intervention $x$ with possible outcome $k$ on a node $A_i$ by $\tau_{A_i}^{k|x}$ (in its CJ form). The edges from the output of one node $A_i$ to (to the input of) another node $A_j$ represent causal relations between potential interventions and are given by the quantum channels (which are CPTP maps as I have said above). The latter can be represented via positive semi-definite operators $\rho_{A_j|A_i} = \rho_{A_j^{input}|A_i^{output}}$ through the CJ isomorphism. When it is written $\rho_{B|DA}\rho_{C|AE}$, what is meant is that $\rho_{B|DA}\rho_{C|AE} = \rho_{B|DA} \otimes \rho_{C|AE} = (\rho_{B|DA} \otimes I_{E^{output}} \otimes I_{C^{input}})(\rho_{C|AE} \otimes I_{B^{input}} \otimes I_{D^{output}})$, where $X^{input}$ and $X^{output}$ are the inputs and outputs of node X. A specific type of quantum channels are unitary channels that map pure states into pure states. This is the case in the example shown in the main text above. Also, above, I am using the following convention, I use the following shorthand notation $Tr_{A_{input},A_{output}}[...]$, where $A_{input}, A_{output}$ is the input system to a node $A$ and the output system to that node. Causal influence/relations in foundations of quantum theory are typically understood by the possibility of "signaling" from one node to the other even when all the relevant systems that participate in causal relations are included (and thus there aren't hidden causes). Signaling here between node X and node Y should be understood as occurring when a variation on the interventions (i.e., variations in the choice of interventions or on the CP maps) performed at node X can vary the probabilities of the outcomes k of interventions/measurements performed at node Y (maintaining the interventions at all other nodes constant). However, in these cases, we may just get rid of the agent-

each system represented via the quantum state $\rho_\Lambda^{UDC}$ is measured by Alice and Bob.[124] These measurements are represented (in the standard way) via the POVM $\tau_A^{x|s\,SDC}$ in the case of Alice, where s is her measurement choice, and x is her outcome or the determinate value of S. Analgously in the case of Bob with the operator $\tau_B^{y|t\,SDC}$. Note that the choices of measurement settings are random. Both the measurement devices of Alice and Bob have value-determining properties of the D*-spin-p. The superscript SDC means that the systems measured by either Alice or Bob will start becoming part of an SDC, even if for a brief moment in time, due to other systems that also belong to SDCs (i.e., the measurement devices of Alice and Bob). So, (cutting a long story short, see footnotes) the expression above tells us that in the EPR-Bell scenario case, we have certain systems S and S' with certain undifferentiated quantum properties at the source

---

dependent talk of signaling and also causation (in a metaphysically loaded sense), and rather consider relations of influence between certain systems qua causation (see the literature on QCM above for more details and see also Perinotti, 2021 for a discussion of why the notion of signalling and causal influence coincide in QT). As I have said in section 4, in this article I didn't provide an account of more general measurements than projective measurements, represented via POVMs. There are different ways of accounting for them from an EnDQT point of view. For instance, such measurements typically involve the coupling of the target system S with an ancilla system A. We can roughly consider that A unstably differentiates to some non-maximal degree D*' the quantum properties D*-P of S via a quantum property D*'-P of A. Then, D*'-P of A is value-determined to obtain information about D*'-P of S. Note that although quantum causal models are formulated in finite-dimensional Hilbert spaces, this is not a principled limitation, and might be surpassed in the future.

[123] The quantum Markov condition is based on a positive semi-definite operator called the process operator, which involves a series of CPTP maps (in the CJ-form, see previous footnote) associated with a DAG that represents a causal structure. Given that $\tau_{A_i}^{x_i} = \sum_k \tau_{A_i}^{k|x_i}$, the so-called process operator $\sigma_{A_1\ldots A_n}$ over nodes $A_1 \ldots A_n$ is a positive semi-definite operator that for all $\tau_{A_i}^{x_i}$ it satisfies the following normalization condition, $Tr_{A_1,\ldots,A_n}[\sigma_{A_1,\ldots,A_n}\tau_{A_1}^{x_1}\otimes\ldots\otimes\tau_{A_n}^{x_n}] = 1$. The causal structure represented by a process operator and the DAG is typically understood as representing the constraints on these signaling relations. So, node X cannot signal to node Y if node X doesn't precede node Y in the graph. However, as I have said in the last footnote, we can get rid of the notion of signaling. A probability distribution $P$ is compatible with a DAG G if and only if it satisfies the classical Markov condition. Analogously to the classical case, a process operator $\sigma_{A_1,\ldots,A_n}$ is compatible with a DAG G with nodes $A_1,\ldots,A_n$, if and only if it obeys the quantum Markov condition (QMC, Barrett et al., 2019) where this condition says that for each i in G, there is a quantum channel such that for all $i,l$ with $i \neq l$, $[\rho_{A_i|Pa(A_i)}, \rho_{A_j|Pa(A_l)}] = 0$, and $\sigma_{A_1,\ldots,A_n} = \prod_i \rho_{A_i|Pa(A_i)}$. The process operator respecting the QMC and the operators representing certain measurements can then serve as input to a certain version of the Born rule (see above) to yield the predictions of certain outcomes, given certain interventions. In the case above, the process operator is $\sigma_{AB\Lambda}^{DM} = \rho_\Lambda^{UDC}\rho_{A|\Lambda}^{UDC}\rho_{B|\Lambda}^{UDC}$, where DM designates the differentiation makeup. The requirement that the above CPTP maps in the CJ-form (or the so-called Choi matrices in a certain basis) commute (i.e., , $[\rho_{A_i|Pa(A_i)}, \rho_{A_j|Pa(A_l)}] = 0$ for all $i,l$ with $i \neq l$) can be understood via the quantum theoretical and probabilistic constraints of having positive operators and the correspondence of QMC with the classical Markov condition "in the classical limit." The CPTP maps in the CJ-form have to commute because the product of two positive operators is positive if and only if they commute. We recover the classical Markov condition from the QMC when the Choi matrices are diagonal. In that case, they will represent classical conditional probabilities, and the QMC reduces to the classical one, where the multiplication of the diagonal choices can be interpreted as multiplying conditional probabilities in the usual sense (Thanks to Aleks Kissinger for helping me understand this). For EnDQT, the classical Markov condition arising from the QMC is seen as being due to the value determination of quantum properties (which is represented via decoherence and helping us explain why the choi matrices become diagonal). This amounts to the UDCs in QCMs becoming SDCs. So, the QMC help us represent and understand the causal relations between systems with quantum properties, the classical Markov condition allow us to represent and understand the causal relations between systems with the determinate values. One relates with the other via value-determination (Pipa, in preparation-a).

[124] The CPTP maps in the EPR-Bell scenario are identity channels that transport systems from the source to regions A and B.

at time t plus certain determinators of S and S' (Alice's and Bob's measurement devices with their different settings) that have a complete influence in S and S' having certain determinate values in regions A and B at certain future times (increasing the probabilities that these determinate values will arise). The above expression will represent what locally concretely occurs in these scenarios.

We can further represent this situation via the following DAG, where in grey we represent the UDC and in black the SDC:

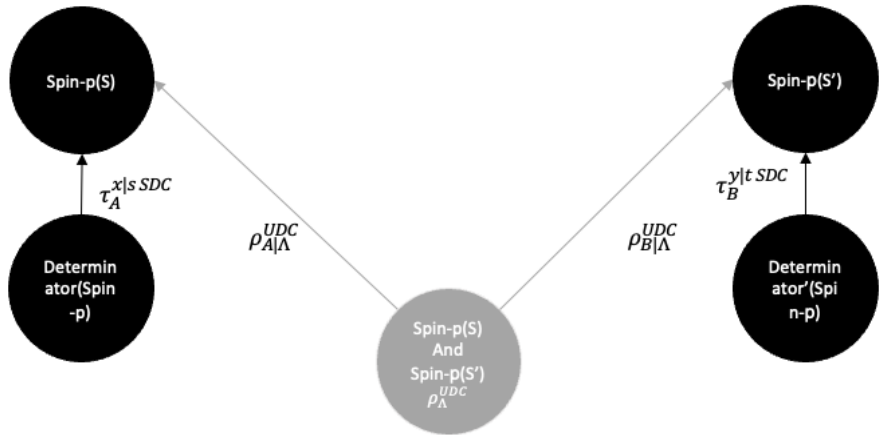

Figure 8: DAG of the common cause structure of Bell correlations, which respects relativity and is local, being represented by quantum causal models as interpreted by EnDQT

Note that by adopting EnDQT's view of quantum states, we don't consider that the measurement of Alice on the system represented by $\rho_\Lambda^{UDC}$ affects the system of Bob and Bob, and vice-versa. Moreover, the channels $\rho_{A|\Lambda}^{UDC}$ and $\rho_{B|\Lambda}^{UDC}$ allow us to represent the local evolution/dynamics of systems with quantum properties between spacetime regions, which are subsequently measured by Alice and Bob; and how systems with certain undifferentiated quantum properties influence certain determinate values arising in the future. More generally, via the quantum Markov condition, channels of this kind and the related interventions allow us to represent and make inferences about concrete systems with quantum properties, how, and why they change due to certain influences.[125]

---

[125] We can also represent relations of influence via quantum structural models mentioned in section 4. Furthermore, these models allow for the derivation of quantum Markov condition, capitalizing on the following no-influence condition applied to the CJ transformation of unitary transformations: given a unitary transformation $\rho_{ZK|XY}^U$ that takes system X and Y, and outputs system Z and K, X *doesn't* influence K if and only if for all inputs $\rho_Y$, we have

As we can see, the differentiation makeup (which includes the SDC and UDC) provides the agent-independent explanandum that QCMs locally explain, allowing for a realist interpretation of QCM (see also footnotes). There is no non-locality at play here. There is just the rejection of the classical Markov condition for reasons mentioned in section 4, and the acceptance that the quantum Markov condition, for reasons also mentioned in section 4, is the proper condition to represent causal relations between systems with certain quantum properties.

I want to end by leaving a brief remark regarding the rejection of the classical Markov condition by EnDQT. This condition allows us to provide causal explanations based on probabilities of events E that factorize conditioned on certain events E' (decorrelating the events E). A particular instance of this principle is factorizability condition (see section 4). However, EnDQT rejects this condition by rejecting the idea that we can perform such conditionalization based on complete causes that in general have determinate values. To further understand why EnDQT rejects the classical Markov condition and the factorizability of probability distributions to provide causal explanation and still allows for local causal explanations, let's see how (assuming EnDQT) we can understand the state non-separability qua non-factorizability of entangled quantum states, such as $\rho_\Lambda^{UDC}$ assigned to each system in Bell scenarios. This state represents the common causes at the source of these correlations. It might be puzzling what this state represents and how it represents quantum properties at the source.

As we have seen, a quantum state like this $\rho_\Lambda^{UDC}$ that is assigned to a single system S and represents systems in an entangled state is understood (in a certain basis) as representing S with a quantum property, which can be correlated with the quantum properties of other systems such as S' in circumstances where determinate values arise. According to EnDQT, these correlations are due to their common past and certain systems with value-determining quantum properties of certain quantum properties of S and S'. Assuming the indirect predictor perspective, these quantum states don't concern the possible violation of locality (as I have argued above) or some ontologically non-

---

some channel $\rho_{K|Y}$ that is independent of $\rho_X$, or equivalently $Tr_Z \rho_{ZK|XY}^U = \rho_{K|Y} \otimes I_X$. From these conditions, we can get the conditions for when X can influence K, which is when the above condition doesn't hold for an X and a K. As it was acknowledged by Barrett et al., (2019) this particular condition was examined, for example, by Schumacher & Westmoreland (2005) and demonstrated that it's equivalent, for example, to the following condition: for an arbitrary choice of $\rho_B$, if we use the product state $\rho_A \otimes \rho_B$ as input for the unitary channel, then $\rho_D = Tr_C (U \rho_A \otimes \rho_B U^\dagger)$, which is the marginal state for system D at the output, will remain the same regardless of the choice of $\rho_A$. Essentially, they conclude, this means that it's impossible to send signals from A to D by changing the input state at A, for any choice of $\rho_B$. Note that as I expressed in a previous footnote, we don't need to appeal to signaling.

separable or holistic reality constituted by space-like separated systems.[126] Instead, since a) we understand systems with quantum properties via their possible determinate values, which are represented via quantum states (being eigenstates of some observables) and observables, and b) since systems don't have determinate values in general, except in certain circumstances, c) it's plausible to assign to systems states that can only be related with determinate values in certain circumstances. More concretely, in certain circumstances, they can represent possible correlations between systems S and S' with certain quantum properties when S and S' become differentiated, where these correlations (as we have seen) are due to their common past and possible interactions with certain other systems in the future.

So, the non-separability of states $\rho_\Lambda^{UDC}$ is unsurprising when adopting EnDQT. It's just a consequence of our way of correctly representing quantum properties and the associated indeterminate values via determinate values (associated with certain eigenstates of observables) and correctly predicting the behavior of systems instantiating these properties. Another way of putting it is that the non-separability for EnDQT is a consequence of top-down heuristic considerations between quantum properties and determinate values. These considerations are typical in science when considering different domains like in our case, and they don't impede $\rho_\Lambda^{UDC}$ of representing common causes with indeterminate values/undifferentiated quantum properties that participate only in local causal relations and non-holistically.

Perhaps it will also be helpful to contrast this state assignment with some empirically inadequate quantum state assignments to the common causes of the correlations. If in an EPR-Bell scenario, we instead assigned to each system a separable state representing a superposition, such as a superposition of different spin directions, this would rather represent wrongly a system with a certain unstably differentiated spin direction as a cause and yield wrong predictions. Even worse if we assigned to them a classical state, which would represent a system with determinate values and a stably differentiated quantum property independently of its interactions with certain systems in its environment.

---

[126] See Healey & Gomes (2022) for a review on these notions. As they explain, "[o]ne may try to avoid the conclusion that experimental violations of Bell inequalities manifest a failure of Local Action by invoking ontological holism for events. The idea would be to deny that these experiments involve distinct, spatiotemporally separate, measurement events, and to maintain instead that what we usually describe as separate measurements involving an entangled system in fact constitute one indivisible, spatiotemporally disconnected, event with no spatiotemporal parts."

Again, as it will also be seen in work in preparation (Pipa, in preparation-a), the simplicity in the application of QCM by EnDQT seems to contrast with the complexity of relationalist interpretations. In general, they are pressed to take into account the multiple existing and incompatible perspectives/worlds if they apply QCMs to analyze Wigner's friend scenarios and sometimes even simpler Bell's experiments (such as in the MWI case). So, standard QCMs for them are very incomplete because they don't take into account these perspectives. According to EnDQT we don't need to take into account these multiple perspectives, not considering them as very incomplete, which is another advantage of EnDQT.